\documentclass[12pt]{article}
\usepackage{graphicx}
\usepackage{epsf,epsfig,amsmath,amssymb,verbatim,mathrsfs}
\usepackage{epstopdf}
\begin{document}
\pagestyle{empty}

\begin{center}
{\Large\bf TASI Lectures on a Holographic View of\\ 
\vspace{0.2cm} Beyond the Standard Model Physics}

\vspace{1cm}

{\sc Tony Gherghetta}\footnote{Based on lectures given at TASI 2009 ``Physics of the 
Large and the Small", June 1-26, 2009, Boulder, Colorado.}

\vspace{0.5cm}

{\it\small School of Physics, University of Melbourne, Victoria 3010, Australia\\
{\tt tgher@unimelb.edu.au}}
\end{center}

\vspace{0.5cm}

\vspace{1cm}
\begin{abstract}
We provide an introduction to the physics of a warped extra dimension 
and the AdS/CFT correspondence. An AdS/CFT dictionary is given which
leads to a 4D holographic view of the 5th dimension. With a particular emphasis
on beyond the standard model physics, this provides a window into 
the strong dynamics associated with either electroweak symmetry breaking 
or supersymmetry breaking. In this way hierarchies associated with either
the electroweak or supersymmetry breaking scale, together with the 
fermion mass spectrum, can be addressed in a consistent framework.
\end{abstract}

\newpage
\setcounter{page}{1}
\setcounter{footnote}{0}
\pagestyle{plain}

\tableofcontents

\section{Introduction}
Extensions of spacetime where fields propagate in a warped extra dimension 
have recently provided an alternative way to naturally generate large hierarchies 
of physical scales. Provided that the extra dimension is suitably stabilized,
a large separation of scales can be fixed and related to the curved geometry of the 
5th dimension. In this way the location or ``geography" of fields in the extra dimension
determines local physical scales. This greatly motivates using the warped dimension
to study hierarchies in the standard model of particle physics.

In the standard model large hierarchies are associated
with the origin of mass. While the Higgs mechanism provides a simple
mass-generating solution it requires an elementary scalar particle, the 
Higgs boson. The spin-0 property of the Higgs boson causes quantum 
corrections to its mass-squared to be quadratically sensitive to the ultraviolet 
(UV) cutoff scale. Given that the logical choice for this cutoff is the (reduced) 
Planck scale, $M_P=2.4\times 10^{18}$ GeV, (associated with the quantum theory 
of gravity such as string theory), there is an inevitable extreme fine-tuning to obtain 
$m_{Higgs} \ll M_P$, as suggested by electroweak precision tests. This quadratic 
sensitivity on the cutoff can be elegantly eliminated by introducing 
supersymmetry, but this shifts the problem into now determining why the scale 
of supersymmetry breaking is much lower than the Planck scale. Even if these 
problems associated with the Higgs boson are solved, there still remains
the question of how to explain the fermion mass hierarchy. The charged leptons and 
quarks require a Yukawa coupling hierarchy ranging from an electron coupling of 
$10^{-6}$ to a top-quark coupling $\sim 1$ and the problem exacerbates with the 
inclusion of neutrino masses. A warped 5th dimension provides a natural setting 
with which to address the gauge hierarchy and fermion mass hierarchy problems 
simultaneously!

A second, perhaps more fascinating motivation for studying the warped 5th dimension 
is provided by the AdS/CFT correspondence~\cite{malda}. This remarkable conjecture, which has its 
origins in string theory, can be used to give a purely 4D ``holographic" description of the 
warped 5th dimension in terms of a strongly-coupled conformal field theory (CFT). In fact 
an AdS/CFT dictionary can be derived that provides a consistent mapping between the 
two descriptions. Using this dictionary, precise calculations of electroweak observables performed rm arx	
on the gravity side can be reinterpreted on the gauge theory side as due to strong dynamics. 
The holographic view even allows for a reinterpretation of the seemingly ``new" geometrical 
solutions of the hierarchy problems in terms of strong dynamics in the gauge theory. So in 
the end the warped 5th dimension need not be real and merely provides a new mathematical 
tool that allows for a precision study of a particular class of models with strong dynamics.

The primary aim of these lectures is to provide a theoretical introduction to the 
warped 5th dimension and the corresponding holographic picture via the AdS/CFT correspondence.  
A major focus will be to explain the AdS/CFT dictionary in some detail so that even though the 
application will be to physics beyond the standard model, the mathematical tools and ideas in this 
review can apply to other gauge/gravity settings. In order to cover this extensive ground
some introductory material will be lightly covered. Fortunately, a number of reviews on extra 
dimensions already exist in the literature and these lectures will rely upon and expand on 
some of the topics already covered. Previous reviews on extra dimensions include those in 
Refs.~\cite{vr, gg, cc,apl, raman, chm, Rattazzi:2003ea}, as well as the TASI 2009 lectures by Hsin-Chia 
Cheng.~\cite{Cheng} Warped dimensions are specifically covered in Refs~\cite{cc, raman, chm, cg, tgreview} 
and the TASI 2009 lectures by Roberto Contino give an introduction to composite Higgs 
models~\cite{Contino:2010rs}. Finally, phenomenological aspects and implications for the LHC are not 
covered extensively here, but can be found in Refs.~\cite{chm,jh,Kribs:2006mq,Davoudiasl:2009cd,Davoudiasl:2009sk,Rizzo}.

\begin{figure}
\begin{center}
\includegraphics[width=0.5\textwidth,height=0.16\textheight]{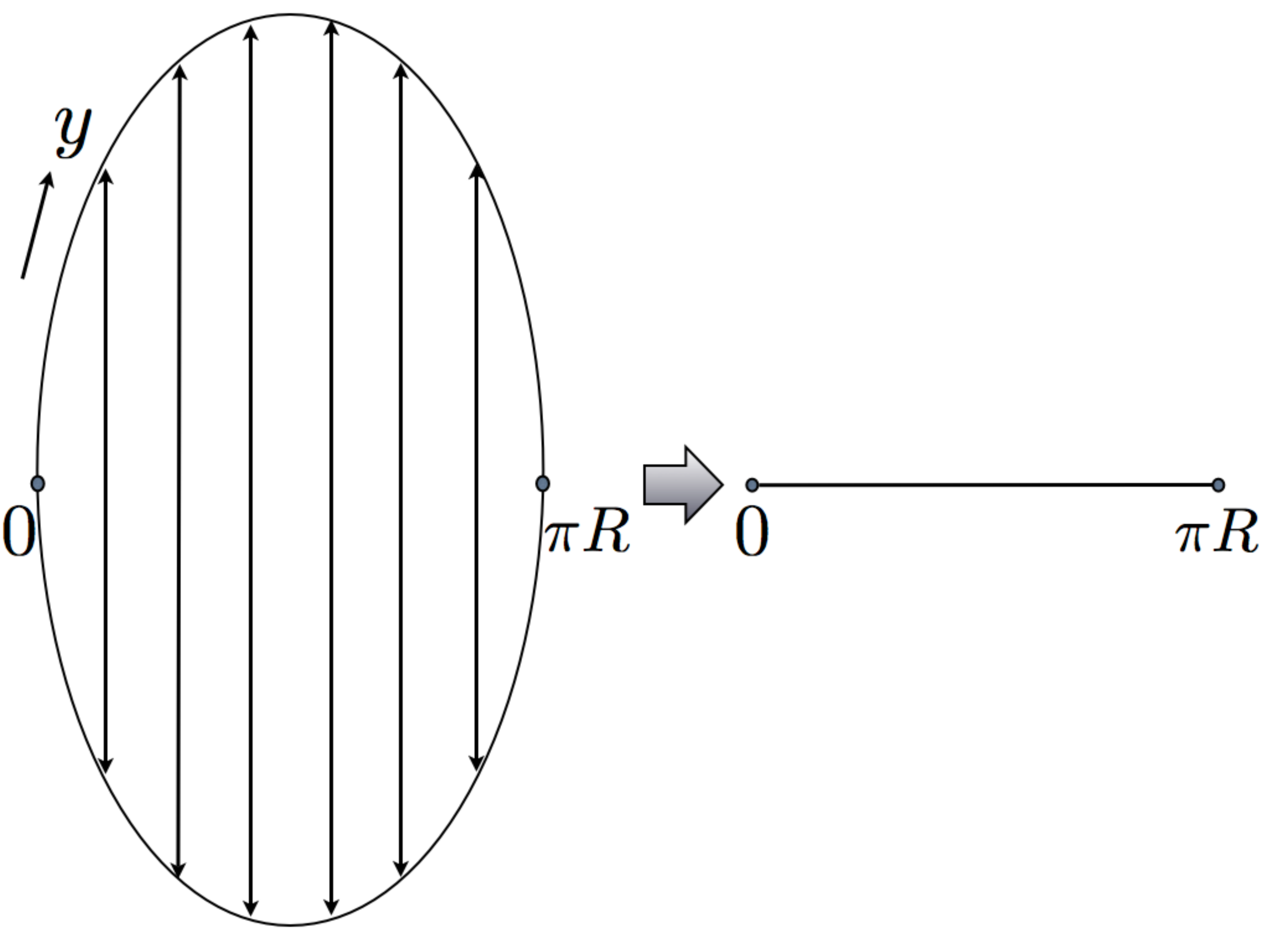}
\end{center}
\caption{ A line segment (or $Z_2$ orbifold) obtained from the identification of 
opposite points on the circle.}
\label{orbifoldfig}
\end{figure}

\section{Warping the 5th Dimension}
Consider a 5D spacetime $x^M=(x^\mu,y)$ where the spacetime indices are labelled
by $M=(\mu,5)$ with $\mu = 0,1,2,3$. To have avoided experimental detection the fifth 
dimension $y$ must have finite extent, and the simplest possibility is to assume a periodic 
geometry such as a circle, $S^1$ of radius $R$ where $y\leftrightarrow y+2\pi R$. However, 
compactifying fermions on a circle does not allow one to describe chiral fermions 
in the Standard Model.  Instead the 5th dimension is compactified on a line segment, which 
can be thought of as resulting from identifying opposite sides of the circle as depicted in Fig~\ref{orbifoldfig}.
This is known as a $Z_2$ orbifold $(S^1/Z_2)$, with $Z_2$ representing the identification 
$y\leftrightarrow -y$. Two 3-branes, known as the UV (IR) brane are located at the endpoints of the 
orbifold $y=0, (\pi R)$.

\begin{figure}
\begin{center}
\includegraphics[width=0.45\textwidth,height=0.28\textheight]{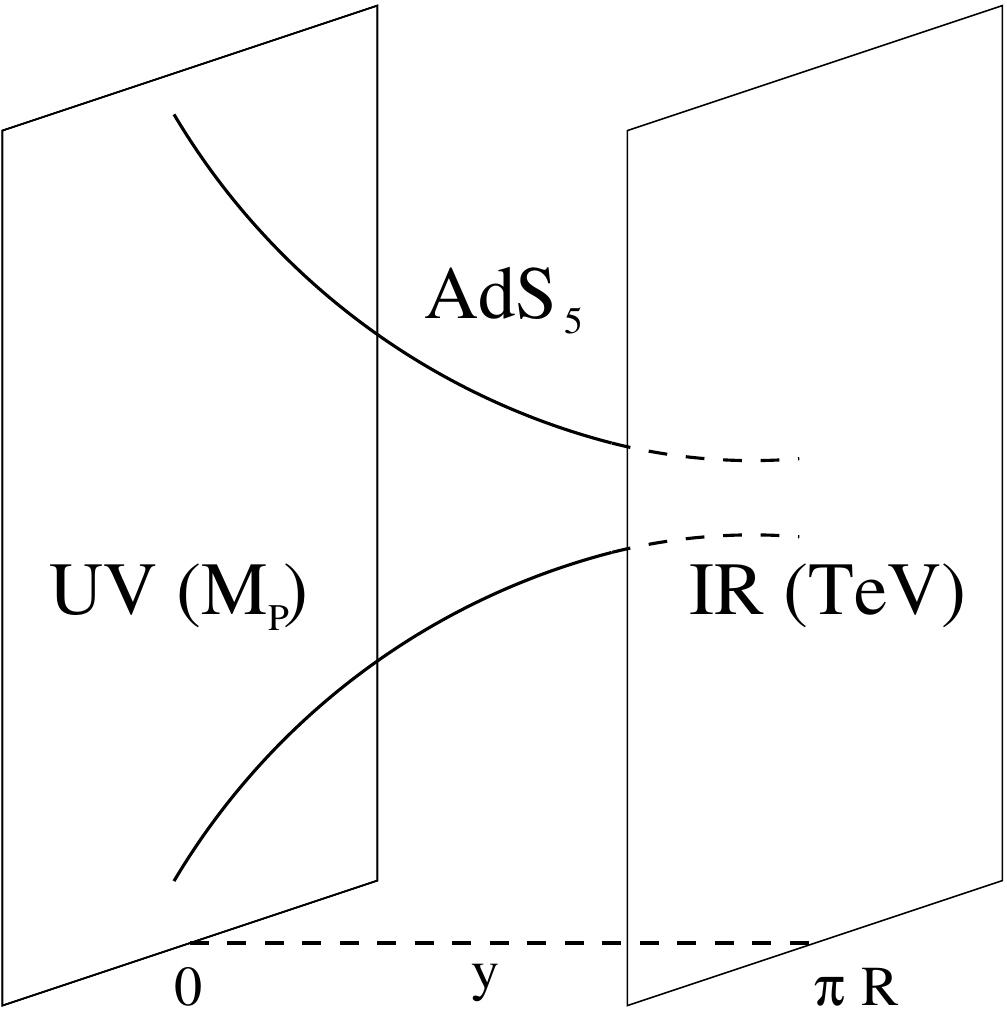}
\end{center}
\caption{ A slice of AdS$_5$: The Randall-Sundrum scenario.}
\label{sliceAdSfig}
\end{figure}

The 5th dimension is then ``warped" by introducing a bulk cosmological constant $\Lambda_5$, 
i.e an energy per unit (spatial) volume in 5D spacetime. A zero 4D cosmological 
constant is obtained by adding brane tensions on the two 3-branes and appropriately tuning
their values with $\Lambda_5$. The solution of Einstein's equations for this configuration 
assumes a negative bulk cosmological constant, $\Lambda_5 <0$ which means that the warped 
geometry is anti-deSitter (AdS) space. The 5D metric solution is given by
\begin{equation}
\label{adsmetric}
    ds^2 = e^{-2ky} \eta_{\mu\nu} dx^\mu dx^\nu + dy^2 \equiv g_{MN} dx^M dx^N,
\end{equation}
where $k$ is the AdS curvature scale and $\eta_{\mu\nu}={\rm diag}(-+++)$ is 
the 4D Minkowski metric.
This slice of AdS$_5$ with $0\leq y \leq \pi R$ is the Randall-Sundrum solution~\cite{rs1} (RS1) 
and is depicted in Fig.\ref{sliceAdSfig}.

The slice of AdS$_5$ provides a simple low-energy effective field theory description
below the Planck scale, but it behooves to contemplate the possible underlying UV theory. 
In string theory, configurations with a similar effective 5D geometry can be obtained
when ten-dimensional string theory is compactified on a six-dimensional manifold. Typically the
internal six dimensions are of order the Planck length $M_P^{-1}$. However, turning on 
various nontrivial configurations of $p$-form fields in the internal manifold leads to quantized 
integrals of the field strengths known as ``fluxes". Compactifying with fluxes can then cause one of 
the internal six dimensions to become much larger than the Planck length~\cite{gkp}, giving rise to
a ``warped throat" geometry which plays a similar role to the bulk spacetime between the UV 
and IR brane. The UV brane is effectively the remaining Planck-sized internal geometry, and 
the smooth tip of the throat represents the IR brane. These flux compactifications are highly nontrivial but
provide a suitable UV completion of the simple brane-world setups that will be considered in these
lectures. It remains a relatively unexplored research area to construct explicit
warped throat solutions that realize the beyond the standard model scenarios 
in these lectures.

\subsection{Gauge hierarchy problem}

To see how the warped geometry can help to explain hierarchies let us consider 
the gauge hierarchy problem, i.e. why $m_{Higgs}\ll M_P$. 
In RS1 the Standard Model particle states are confined to the IR brane. In particular let 
$H$ be a complex scalar field, representing the Higgs doublet, with the action:
\begin{eqnarray}
      \label{Higgsaction1}
      S_H&=& -\int d^5x \sqrt{-g} \left[ g^{\mu\nu} \partial_\mu H^\dagger \partial_\nu H 
      - M_5^2 |H|^2+\lambda |H|^4\right] \delta(y-\pi R),\\
      &=&  -\int d^4x \left[ e^{-2\pi k R} \eta^{\mu\nu} \partial_\mu H^\dagger \partial_\nu H 
      - M_5^2 e^{-4\pi k R} |H|^2 + \lambda e^{-4\pi k R} |H|^4\right]~.
      \label{Higgsaction2}
\end{eqnarray}
A factor of $\sqrt{-g}$, where $g\equiv{\rm det}\, g_{MN}$, has been included in the Lagrangian 
(\ref{Higgsaction1}) because for a curved background $d^5 x\sqrt{-g}$ is the invariant volume element
under general coordinate transformations.
In the slice of AdS$_5$ the Higgs mass, $M_5$ represents a value near the 5D cutoff scale, as 
expected for a scalar field. The second line in (\ref{Higgsaction2}) is obtained by using the metric 
(\ref{adsmetric}) and performing the $y$ integration. The result is the usual 4D action for the Higgs 
field except that the kinetic term is not canonically normalized to one. This can be achieved by 
rescaling the field $H\rightarrow e^{\pi kR} H$ leading to 
\begin{equation}
	S_H = -\int d^4x \left[\eta^{\mu\nu} \partial_\mu H^\dagger \partial_\nu H
      - (M_5 e^{-\pi k R})^2 | H |^2 +\lambda |H|^4 \right]~.
\end{equation}
The Higgs mass parameter is now identified as $M_5 e^{-\pi k R}$. The original mass parameter 
has been scaled down or redshifted by an amount $e^{-\pi k R}$ which is due to the fact that the Higgs 
boson is confined to the IR brane at $y=\pi R$. Instead if the Higgs were confined to the UV brane at 
$y=0$ there would be no redshift factor. So we see that location in the warped dimension determines 
the local physical scales. The hierarchy problem is now easily solved. The physical Higgs mass
$m_{Higgs}\propto M_5 e^{-\pi k R}$. Assuming $M_5\simeq k \simeq M_P$, 
the radius $R$ can be chosen so that $m_{Higgs}\simeq$ TeV. This requires $\pi k R \simeq 35$ or 
$R\simeq 10 M_P^{-1}$.

It is important to note that generating the large hierarchy relied on the exponential warp factor in the 
metric (\ref{adsmetric}). However, this factor can be eliminated by a simple change of coordinates such as 
$z=e^{ky}/k$, seemingly nullifying the large hierarchy. Of course there is no contradiction since the 
generation of the hierarchy relies on an underlying stabilization mechanism for the fifth dimension. 
In the slice of AdS$_5$, the Goldberger-Wise mechanism~\cite{gw} provides a suitable way to stabilize 
the separation between the UV and IR branes. This separation is governed by a modulus field-- a scalar 
field with zero potential. To fix the separation of the branes, a potential for the modulus field is 
generated by a bulk scalar field with quartic interactions localized on the UV and IR branes. The 
minimum of this potential then yields a compactification scale, $R$ that solves the hierarchy problem 
without any severe fine-tuning of parameters. Clearly a change of coordinate systems will not alter the 
minimum of the potential. We will not delve further into the stabilization mechanism since more details 
can be found in Ref.~\cite{gw}. The main point is that the slice of AdS 
is always stabilized by the Goldberger-Wise (or other) mechanism. Incidentally in the string theory 
warped throat construction the length of the throat is stabilized by fluxes. This effectively 
behaves like a Goldberger-Wise mechanism~\cite{bht}.

More generally we see that any mass scale on the IR brane is redshifted by an 
amount $e^{-\pi k R}$. Since the SM is confined to the IR brane in RS1, this affects 
higher-dimension operators with dimension greater than four, such as those associated with proton 
decay, flavor changing neutral currents (FCNC) and neutrino masses which are now suppressed 
by the warped-down scale
\begin{eqnarray}
       \frac{1}{M_5^2} \bar\Psi_i\Psi_j\bar\Psi_k\Psi_l& \rightarrow&
        \frac{1}{(M_5 e^{-\pi kR})^2} \bar\Psi_i\Psi_j\bar\Psi_k\Psi_l ~,\\
        \frac{1}{M_5}\nu\nu H H &\rightarrow& \frac{1}{M_5 e^{-\pi k R}} 
    \nu\nu H H~,
\end{eqnarray}
where $\Psi_i$ is a Standard Model fermion and $\nu$ is the neutrino. This leads to generic 
problems with proton decay, FCNC and also neutrino masses without further complicating 
the model by introducing discrete symmetries to forbid these terms.

Instead a much simpler solution is to note that only the Higgs field needs to be localized on the 
IR brane in order to address the gauge hierarchy problem. Therefore the Standard
Model fermions and gauge fields can actually propagate in the bulk.~\cite{gf1,gf2,gn,chnoy,gp,dhr}
In this way the UV brane can be used to provide a sufficiently high
scale to help suppress higher-dimension operators while simultaneously solving 
the gauge hierarchy and fermion mass hierarchy problems~\cite{gp}. But before explaining how 
this is done we need to consider fermionic and bosonic fields propagating in the warped 
dimension. Like a flat extra dimension, a 5D field propagating in a warped extra 
dimension leads to a Kaluza-Klein tower of 4D fields except that we need to determine 
the spacing of the Kaluza-Klein tower and the wavefunction profile of the 4D mode in the extra dimension. 
This mass spectrum of 4D modes will be the experimental signature of a warped extra dimension

\subsection{Bulk Fields in a Slice of AdS}
We will consider scalar, fermion and vector bulk fields propagating in a slice of AdS$_5$. 
These fields are assumed to have negligible backreaction on the background geometry 
$(\ref{adsmetric})$. In general the equation of motion for the bulk fields is obtained by
requiring that $\delta S_5 =0$ where $S_5$ is the bulk action. This variation of the action 
can be written in the generic form 
\begin{equation}
\label{var5daction}
      \delta S_5 = \int  d^5 x~\delta\phi~({\cal D}\phi)
      +\int  d^4 x~ \delta\phi~({\cal B}\phi) \big |_{y^\ast}~,
\end{equation}
where $\phi$ is any bulk field. Requiring the first term in (\ref{var5daction})
to vanish gives the equation of motion ${\cal D}\phi=0$. However the second 
term in (\ref{var5daction}) is evaluated at the boundaries $y^\ast$ of the 
fifth dimension $y$. The vanishing of the second term thus leads to the 
boundary conditions $\delta\phi|_{y^\ast}=0$ or ${\cal B}\phi|_{y^\ast}=0$. 
The solution of the equation of motion will need to satisfy either of these conditions.
Note that there are also boundary terms arising from the orthogonal directions $x^\mu$, 
but these are automatically zero because $\phi$ is assumed to vanish at the 
4D boundary $x^\mu \rightarrow \pm \infty$.

\subsubsection{Scalar fields}
Consider a bulk complex scalar field $\Phi$ whose action to quadratic order 
is given by
\begin{equation}
     S_\Phi =-\int d^5 x\sqrt{-g}\, \left[ |\partial_M \Phi|^2 +m_\Phi^2 |\Phi|^2 \right],
     \label{scalaraction}
\end{equation}
where $m_\Phi^2 \equiv a k^2$ is a bulk mass parameter defined in units 
of the curvature scale $k$ with dimensionless coefficient $a$. 
The equation of motion derived from the variation of the action (\ref{scalaraction}) is
\begin{equation}
        \Box \Phi + e^{2 ky}\partial_5 (e^{-4k y} \partial_5 \Phi) -m_\phi^2 e^{-2 ky}\Phi = 0,
    \label{scalareqn}
\end{equation}
where $\Box = \eta^{\mu\nu}\partial_\mu\partial_\nu$ and $\partial_5=\partial/\partial y$. 
The boundary terms vanish provided
\begin{equation}
       (\delta\Phi^* \partial_5\Phi) \big|_{0,\pi R}=0.
       \label{scalarBC}
\end{equation}
To solve (\ref{scalareqn}) we assume a separation of variables 
\begin{equation}
     \Phi(x^\mu,y)=\sum_{n=0}^{\infty} \Phi^{(n)}(x^\mu) f_\Phi^{(n)}(y)~,
     \label{scalardecomp}
\end{equation}
where $\Phi^{(n)}(x^\mu)$ are the 4D Kaluza-Klein modes satisfying the Klein-Gordon
equation $\Box \Phi^{(n)}= m_n^2 \Phi^{(n)}$ with masses $m_n$, and $f_\Phi^{(n)}(y)$ is the 
bulk profile of the Kaluza-Klein mode. Substituting (\ref{scalardecomp}) into (\ref{scalareqn}) 
leads to an equation for the profile 
\begin{equation}
      -\partial_5 (e^{-4k y} \partial_5 f_\Phi^{(n)}) +m_\Phi^2 e^{-4 ky}f_\Phi^{(n)} = 
      m_n^2 e^{-2 ky}f_\Phi^{(n)}.
      \label{scalarprofileeqn}
\end{equation}
The differential equation (\ref{scalarprofileeqn}) actually has the form of a Sturm-Liouville equation 
\begin{equation}
   -\frac{d}{dy}\left(p(y) \frac{d f_\Phi^{(n)}}{dy}\right) + q(y)f_\Phi^{(n)}= \lambda_n w(y)f_\Phi^{(n)}~,
   \label{SLeqn}
\end{equation}
where $p(y) = e^{-4ky}$, $q(y)= m_\Phi^2 e^{-4ky}$, $w(y)=e^{-2ky}$ and the eigenvalues 
$\lambda_n=m_n^2$. From general results in Sturm-Liouville theory we know that since 
$p(y)$ is differentiable, $q(y)$ and $w(y)$ are continuous, $p(y)>0$ and $w(y)>0$ over 
the interval $[0,\pi R]$, the eigenvalues $\lambda_n$ are real and well-ordered i.e. 
$\lambda_0<\lambda_1<\cdots <\lambda_n<\cdots\rightarrow \infty$. Furthermore, the 
eigenfunctions $f_\Phi^{(n)}(y)$ form a complete set and satisfy the orthonormal relation
\begin{equation}
   \int_0^{\pi R} dy\, w(y)\,f_\Phi^{(n)}f_\Phi^{(m)} = \delta_{nm}~.
   \label{normcondition}
\end{equation}
Let us now consider the possible solutions to (\ref{scalarprofileeqn}). The boundary conditions 
(\ref{scalarBC}) can be satisfied if either Neumann, $\partial_5\Phi |_{0,\pi R}=0$ or Dirichlet, 
$\Phi |_{0,\pi R}=0$ conditions are imposed. Since the differential equation is second order the 
general solution will contain two arbitrary constants. The normalization condition (\ref{normcondition}) 
will determine one of these constants while the boundary conditions at $y=0$ and $\pi R$ will 
determine the remaining constant and fix the eigenvalues $m_n^2$.
\\
\\
\noindent
$\bullet$ {\sc Scalar: $m_0=0$} 

\noindent
The general solution for a massless mode $(m_0=0)$ is given by
\begin{equation}
    f_\Phi^{(0)}(y) = c_1^{(0)}e^{(2-\sqrt{4+a})k y}+c_2^{(0)}e^{(2+\sqrt{4+a})k y}~,
\end{equation}
where $c_1^{(0)},c_2^{(0)}$ are arbitrary constants. In general for $a\neq 0$, there is no 
massless mode solution for either Neumann or Dirichlet boundary conditions.
Instead to obtain a massless mode we need to modify the boundary 
action and include boundary mass terms of the form~\cite{gp}
\begin{equation}
\label{sbdy}
     S_{\partial\Phi}= -\int d^5 x\sqrt{-g}\, 2\,b\,k \left[ \delta(y) - \delta(y-\pi R) \right] |\Phi|^2~,
\end{equation}
where $b$ is a dimensionless constant parametrising the boundary mass in 
units of $k$. The Neumann boundary conditions are now modified to
$(\partial_5~-~b k) f_\Phi^{(0)}\big|_{0,\pi R}=0$ and lead to a zero mode solution
\begin{equation}
        f_\Phi^{(0)}(y) \propto e^{b k y}~,
\end{equation}
where the boundary mass parameter must be tuned to satisfy $b=2\pm \sqrt{4+a}$ (which can
be enforced by supersymmetry~\cite{gp}). Assuming $a\geq -4$, in accord with the Breitenlohner-Freedman 
bound~\cite{bf} for the stability of AdS space, the parameter $b$ has a range 
$-\infty < b < \infty$. The localisation features of the zero mode follows 
from considering the kinetic term in (\ref{scalaraction})
\begin{eqnarray}
  && -\int d^5 x\,\sqrt{-g}\,g^{\mu\nu}\partial_\mu \Phi^\ast \partial_\nu \Phi + \dots \nonumber\\
   &=&-\int d^5 x\,e^{2(b-1) k y}\eta^{\mu\nu}\partial_\mu \Phi^{(0)\ast}(x)\partial_\nu \Phi^{(0)}(x)+\dots
\end{eqnarray}
Hence with respect to the 5D flat metric the zero mode profile is given by
\begin{equation}
     \widetilde f_\Phi^{(0)}(y) \propto e^{(b-1) k y} = e^{(1\pm\sqrt{4+a})ky}~.
     \label{zmflat}
\end{equation}
We see that for $b<1~(b >1)$ the zero mode is localized towards the UV (IR) 
brane and when $b=1$ the zero mode is flat. Therefore with the free parameter, 
$b$ the scalar zero mode can be localized anywhere in the bulk.
\\
\\
\noindent
$\bullet$ {\sc Scalar: $m_n\neq 0$} 

The general solution of the  Kaluza-Klein modes for $m_n\neq 0$ corresponding to
$b=2-\alpha$ with $\alpha\equiv \pm \sqrt{4+a}$ is given by
\begin{equation}
    f_\Phi^{(n)}(y) = N_\Phi^{(n)} e^{2 k y}\left[J_{\alpha} \left(\frac{m_n}{k e^{-k y}}
    \right) + b_\Phi^{(n)} Y_{\alpha} \left(\frac{m_n}{k e^{-k y}}\right) \right]~,
    \label{scalarKKprofile}
\end{equation}
where $J_\alpha,Y_\alpha$ are Bessel functions of order $\alpha$ and 
$N_\Phi^{(n)}, b_\Phi^{(n)}$ are arbitrary constants. The normalization constants $N_\Phi^{(n)}$
are obtained from the orthonormal relation (\ref{normcondition}). The Kaluza-Klein masses are 
determined by imposing the boundary conditions and in the limit $\pi kR \gg1$ lead to the 
approximate values
\begin{equation}
\label{scKK}
    m_n\approx \left(n+\frac{1}{2}\sqrt{4+a}-\frac{3}{4}\right)\pi\, k\,e^{-\pi k R}~,
\end{equation}
for $n=1,2,\dots$. We see here the first nontrivial result of the warped dimension. Even though
the branes are separated by a distance $\pi R$ the mass scale of the Kaluza-Klein modes is not
$1/R$ but instead $k e^{-\pi kR}$. The fact that the Kaluza-Klein mass scale is associated with the IR scale 
($k e^{-\pi kR}$) suggests that the Kaluza-Klein modes are sensitive to the warp factor and therefore
must be localized near the IR brane--a fact confirmed by plotting the wavefunction profiles 
(\ref{scalarKKprofile}). However, unlike the zero mode they can not be arbitrarily localized in the bulk.

\subsubsection{Fermions}
Let us next consider bulk fermions in a slice of AdS$_5$~\cite{gn,gp}. The 5D Dirac algebra
in a curved geometry is given by
\begin{equation}
\{ \Gamma^M,\Gamma^N \}=2\,g^{MN}.
\end{equation}
To deal with gamma matrices in curved spacetime, we define $\Gamma^M = e^M_A \gamma^A$ 
where we have introduced the vielbein $e^M_A$, defined via the relation $g^{MN}=e^M_A e^N_B \eta^{AB}$. 
The gamma matrices $\gamma^A$ then satisfy the usual 5D Dirac algebra in Minkowski space, namely
\begin{equation}
\{ \gamma^M,\gamma^N \}=2\,\eta^{MN}=2\,{\rm diag}(-,+,+,+,+).
\label{5Dalg}
\end{equation}
A convenient representation of the gamma matrices satsifying (\ref{5Dalg}) is given by
\begin{eqnarray}
\gamma^\mu = -i \left(
\begin{array}{cc}
0 & \sigma^\mu \\
{\bar\sigma}^\mu & 0
\end{array}\right),
& &
\gamma^5 =  \left(
\begin{array}{cc}
1 & 0 \\
0 & -1
\end{array}
\right),
\end{eqnarray}
where $\sigma^\mu = (1, \sigma^i)$, ${\bar\sigma}^\mu = (1, -\sigma^i)$ and $\sigma^i$ are the usual 
Pauli matrices. In four dimensions the irreducible spinor representation is the two-component Weyl 
spinor that can be either left or right-handed. However this is not the case in five dimensions because, 
unlike in four dimensions, $\gamma^5$ is now part of the 5D Dirac algebra. Lorentz invariant terms 
cannot just depend on $\gamma^5$ (as in 4D), but must necessarily involve both left and right-handed components. So fermions in five dimensions must be represented by four-component Dirac spinors $\Psi$.

Therefore we will consider a bulk Dirac fermion with action
\begin{equation}
	S_\Psi= -\int d^5 x \sqrt{-g}\, \left[\frac{1}{2}\left({\bar\Psi} \Gamma^M D_M \Psi -
	{D_M\bar\Psi} \Gamma^M \Psi\right) + m_\Psi \bar\Psi\Psi \right]~,
	\label{fermionAction}
\end{equation}
where $\bar\Psi \equiv\Psi^\dagger i\gamma^0$ and we have included a bulk mass $m_\Psi$. The covariant derivative is defined as $D_M = \partial_M +\omega_M$, where $\omega_M$ is the spin connection:
\begin{equation}
\omega_M = \frac{i}{2}{\cal J}_{AB}~\omega^{AB}_M =\frac{1}{8}\omega_{M A B}~ \left[\gamma^A,\gamma^B\right],
\end{equation}
with the Lorentz generators, ${\cal J}_{AB}=-\frac{i}{4}~\left[\gamma_A,\gamma_B\right]$.
The coefficients ${{\omega_M}^A}_B$ are determined by
\begin{equation}
{{\omega_M}^A}_B= e^A_R ~ e^S_B ~\Gamma^R_{MS} - e^R_B ~\partial_M e^A_S,
\end{equation}
where $\Gamma^R_{MS}$ is the Christoffel symbol. Specializing to the case of the AdS metric (\ref{adsmetric})
the vielbien becomes $e^M_A=(e^{k y}\delta_\alpha^\mu,1)$, where $\delta^\mu_\alpha$ is the Kronecker delta.
This corresponds to a spin connection 
\begin{equation}
\omega_M = \left(-\frac{k}{2}e^{-k y}\gamma_\mu \gamma^5, 0 \right).
\end{equation}

To obtain the fermion equation of motion we will decompose the Dirac spinor into two
Weyl spinors $\psi_\pm$ by writing
\begin{equation}
       \Psi= \left(\begin{array}{c}\psi_+\\ \psi_- \end{array}\right),\quad
       \Psi_+=\left(\begin{array}{c}\psi_+\\ 0 \end{array}\right),\quad
       \Psi_- =\left(\begin{array}{c} 0\\ \psi_- \end{array}\right),
\end{equation}
where $\Psi= \Psi_+ + \Psi_-$ with $\Psi_{\pm}=\pm\gamma_5 \Psi_{\pm}$
denoting left and right-handed components, respectively.
The corresponding fermion equation of motion, resulting from varying the action 
(\ref{fermionAction}), then becomes
\begin{eqnarray}
\label{feom1}
   e^{ky} \gamma^\mu\partial_\mu \Psi_- +
   \partial_5 \Psi_+ + (c-2)k \Psi_+ &=& 0~,\\
   e^{ky}\gamma^\mu\partial_\mu \Psi_+ -
   \partial_5 \Psi_- + (c+2)k \Psi_- &=& 0~,
   \label{feom2}
\end{eqnarray}
where the bulk mass $m_\Psi = c k$ is parametrized in units of $k$ 
with dimensionless coefficient $c$. Note that the equation of motion is 
now a first-order coupled equation between the components of the Dirac 
spinor $\Psi$. The boundary variation vanishes provided that
\begin{equation}
       (\delta {\bar\Psi}_+ \Psi_-) \big|_{0,\pi R}  =( \delta {\bar\Psi}_- \Psi_+) \big|_{0,\pi R} = 0.
       \label{fermbc}
\end{equation}
The solutions of the bulk fermion equations of motion (\ref{feom1}) and 
(\ref{feom2}) are again obtained by assuming a separation of variables 
\begin{equation}
     \Psi_{\pm}(x^\mu,y)=\sum_{n=0}^\infty \Psi_{\pm}^{(n)}(x^\mu) f_{\pm}^{(n)}(y)~,
\end{equation}
where $\Psi_{\pm}^{(n)}$ are the 4D Kaluza-Klein modes satisfying 
the Dirac equation $\gamma^\mu\partial_\mu \Psi_{\pm}^{(n)} = -m_n \Psi_{\mp}^{(n)}$.
The equations of motion for the profile functions $f_{\pm}^{(n)}$ become
\begin{eqnarray}
\label{fprofile1}
     \partial_5 f_+^{(n)} + (c-2)k f_+^{(n)} &=& m_n e^{ky}f_-^{(n)} ~,\\
     -\partial_5 f_-^{(n)} + (c+2)k f_-^{(n)} &=& m_n e^{ky}f_+^{(n)}~,
   \label{fprofile2}
\end{eqnarray}
These equations can be solved subject to the boundary conditions (\ref{fermbc}).
\\
\\
\noindent
$\bullet$ {\sc Fermion: $m_0=0$} 

The solutions of $\gamma^\mu\partial_\mu \Psi_{\pm}^{(0)} = 0$, are states
of definite helicity, consistent with the fact that $+ (-)$ denotes the left (right)-handed 
components. The profile equations of motion (\ref{fprofile1}) and (\ref{fprofile2}) are easy to 
solve when $m_0=0$.  The equations decouple and the general solution is given by
\begin{equation}
     f_\pm^{(0)}(y) = d_\pm^{(0)}e^{(2\mp c) ky}~,
     \label{f0soln}
\end{equation}
where $d_{\pm}^{(0)}$ are arbitrary constants. Applying the boundary conditions 
(\ref{fermbc}) we see that they are satisfied if either $\Psi_-$ is fixed on the 
boundaries with $\Psi_-|_{0,\pi R}=0$ or instead $\Psi_+$ is fixed.  This is simply
a Dirichlet condition for one of the components but it implies that one solution in 
(\ref{f0soln}) is always killed by the boundary conditions. Thus we can either have 
a left or right-handed massless mode but not both!
In fact this is how 4D chirality is recovered from the vectorlike 5D bulk 
and is the result of compactifying on the orbifold $S^1/Z_2$. This property
will be very useful to describe the standard model fermions since left and right-handed
fermions transform differently under the electroweak gauge group.

For concreteness, to check the localization features of the massless mode,
let us choose $\Psi_-$ to satisfy Dirichlet boundary conditions. The only 
nonvanishing component of $\Psi^{(0)}$ is then the 
left-handed component, $\Psi_+^{(0)}$ with profile $f_+^{(0)}$.
Again the localisation property of this mode is obtained by considering the kinetic term
\begin{equation}
    \int d^5 x\sqrt{-g} \, \bar\Psi \Gamma^\mu \partial_\mu 
   \Psi + \dots = \int d^5 x \, e^{2(\frac{1}{2}-c) k y} \, \bar\Psi_+^{(0)}(x)
    \gamma^\mu \partial_\mu \Psi_+^{(0)}(x)+\dots~.
\end{equation}
Hence with respect to the 5D flat metric the fermion zero mode profile is
\begin{equation}
     \widetilde f_+^{(0)}(y) \propto e^{(\frac{1}{2}-c)ky}~.
\end{equation}
When $c> 1/2~(c<1/2)$ the fermion zero mode is localized towards
the UV (IR) brane while the zero mode fermion is flat for $c=1/2$.
If instead $\Psi_+$ were chosen to satisfy the Dirichlet boundary conditions then
the massless mode ($f_-^{(0)}$) is right-handed and similar results are obtained 
with the substitution $c\leftrightarrow -c$. So in either case, just like the scalar field 
zero mode, the fermion zero mode can be localized anywhere in the 5D bulk. 
\\
\\
\noindent
$\bullet$ {\sc Fermion: $m_n\neq 0$} 

The nonzero Kaluza-Klein fermion modes can be obtained by solving the
first-order coupled equations of motion for the Dirac component profiles $f_\pm^{(n)}$.
The simplest way to proceed is to derive a pair of decoupled second-order 
equations. Each equation is then equivalent to a Sturm-Liouville equation 
(\ref{SLeqn}) for $\widehat f_\pm^{(n)}\equiv e^{-2k y}f_\pm^{(n)}$ 
with $p(y)=e^{-k y}$, $q(y)=c(c\pm 1)k^2 e^{-k y}$ and $w(y)= e^{ky}$.
The general solution is given by
\begin{equation}
f_\pm^{(n)} (y) = N_\psi^{(n)} e^{\frac{5}{2} k y}\left[J_{c\pm \frac{1}{2}} \left(\frac{m_n}{k e^{-k y}}
    \right) + b_\psi^{(n)} Y_{c\pm \frac{1}{2}} \left(\frac{m_n}{k e^{-k y}}\right) \right]~,
    \label{fermionKKprofile}
\end{equation}
where $N_\psi^{(n)}, b_\psi^{(n)}$ are arbitrary constants. Note that these 
constants are the same for both Dirac components since (\ref{fermionKKprofile}) must 
also satisfy the first-order equations (\ref{fprofile1}) and (\ref{fprofile2}).
The boundary conditions will determine $b_\psi^{(n)}$ and $m_n$, while $N_\psi^{(n)}$
is obtained from the orthonormal condition
\begin{equation}
      \int_0^{\pi R} dy \, e^{-3ky} \,f_\pm^{(n)}f_\pm^{(m)} = \delta_{nm}.
      \label{fermionnorm}
\end{equation}
The 4D Kaluza-Klein mass spectrum for $n=1,2,\dots$ is given by
\begin{equation}
     m_n\simeq \left(n + \frac{|\alpha|}{2}-\frac{1}{4}\right) \pi k e^{-\pi kR},
\end{equation}
where $\alpha=c\pm \frac{1}{2}$ for $\Psi_\pm$ obeying Dirichlet boundary conditions.
Note that substituting the solutions (\ref{fermionKKprofile}) back into the action
(\ref{fermionAction}) and peforming the $y$ integration gives rise to
\begin{equation}
     S_\Psi =-\int d^4x \,\sum_{n=0}^\infty \left[{\bar\Psi}^{(n)} \gamma^\mu \partial_\mu \Psi^{(n)}
      + m_n \bar\Psi^{(n)}\Psi^{(n)} \right],
\end{equation}
where $\Psi^{(n)}=\Psi_+^{(n)}+\Psi_-^{(n)}$. Thus for $\Psi_- (\Psi_+)$ obeying Dirichlet boundary conditions, 
the 4D Kaluza-Klein modes consist of a left (right)-handed massless mode ($m_0=0$), together with a set of massive Dirac states
with mass $m_n$.

\subsubsection{Gauge fields}
Consider a bulk gauge field $A_M$ in a slice of AdS~\cite{gf1,gf2}. Without loss of generality we
will consider a U(1) gauge field with the action
\begin{equation}
     S_A =\int d^5 x\sqrt{-g}\, \left[-\frac{1}{4 g_5^2} F_{MN}F^{MN}\right],
     \label{gaugeaction}
\end{equation}
where $g_5$ is the 5D gauge coupling and $F_{MN} = \partial_M A_N-\partial_N A_M$.
Working in the gauge $A_5=0$, together with the constraint $\partial_\mu A^\mu=0$, the
equation of motion is
\begin{equation}
      \eta^{\mu\rho}\eta^{\nu\sigma}\partial_\mu F_{\rho\sigma} + \eta^{\nu\sigma}\partial_5( e^{-2k y}
      \partial_5 A_\sigma) =0~,
      \label{vectoreom}
\end{equation}
and the boundary terms satisfy
\begin{equation}
       (\delta A^\mu \partial_5 A_\mu) \big|_{0,\pi R}=0~.
       \label{vectorBC}
\end{equation}
To solve (\ref{vectoreom}) we assume a separation of variables 
\begin{equation}
     A_\mu(x^\nu,y)= \sum_{n=0}^\infty A_\mu^{(n)}(x^\nu) f_A^{(n)}(y)~,
     \label{vectordecomp}
\end{equation}
where $A_\mu^{(n)}(x^\nu)$ are the 4D Kaluza-Klein modes satisfying the Proca
equation $\eta^{\mu\rho} \partial_\mu F_{\rho\sigma}^{(n)}= m_n^2 A_\sigma^{(n)}$ with 
masses $m_n$, and bulk profile $f_A^{(n)}(y)$. Substituting (\ref{vectordecomp}) into (\ref{vectoreom}) 
leads to an equation for the profile 
\begin{equation}
      -\partial_5 (e^{-2 k y} \partial_5 f_A^{(n)}) = m_n^2 f_A^{(n)}.
      \label{vectorprofileeqn}
\end{equation}
This is a Sturm-Liouville  equation with $p(y)=e^{-2 k y}$, $q(y)=0$ and $w(y)=1$. Thus the 
modes $f_A^{(n)}$ form a complete set and satisfy the orthonormal relation
\begin{equation}
   \int_0^{\pi R} dy\, f_A^{(n)} f_A^{(m)} = \delta_{nm}~.
   \label{vectornorm}
\end{equation}
The boundary conditions (\ref{vectorBC}) can be satisfied if either Neumann, $\partial_5A_\mu \big|_{0,\pi R}=0$ or Dirichlet, $A_\mu \big|_{0,\pi R}=0$ conditions are imposed. Just like the scalar case 
the normalization and boundary conditions will determine the constants in the general solution as well as 
the eigenvalues $m_n^2$.
\\
\\
\noindent
$\bullet$ {\sc Gauge boson: $m_0=0$} 

\noindent
The general solution for a massless mode $(m_0=0)$ is given by
\begin{equation}
    f_A^{(0)}(y) = c_0^{(0)}+ c_1^{(0)}e^{2 k y}~,
\end{equation}
where $c_1^{(0)},c_2^{(0)}$ are arbitrary constants. There is no massless mode solution when
Dirichlet boundary conditions are imposed. However, Neumann boundary conditions lead to
$c_1^{(0)}=0$ so that the massless mode becomes
\begin{equation}
    f_A^{(0)}(y) = \frac{1}{\sqrt{\pi R}},
    \label{gaugeprofile}
\end{equation}
where we have used the normalization condition (\ref{vectornorm}). Unlike the scalar and fermion
massless modes the localization of this mode is fixed. From the kinetic term
\begin{equation}
 \int d^5 x\sqrt{-g}~g^{\mu\rho} g^{\nu\sigma} F_{\mu\nu}F_{\rho\sigma} + \dots
  =\int d^5 x \frac{1}{\pi R}\, \eta^{\mu\rho}\eta^{\nu\sigma} F_{\mu\nu}^{(0)}(x) F_{\rho\sigma}^{(0)}(x)+\dots
\end{equation}
we see that the massless mode is not localized in the warped bulk. This feature plays a prominent
role when we consider the standard model in the bulk. It is possible to change the localization of the 
zero mode but this involves adding a dilaton coupling, which is equivalent to 
bulk and boundary masses for the gauge field~\cite{bg}.
\\
\\
\noindent
$\bullet$ {\sc Gauge boson: $m_n\neq 0$} 

The general solution of the  Kaluza-Klein modes corresponding to $m_n\neq 0$ is given by
\begin{equation}
    f_A^{(n)}(y) = N_A^{(n)} e^{k y}\left[J_1 \left(\frac{m_n}{k e^{-k y}}\right) + 
    b_A^{(n)} Y_1 \left(\frac{m_n}{k e^{-k y}}\right) \right]~,
    \label{vectorKKprofile}
\end{equation}
where $N_A^{(n)}, b_A^{(n)}$ are arbitrary constants. The Kaluza-Klein masses (or eigenvalues) are 
determined by imposing the boundary conditions. In the limit $\pi kR \gg1$ the masses for Neumann (Dirichlet) boundary conditions are given by
\begin{equation}
\label{vectorKK}
    m_n\approx \left(n\mp \frac{1}{4}\right)\pi\, k\,e^{-\pi k R}~,
\end{equation}
for $n=1,2,\dots$. Even though the massless mode is not localized we see that the Kaluza-Klein modes
are again localized near the IR brane.

\subsubsection{Graviton}
For completeness we also present the analysis for the graviton. In this case we must 
consider just tensor fluctuations of the metric $g_{MN}$ (\ref{adsmetric}) which 
have the form
\begin{equation}
ds^2= e^{-2ky} \left(\eta_{\mu\nu}+h_{\mu\nu}(x^\mu,y)\right)dx^\mu dx^\nu +dy^2,
\end{equation}
where $h_{\mu\nu}$ is the graviton fluctuation. In the transverse-traceless gauge, 
$\partial_\mu h^{\mu\nu}= h^\mu_\mu = 0$, the 5D gravitational action becomes
\begin{eqnarray}
S&=&\int d^5x \sqrt{-g}\,(M_5^3 R +\Lambda_5), \\
&\rightarrow& M_5^3\int d^5x \,e^{-2ky}\left(-\frac{1}{4}\partial_\rho 
h_{\mu\nu}\partial^\rho h^{\mu\nu} -\frac{1}{4}e^{-2k y}\partial_5 h_{\mu\nu}\partial_5 h^{\mu\nu}\right).
\label{EHaction}
\end{eqnarray}
The variation of this action leads to the equation of motion
\begin{equation}
        \Box h_{\mu\nu} + e^{2 ky}\partial_5 (e^{-4k y} \partial_5 h_{\mu\nu}) = 0,
    \label{gravityeqn}
\end{equation}
and boundary condition
\begin{equation}
    \delta h^{\mu\nu}\partial_5 h_{\mu\nu} \big|_{0,\pi R} = 0.
    \label{gravitybc}
\end{equation}
The bulk graviton is expanded in Kaluza-Klein modes
\begin{equation}
h_{\mu\nu}(x^\mu,y)=\sum_{n=0}^\infty h^{(n)}_{\mu\nu}(x^\mu)f_h^{(n)}(y),
\end{equation}
where the wavefunctions $f_h^{(n)}$ obey the equation of motion 
\begin{equation}
-\partial_5 (e^{-4k y}\partial_5 f_h^{(n)}) = m_n^2 e^{-2k y}f_h^{(n)},
\end{equation}
with $\Box h_{\mu\nu}^{(n)}=m_n^2 h_{\mu\nu}^{(n)}$.
This equation for the profile $f_h^{(n)}$ is in fact the same as a massless scalar field
(\ref{scalarprofileeqn}). It is a Sturm-Liouville equation and gives rise to a similar 
orthonormal condition
\begin{equation}
\int_0^{\pi R} dy \,e^{-2k y} f_h^{(n)}f_h^{(m)} = \delta_{nm}.
\label{normh}
\end{equation}
The boundary conditions (\ref{gravitybc}) can again be satisfied by either imposing
Dirichlet $f_h^{(n)}\big|_{0,\pi R}=0$ or Neumann $\partial_5 f_h^{(n)}\big|_{0,\pi R}=0$ conditions.
\\
\\
\noindent
$\bullet$ {\sc Graviton: $m_0=0$} 

\noindent
The general solution for the massless mode $(m_0=0)$ is given by
\begin{equation}
    f_h^{(0)}(y) = c_0^{(0)}+ c_1^{(0)}e^{4 k y}~,
\end{equation}
where $c_1^{(0)},c_2^{(0)}$ are arbitrary constants. The Dirichlet conditions do not 
allow a massless mode, but Neumann boundary conditions lead to a constant 
massless mode
\begin{equation}
    f_h^{(0)}(y) = c_0^{(0)}.
    \label{gravity0}
\end{equation}
This represents a 4D graviton and to see where it is localized in the bulk we substitute
(\ref{gravity0}) back into the action (\ref{EHaction}). The kinetic term becomes
\begin{equation}
 \int d^5x\, e^{-2ky} \partial_\rho h_{\mu\nu}\partial^\rho h^{\mu\nu} + \dots
  =\int d^5 x\, e^{-2ky} (c_0^{(0)})^2 \partial_\rho h_{\mu\nu}^{(0)}\partial^\rho h^{(0)\mu\nu}+\dots
\end{equation}
which shows that with respect to the flat 5D metric the massless mode
\begin{equation}
         {\tilde f}_h^{(0)}(y) \propto e^{-k y}~.
         \label{gravprofile}
\end{equation}
Thus the 4D graviton is localized on the UV brane~\cite{rs2}. Note that just like the
massless gauge field, the profile
can be changed by adding bulk and boundary masses for the graviton~\cite{gpp}.
\\
\\
\noindent
$\bullet$ {\sc Graviton: $m_n\neq 0$} 

The general solution of the  Kaluza-Klein modes corresponding to $m_n\neq 0$ is given by
\begin{equation}
    f_h^{(n)}(y) = N_h^{(n)} e^{2k y}\left[J_2 \left(\frac{m_n}{k e^{-k y}}
    \right) + b_h^{(n)} Y_2 \left(\frac{m_n}{k e^{-k y}}\right) \right]~,
    \label{gravityKKprofile}
\end{equation}
where $N_h^{(n)}, b_h^{(n)}$ are arbitrary constants. The approximate Kaluza-Klein masses
in the limit $\pi kR \gg1$, for Neumann (Dirichlet) boundary conditions is
\begin{equation}
\label{gravityKK}
    m_n\approx \left(n+\frac{1}{2} \mp \frac{1}{4}\right)\pi\, k\,e^{-\pi k R}~,
\end{equation}
for $n=1,2,\dots$. Even though the massless mode is not localized we see that the 
Kaluza-Klein modes are again localized near the IR brane.

\subsubsection{SUMMARY}
\label{profilesummary}
We have the following possible behaviour for 4D massless mode ($m_0=0$) profiles of bulk fields:
\begin{center}
\begin{tabular}{|c|c|}
\hline
Field  & Profile \\
\hline
scalar $\phi^{(0)}$ &  $e^{(1\pm\sqrt{4+a})ky}$ \\
fermion  $\psi^{(0)}_\pm$ & $e^{\left(\frac{1}{2} \mp c\right)ky}$\\
vector $A_\mu^{(0)}$ & 1\\
graviton $ h_{\mu\nu}^{(0)}$ & $e^{-ky}$ \\
\hline
\end{tabular}
\\
\end{center}
Similarly, the Kaluza-Klein mode ($m_n\neq 0$) solutions can be obtained for 
all types of bulk fields and combined into one general expression~\cite{gp}
\begin{equation}
\label{genwf}
    f^{(n)}(y) = N^{(n)} e^{(2-s)k y} \left[ J_\alpha \left(\frac{m_n}{k e^{-k y}}\right) 
    + b^{(n)} Y_\alpha \left(\frac{m_n}{k e^{-k y}}\right) \right]~,
\end{equation}
for $f^{(n)}=(f_\Phi^{(n)},f_{\pm}^{(n)}, f_A^{(n)})$ with $s=(0,-\frac{1}{2},1)$, and 
$\alpha=(\pm\sqrt{4+a}, c\pm \frac{1}{2}, 1)$. The graviton profiles $f_h^{(n)}$ 
are identical to the scalar modes $f_\Phi^{(n)}$ with $a=b=0$.
The normalization constants $N^{(n)}$ are determined from the orthonormal relation
\begin{equation}
      \int_0^{\pi R} dy\, e^{2(s-1) ky} f^{(n)} f^{(m)} = \delta_{nm}.
\end{equation}
The constants $b^{(n)}$ for Kaluza-Klein mode solutions with zero modes (corresponding
to Neumann boundary conditions for bosons, and either Neumann or Dirichlet boundary
conditions for fermions) are given by
\begin{equation}
\label{balpha}
    b^{(n)} = \begin{cases}
    -\frac{J_{\alpha-1}\left(\frac{m_n}{k}\right)}{Y_{\alpha-1}\left(\frac{m_n}
    {k}\right)}= -\frac{J_{\alpha-1}\left(\frac{m_n}{k e^{-\pi k R}}\right)}{Y_{\alpha-1}\left(\frac{m_n}
    {k e^{-\pi k R}}\right)} \quad\quad~{\rm bosons}~,\\
     -\frac{J_{\alpha}\left(\frac{m_n}{k}\right)}{Y_{\alpha}\left(\frac{m_n}
    {k}\right)}= -\frac{J_{\alpha}\left(\frac{m_n}{k e^{-\pi k R}}\right)}{Y_{\alpha}\left(\frac{m_n}
    {k e^{-\pi k R}}\right)}  \quad\quad\quad\quad~{\rm fermions}~.
\end{cases}    
\end{equation}
In the limit $\pi k R\gg 1$ the Kaluza-Klein mass spectrum, obtained by solving the equations
in (\ref{balpha}), is approximately given by
\begin{equation}
\label{KKspect}
m_n\simeq
\begin{cases}
    \left(n+\frac{|\alpha|}{2}-\frac{3}{4}\right)\pi k \,e^{-\pi kR} \quad\quad~{\rm bosons}~,\\
   \left(n+\frac{|\alpha|}{2}-\frac{1}{4}\right)\pi k \,e^{-\pi kR} \quad\quad~{\rm fermions}~,
\end{cases} 
\end{equation}
for $n=1,2,\dots$. Note that the Kaluza-Klein modes for all types of bulk fields are 
always localized near the IR brane. Unlike the zero mode there is no freedom to 
delocalize the Kaluza-Klein (nonzero) modes away from the IR brane.

\section{The Standard Model in the Bulk}

We can now use the freedom to localize scalar and fermion zero mode 
fields anywhere in the warped bulk to construct a bulk Standard Model. 
Recall that the hierarchy problem only affects the Higgs boson.
Hence to solve the hierarchy problem the Higgs scalar field must
be localized very near the TeV brane, and for simplicity we will assume that
the Higgs is confined to the TeV brane (as in RS1). However we will now
consider the possible effects of allowing fermions and gauge bosons 
to propagate in the warped bulk.

\subsection{Yukawa couplings}

\begin{figure}
\begin{center}
\includegraphics[width=0.5\textwidth,height=0.27\textheight]{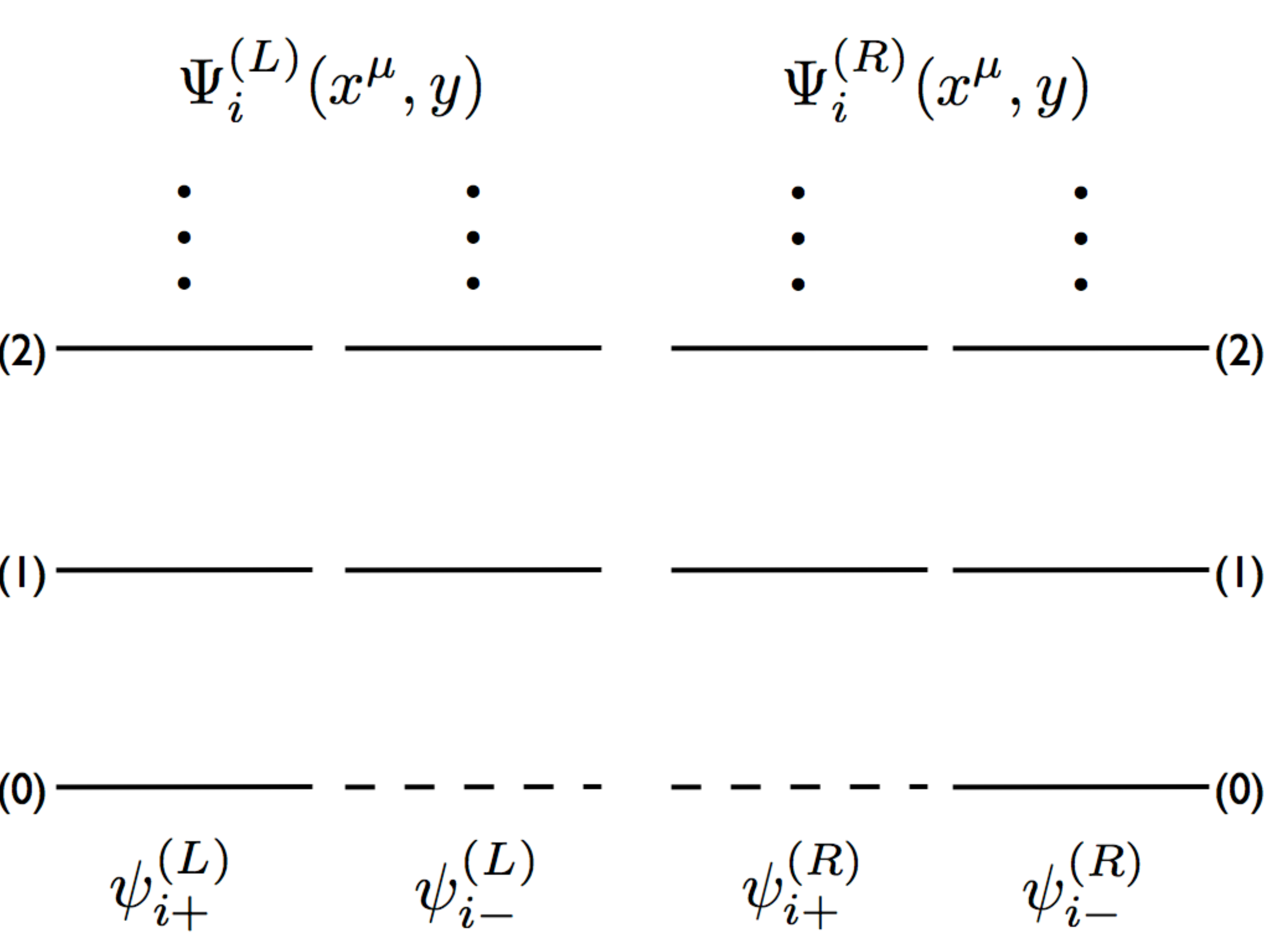}
\end{center}
\caption{\it Embedding the Standard Model Weyl fermions into 5D Dirac spinors. The dashed lines
indicate an absent Kaluza-Klein mode.}
\label{fermionfig}
\end{figure}

One consequence of allowing fermions to be localized anywhere in the bulk
is that Yukawa coupling hierarchies are naturally generated by separating 
the fermions from the Higgs boson that is confined on the IR brane. In the Standard Model 
the weak interactions do not conserve parity, and consequently left and right-handed
fermions transform differently under the electroweak gauge group. However we have seen that
the massless zero mode of a bulk Dirac spinor can either be left or right-handed.
Hence, for every Standard Model Weyl fermion $\psi_i$ we introduce a corresponding 5D 
Dirac spinor $\Psi_i$. Boundary conditions are then chosen so that left-handed spinors, 
$\psi_{i+}$ are identified with the massless zero mode of $\Psi_i^{(L)}$ and similarly right-handed 
spinors, $\psi_{i-}$ are identified with the zero mode of $\Psi_i^{(R)}$, where $i$ is a flavor index. 
This embedding is depicted in Figure~\ref{fermionfig}.

The 4D Standard Model Yukawa interactions, ${\bar \Psi}_i^{(L)} \Psi_j^{(R)} H$, are then 
promoted to 5D interactions in the warped bulk in the following way
\begin{eqnarray}
&& \int d^5 x \sqrt{-g}~\lambda_{ij}^{(5)}\left[ \bar\Psi_i^{(L)}(x^\mu,y) \Psi_j^{(R)}(x^\mu,y) + h.c.\right]
      H(x^\mu)\delta(y-\pi R\nonumber)\\
&&      \equiv\int d^4 x~\lambda_{ij}\left(\bar\Psi_{iL+}^{(0)}(x^\mu) 
      \Psi_{jR-}^{(0)}(x^\mu)H(x^\mu) + h.c. + \dots\right)~,
      \label{IRYukawa}
\end{eqnarray}
where $i,j$ are flavor indices, $\lambda_{ij}^{(5)}$ is the dimensionful (mass dimension $=-1$) 
5D Yukawa coupling and $\lambda_{ij}$ is the dimensionless 4D Yukawa coupling. Assuming 
$c_{iL}=-c_{iR}\equiv c_i$, the normalized zero mode profile is
\begin{equation}
   f^{(0)}_{iL+,R-}(y)= \sqrt{\frac{(1-2 c_i)k}{ e^{(1- 2c_i)\pi kR}-1}} \, e^{(2-c_i)k y}.
   \label{normfermion0}
\end{equation}
When $c_i>1/2$ the 4D Yukawa couplings exponentially depend on the mass parameters $c_i$ 
and are approximately given by~\cite{gp}
\begin{equation}
\label{yukcoup}
   \lambda_{ij} \simeq \lambda_{ij}^{(5)} k\, \left(c_i-\frac{1}{2}\right)\,e^{(1-2c_i)\pi k R}~.
\end{equation}
This is consistent with the fermions being localized near the UV brane and therefore having 
a small wavefunction overlap with the IR confined Higgs boson. Assuming $\lambda_{ij}^{(5)} k \simeq 1$,
the parameters $c_i$ can now be chosen to match the fermion mass hierarchy. Working in a basis 
where the Yukawa coupling matrix is diagonal for the charged leptons, the electron Yukawa coupling 
$\lambda_e\sim 10^{-6}$ is obtained for $c_e\simeq 0.64$. Instead when $c_i <1/2$, both left 
and right-handed fermions are localized near the IR brane giving 
\begin{equation}
   \lambda_{ij} \simeq \lambda_{ij}^{(5)} k \,\left(\frac{1}{2}-c_i\right)~,
\end{equation}
with no exponential suppression. Hence 
the top Yukawa coupling $\lambda_t\sim 1$ is obtained for $c_t \simeq -0.5$. 
The remaining fermion Yukawa couplings are then obtained with $c_i$ in the range
$c_t \lesssim c_i \lesssim c_e$~\cite{gp,hs}. Thus, we see that for bulk mass parameters 
$c_i$ of ${\cal O}(1)$ the fermion mass hierarchy is explained without invoking any 
bulk flavor symmetries.

It is important to note that the boundary Yukawa interaction (\ref{IRYukawa}) changes the 
fermion boundary conditions, causing a modification of the fermion orthonormal condition 
(\ref{fermionnorm}). The orthonormal relation is now generalized to~\cite{cghnp}
\begin{equation}
      \int_0^{\pi R} dy \, e^{-3ky} \,f_\pm^{(n)}f_\pm^{(m)} = \delta_{nm}+\Delta_{nm}^\pm~,
      \label{newfermionnorm}
\end{equation}
where $\Delta_{mn}^\pm$ is determined from the equation
\begin{equation}
      m_m \Delta_{mn}^+ -m_n \Delta_{mn}^- = \pm 2 f_+^{(n)}(\pi R) f_+^{(m)}(\pi R)~.
\end{equation}
However it turns out that these corrections do not substantially affect the determination of the 
masses (although it does affect phases of mixing matrices) 
and using the zero mode profiles (\ref{normfermion0}) remains a good approximation to order
$v^2/m_{KK}^2$. Furthermore a more comprehensive analysis of the fermion masses and mixings 
can be done assuming three independent $c$ parameters for each fermion generation, one for the 
SU(2)$_L$ doublet and two parameters for each of the right-handed singlets. Under the assumption 
of anarchic 5D Yukawa coupling matrices, detailed fits of the $c$ parameters can be found in 
Ref.~\cite{huber,cghnp} where the hierarchies in the fermion masses and CKM matrix are 
naturally explained by the overlap of profiles. 
In addition, at order $v^2/m_{KK}^2$, the modified fermion boundary conditions lead to anomalous 
right-handed charged currents, tree-level FCNC couplings of the $Z$ and Higgs boson, rare
top-quark decays and non-unitarity of the CKM matrix. A complete analysis of these effects is given 
in Ref.~\cite{cghnp}.

The warped bulk can also be used to obtain naturally small neutrino masses.
Various scenarios are possible. If the right (left) handed neutrino is 
localized near the UV (IR) brane then a tiny Dirac neutrino mass is 
obtained~\cite{gn}. However this requires that lepton number is 
conserved on the UV brane. Alternatively in the ``reversed'' scenario one can 
place the right (left) handed neutrino near the IR (UV) brane. In this case 
even though lepton number is violated on the UV brane, the neutrinos will 
still obtain naturally tiny Dirac masses~\cite{tg}. In either case Dirac neutrino masses of the 
right order of magnitude are obtained without invoking a seesaw mechanism.
The warped bulk provides a natural setting to generate tiny Yukawa couplings.
However the non-hierarchical mixings in the neutrino sector typically require an 
extra bulk flavor symmetry~\cite{pr, cdgg}, although a bulk Higgs can lead to 
no large flavor-dependent hierarchies~\cite{aos}.

\subsection{Gauge couplings}

Since fermions are located at different places in the warped extra dimension it may appear
that gauge-coupling universality is lost i.e. that all fermion flavors couple to the SM gauge bosons
with a universal coupling. Of course this does not happen because 4D gauge invariance is preserved,
but let's see how this happens. For simplicity consider the U(1) coupling (such as hypercharge).
The 4D gauge coupling is obtained from the 5D fermion kinetic term
\begin{eqnarray}
&& \int d^5 x \sqrt{-g}\, g_5 \left[ \bar\Psi_i(x^\nu,y) \Gamma^\mu A_\mu(x^\nu,y) \Psi_i(x^\nu,y) \right],\nonumber\\
&& \equiv\int d^4 x~g_4 \left[\bar\Psi_i^{(n)}(x^\nu) \gamma^\mu A_\mu^{(0)}(x^\nu) \Psi_i^{(m)}(x^\nu)+ 
\dots\right]~,
\label{4Dgc}
\end{eqnarray}
where the zero mode gauge boson couples to the Kaluza-Klein fermion modes.
By substituting the Kaluza-Klein mode profiles in (\ref{4Dgc}) we obtain
\begin{equation}
     g_4= g_5 \int_0^{\pi R} dy \,e^{-3 k y} f_A^{(0)} f_{i\pm}^{(n)} f_{i\pm}^{(m)} = 
     \frac{g_5}{\sqrt{\pi R}}\delta_{nm}~,
\end{equation}
where we have used the fact that the gauge boson zero mode profile is constant (\ref{gaugeprofile}), thereby
allowing the fermion orthonormal condition (\ref{fermionnorm}) to be used. Thus the 4D gauge coupling
is flavor universal for all Kaluza-Klein fermions and we see how 4D gauge invariance is preserved.
This expression also shows that for the Standard Model couplings of order one requires 
$g_5 \sqrt{k} \sim \sqrt {\pi k R} \simeq 6$.

Incidentally, similiar arguments also apply for gravity where fermions located at different places in the 
warped extra dimension would seem to lead to a non-universal coupling to gravity causing a violation 
of the equivalence principle. But again this does not occur because 4D general coordinate invariance
is not broken and just like the gauge boson the graviton zero mode profile is constant (\ref{gravity0}). 
The gravitational coupling of matter in the bulk is given by
\begin{equation}
      \int d^5 x \sqrt{-g}\, \frac{1}{M_5} h_{\mu\nu}T^{\mu\nu}(x^\rho,y)
      \equiv\int d^4 x~\frac{1}{M_P} h_{\mu\nu}^{(0)}T^{\mu\nu}(x^\rho)+\dots~,
\label{4Dgravitycoupling}
\end{equation}
where $T^{\mu\nu}$ is the matter energy-momentum tensor. Since the graviton zero-mode 
profile is constant and $T^{\mu\nu}(x^\rho)$ is quadratic in the Kaluza-Klein 
fermion fields, one can use the orthonormal condition (\ref{fermionnorm}) to derive the universal 
coupling to gravity (\ref{4Dgravitycoupling}). This will also be true for other bulk fields. 

\begin{figure}
\begin{center}
\includegraphics[width=0.6\textwidth,height=0.3\textheight]{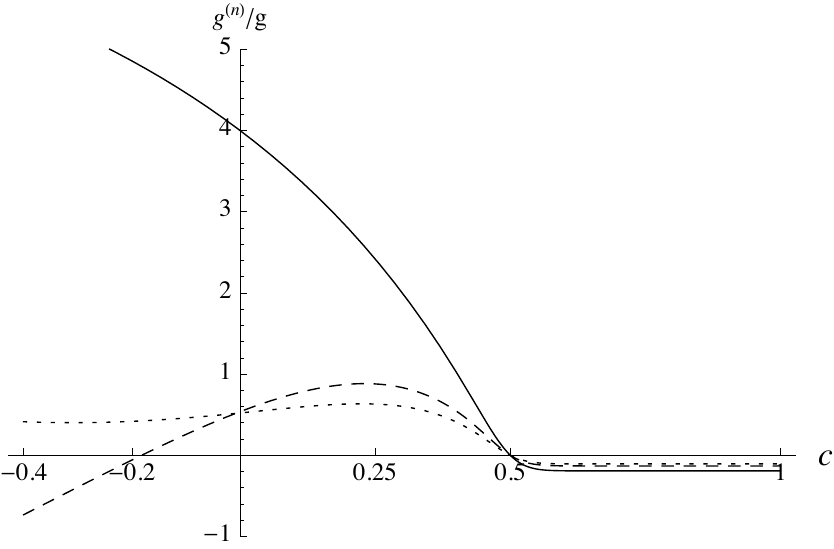}
\end{center}
\caption{ The gauge coupling of zero mode fermions to Kaluza-Klein gauge bosons for $n=1$ (solid),
$n=2$ (dashed) and $n=3$ (dotted).}
\label{gcfig}
\end{figure}

It is also interesting to consider the coupling of the zero-mode fermions to the Kaluza-Klein gauge bosons.
The coupling in this case becomes
\begin{eqnarray}
     g_i^{(n)}&=&g_5 \int_0^{\pi R} dy \,e^{-3 k y} f_A^{(n)} f_{i\pm}^{(0)} f_{i\pm}^{(0)},\nonumber\\
     &=&g_5 N_{\psi_i}^{(0)2} N_A^{(n)} \int_0^{\pi R} dy\, e^{2(1-c_i)k y} \left[J_1 \left(\frac{m_n}{k e^{-k y}}\right) -
    \frac{J_0(\frac{m_n}{k})}{Y_0(\frac{m_n}{k})} Y_1 \left(\frac{m_n}{k e^{-k y}}\right)\right]~.\nonumber\\
    \label{fermionKKgc}
\end{eqnarray}
These couplings are plotted in Fig.~\ref{gcfig}, assuming $k e^{-\pi k R} = {\rm TeV}$.
When $c_i$ is large and negative, the fermions are localized near the IR brane
and the ratio $g^{(1)}/g$ approaches the asymptotic limit 
$g^{(1)}/g\simeq\sqrt{2\pi kR}\simeq 8.4$, corresponding to an IR brane confined 
fermion~\cite{dhr,pomarol}. 
Interestingly for $c_i > 1/2$, the coupling quickly becomes universal for all fermion flavors. 
This is because the fermions are now UV localized, where the wavefunction of the 
Kaluza-Klein gauge bosons is constant and the fermion 
orthonormal condition (\ref{fermionnorm}) leads to a universal coupling.
We will see that this property helps to ameliorate bounds from FCNC processes.

\subsection{A GIM-like mechanism and higher-dimension operators}

A generic four-fermion operator that arises from proton decay and FCNC processes
is given by
\begin{equation}
   \int d^5 x \sqrt{-g}~\frac{1}{M_5^3} \bar\Psi_i\Psi_j
     \bar\Psi_k\Psi_l \equiv \int d^4 x \frac{1}{M_4^2} 
     \bar\Psi_i^{(0)}\Psi_j^{(0)}\bar\Psi_k^{(0)}\Psi_l^{(0)}~,
\end{equation}
where the effective 4D mass scale $M_4$ for $1/2\lesssim c_i \lesssim 1$ is approximately 
given by\cite{gp}
\begin{equation}
\label{4dscale}
    \frac{1}{M_4^2} \simeq \frac{k}{M_5^3} e^{(4-c_i-c_j-c_k-c_l)\pi kR}~.
\end{equation}
If we want the suppression scale for higher-dimension proton decay operators 
to be $M_4\sim M_P$ then (\ref{4dscale}) requires $c_i\simeq 1$
assuming $k\sim M_5\sim M_P$. Unfortunately for these values of $c_i$ the
corresponding Yukawa couplings would be too small. Nevertheless, 
the values of $c$ needed to explain the Yukawa coupling hierarchies still 
suppresses proton decay by a mass scale larger than the TeV 
scale~\cite{gp,huber}. Thus a discrete symmetry is required but there is no need 
to forbid very large higher-dimension operators.

On the other hand the suppression scale for FCNC processes only needs to be 
$M_4 \gtrsim 1000$ TeV. This can easily be achieved for the values of $c$ 
that are needed to explain the Yukawa coupling hierarchies of light fermions.
However for the third generation, the FCNC processes are not so suppressed, leading to 
larger effects in B-physics and top decays.

In fact the FCNC constraints can be used to obtain a lower bound on the Kaluza-Klein 
mass scale $m_{KK}$. For example, consider $K-{\bar K}$ mixing which is the 
intraconversion of neutral kaons, $K^0$ and ${\bar K}^0$ via a strangeness-changing 
$\Delta S=2$ process. In the Standard Model this intraconversion could proceed at tree-level 
if the $Z$-boson could change flavor. This absence of tree-level FCNC processes is due to the 
GIM mechanism which cancels the strangeness changing neutral current by simply introducing 
the charm quark. But if fermions are located at different places in the extra dimension, tree-level 
FCNC processes can be mediated by Kaluza-Klein gauge bosons. In flat space with split fermions 
this leads to strong constraints $m_{KK} \gtrsim 25-300$ TeV (with the range depending on 
whether FCNC processes violate CP)~\cite{dpq}. 

It turns out that the bound in warped space is ameliorated\cite{gp}. We will consider 
Kaluza-Klein gluons since they provide the strongest constraint. Assuming for simplicity
just two families, the flavor-violating couplings in the mass eigenstate basis are~\cite{dpq}
\begin{eqnarray}
   -{\cal L} &=& ({\bar d}_{R}, {\bar s}_{R})
   \left( \begin{array}{cc}
m_d & 0 \\
0 & m_s
\end{array}\right)
\left( \begin{array}{c}
d_{L}\\
s_{L}
\end{array}\right)
+\frac{g}{\sqrt{2}} W_\mu ({\bar u}_L, {\bar c}_L) \gamma^\mu V_{CKM} 
\left( \begin{array}{c}
d_{L}\\
s_{L}
\end{array}\right)
\nonumber \\
&&+\sum_{n=1}^\infty\left[ \sqrt{2}g_s G_\mu^{A(n)} ({\bar d}_{L}, {\bar s}_{L})
\gamma^\mu T^A U_L^d
\left( \begin{array}{c}
d_{L}\\
s_{L}
\end{array}\right)
+(L\leftrightarrow R)\right],
\label{fcncLag}
\end{eqnarray}
where $d_{L,R}, s_{L,R}$ represent the respective zero mode fermion fields and 
\begin{equation}
    U_L^d\equiv V_L^d
    \left( \begin{array}{cc}
    \epsilon_1^{(n)} & 0 \\
0 & \epsilon_2^{(n)}
\end{array}\right)
V_L^{d\dagger},
\label{Udefn}
\end{equation}
for generic unitary matrices $V_{L,R}^d$. 
In (\ref{fcncLag}) we have depicted two sources of flavor violation, the usual violation
via the CKM matrix ($V_{CKM}$) and the other mediated by Kaluza-Klein gluons via the matrix $U_L^d$
where in (\ref{Udefn}), $\epsilon_{1,2}^{(n)}$ denotes the overlap integral between fermion and gauge 
boson wavefunction profiles as in (\ref{fermionKKgc}). The flavor violating coupling  mediated by the 
Kaluza-Klein gluons occurs at tree-level and leads to the effective $\Delta S=2$ Lagrangian
\begin{equation}
      {\cal L}_{\Delta S=2} = \sum_{n=1}^\infty \frac{2 g_s^2}{3 n^2 M_{KK}^2}
      \left[ U_{L\{12\}}^d {\bar d}_L\gamma^\mu s_L + (L\leftrightarrow R) + h.c. \right]^2,
\end{equation}
where $U_{L\{ij\}}^d$ denotes the $\{ij\}$ element of the matrix $U_L^d$.
Using the unitarity of $V_L^d$ we obtain $U_{L\{12\}}^d = (\epsilon_1^{(n)}-\epsilon_2^{(n)}) V_{11}^d
V_{21}^{d\ast}$ so that the amount of tree-level flavor violation is proportional to the difference of the effective 
coupling overlap integrals. 

Now in warped space for $c_i\gtrsim 1/2$, we have seen that the Kaluza-Klein gauge boson coupling to 
fermions is universal so that $\epsilon_1^{(n)} \simeq \epsilon_2^{(n)}$.
Therefore the tree-level flavor violation is essentially cancelled giving rise
to a GIM-like mechanism in the 5D bulk. The corresponding bound for warped dimensions then
becomes $m_{KK} \gtrsim 2$ TeV (assuming no CP violation), greatly ameliorating the bounds on the 
Kaluza-Klein scale compared to the flat extra dimension case~\cite{gp}.

\begin{figure}
\begin{center}
\includegraphics[width=0.45\textwidth,height=0.28\textheight]{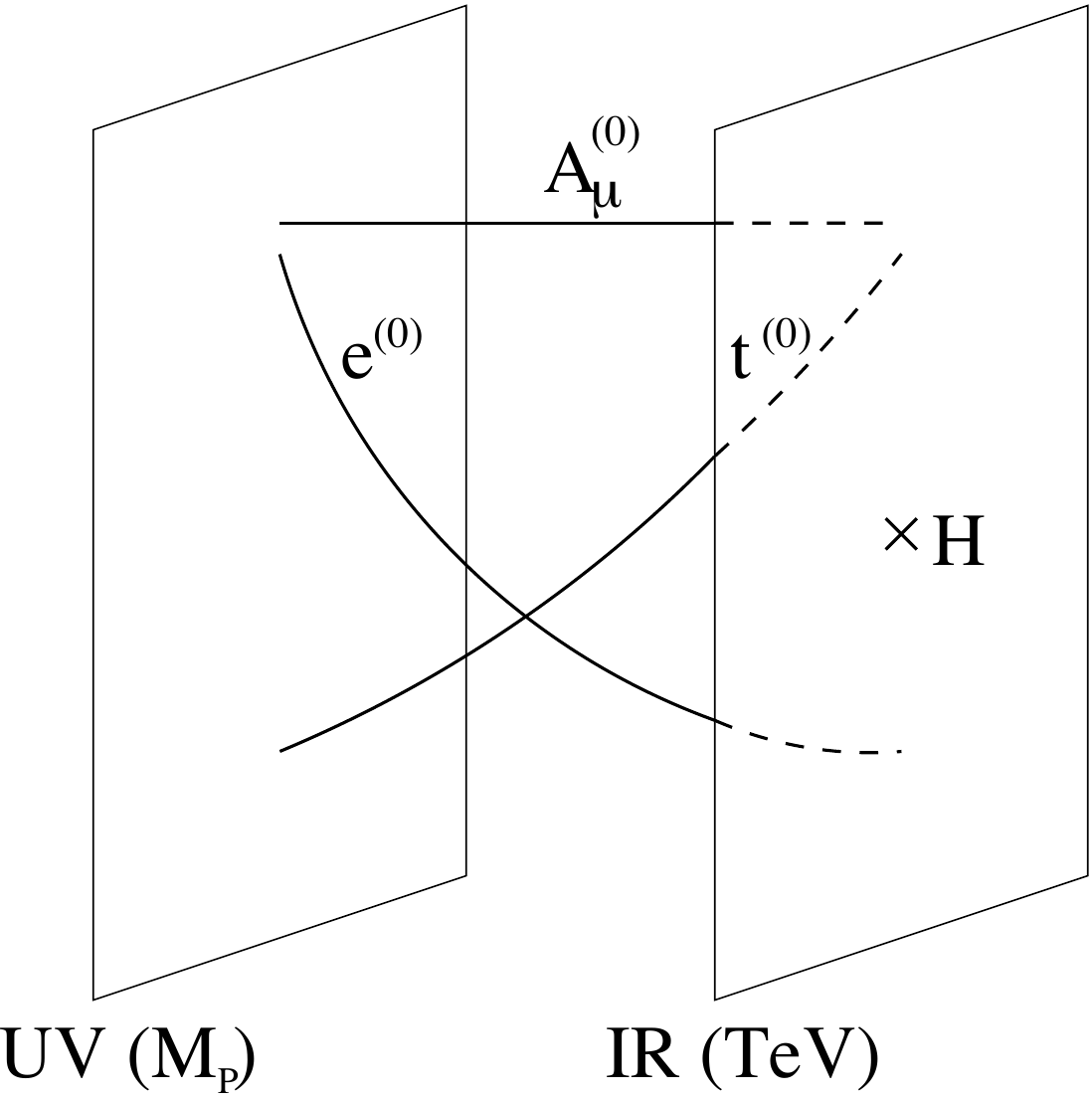}
\end{center}
\caption{\it The Standard Model in the warped five-dimensional bulk.}
\label{bulkSMfig}
\end{figure}

\newpage
\subsection{SUMMARY}

To solve the gauge hierarchy problem only the Higgs boson needs to be confined on the 
IR brane. This allows fermions and gauge bosons to propagate in the bulk. From Table 1 
the gauge field zero mode is flat whereas fermion zero modes can be localized anywhere 
in the bulk so that wavefunction overlap with the Higgs naturally leads to Yukawa 
coupling hierarchies. The picture that emerges is a Standard Model in the warped bulk as depicted 
in Figure~\ref{bulkSMfig}. The fermions are localized to varying degrees in the bulk 
with the electron, being the lightest charged fermion, furthest away from the IR-confined 
Higgs while the top quark, being the heaviest, is closest to the Higgs. Dirac neutrino masses
are also naturally incorporated. Thus the warped dimension not only solves the gauge hierarchy 
problem but also addresses the Yukawa coupling hierarchies.

\section {AdS/CFT and Holography}
Remarkably 5D models in a slice of AdS can be given a purely 4D description. This relation 
between a 5D theory and a field theory in one less dimension is holographic and 
originates from the AdS/CFT correspondence in string theory. In 1997 Maldacena conjectured 
that~\cite{malda}
\begin{equation}
\begin{tabular}{c}
   {\rm type IIB string theory}\\
     on  AdS$_5\times S^5$     
\end{tabular}
\begin{tabular}{c}
{\footnotesize\sc DUAL}\\[-2mm]
$\quad\Longleftrightarrow\quad$
\end{tabular}
\begin{tabular}{c}
${\cal N}=4$~~SU(N) 4D gauge theory 
\end{tabular}
\nonumber
\end{equation}
where ${\cal N}$ is the number of supersymmetry generators and $S^5$ is the five-dimensional sphere. 
The parameters of the correspondence were found to be related by 
\begin{equation}
      \frac{R_{AdS}^4}{l_s^4} = 4\pi g_{YM}^2 N,
      \label{adscftrelation}
\end{equation}
where the AdS$_5$ curvature length $R_{AdS}\equiv 1/k$, $l_s$ is the string length and 
$g_{YM}$ is the SU(N) Yang-Mills gauge coupling. Furthermore symmetries on both sides of 
the correspondence are also related. The isometry of $S^5$ is the rotation group SO(6) $\cong$ 
SU(4), which is the same as the R-symmetry group of the supersymmetric gauge theory. 
Similarly, the ${\cal N}=4$ gauge theory is a conformal field theory (CFT) because the 
isometry group of AdS$_5$ is precisely the conformal group in four dimensions. 
In particular this means that gauge couplings do not receive quantum corrections and 
therefore do not run with energy. Thus we see that a very special 4D gauge theory is conjectured 
to be equivalent to strings propagating on a particular ten-dimensional curved 
background AdS$_5\times S^5$.

What are the consequences of the AdS/CFT correspondence for simple 5D 
gravitational models? We have only considered gravity in the warped bulk which
represents the effective low-energy description of the full string theory. In 
order to neglect the string corrections, so that the bulk gravity description 
is valid, we require that $R_{AdS}\gg l_s$. Using (\ref{adscftrelation}) this leads to 
the condition that $g_{YM}^2 N \gg 1$, which means that the 4D dual CFT is strongly 
coupled!\footnote{In addition one also requires that the string coupling, 
$g_s\rightarrow 0$ so that nonperturbative string states with masses $\sim 1/g_s$ 
remain heavy. Since $g_s\sim1/N$ this separately requires that $N\rightarrow \infty$.} 
Thus for our purposes the correspondence takes the form of a duality in which 
the weakly coupled  5D gravity description is dual to a strongly coupled 
4D CFT. This remarkable duality means that any geometric configuration of 
fields in the bulk can be given a purely 4D description in terms of a 
strongly coupled gauge theory. Therefore warped models provide a new way 
to study strongly-coupled gauge theories.

While there is no rigorous mathematical proof of the AdS/CFT conjecture,
it has passed many nontrivial tests and an AdS/CFT dictionary to relate the 
two dual descriptions can be established~\cite{magoo}. Let us begin with the basic objects
of the two theories. The 5D bulk description is characterized
by set of bulk fields, while a CFT is characterized by a set of operators, $\cal O$. 
Therefore for every 5D bulk field $\Phi$ there is an associated operator $\cal O$ of 
the CFT
\begin{equation}
      \Phi(x^\mu,y) \quad\Longleftrightarrow\quad {\rm CFT~operator,}~~ {\cal O}
\end{equation}
where the boundary value of the bulk field
\begin{equation}
      \Phi(x^\mu,y)\bigg |_{\rm AdS~boundary} \equiv \phi_0(x^\mu)
\end{equation}
acts as a source field for the CFT operator ${\cal O}$.
For the AdS$_5$ metric (\ref{adsmetric}) the boundary of AdS space is 
located at $y=-\infty$. 
The AdS/CFT correspondence can then be quantified in the following way by defining the 
generating functional to be~\cite{gkp,witten}
\begin{equation}
\label{adsrel}
     Z[\phi_0]=\int {\cal D}\phi_{CFT} ~e^{-S_{CFT}[\phi_{CFT}] - \int d^4 x \,\phi_0  
    {\cal O}} = \int_{\phi_0}{\cal D}\phi~ e^{-S_{bulk}[\phi]}
     \equiv e^{i S_{eff}[\phi_0]}~,
\end{equation}
where $S_{CFT}$ is the CFT action with $\phi_{CFT}$ generically denoting the
CFT fields and $S_{bulk}$ is the bulk 5D action. Note that a source term $\phi_0{\cal O}$ 
has been added to the CFT action. The on-shell gravity action, 
$S_{eff}$ is obtained by integrating out the bulk degrees of freedom for 
suitably chosen IR boundary conditions.
In general $n$-point functions can be calculated via
\begin{equation}
    \langle{ \cal O} \dots {\cal O}\rangle = \frac{\delta^n S_{eff}}
      {\delta\phi_0\dots\delta\phi_0}~.
      \label{npointdef}
\end{equation}
In this way we see that the on-shell bulk action is the generating 
functional for connected Greens functions in the CFT. In other words
$n$-point functions for the strongly-coupled CFT can now be computed
from knowing the 5D on-shell bulk action!

So far the correspondence has been formulated purely in AdS$_5$ without the 
presence of the UV and IR branes. In particular notice from (\ref{adsrel}) that 
the source field $\phi_0$ is a nondynamical field with no kinetic term. 
However since we are interested in the 4D dual of a slice of AdS$_5$ (and not
the complete AdS space) we will need the corresponding dual description in the 
presence of two branes.

\subsection{A Slice of AdS/CFT}
There are no mass scales in a CFT because it is invariant under conformal 
transformations. In the complete AdS space where $-\infty < y < \infty$ this 
corresponds to having no branes present. But if branes are added we expect that the 
conformal symmetry will be broken. In fact the position $y$ in the extra dimension 
is related to the 4D energy scale $E$ of the CFT. 

Consider first a UV brane that is placed at $y=0$ (we assume that the 
$-\infty<y<0$ part of AdS space is chopped off and the remaining $0< y< \infty$ 
part is reflected about $y=0$ with a $Z_2$ symmetry). The presence of the UV 
brane with an associated UV scale $\Lambda_{UV}$ corresponds to a CFT with a UV 
cutoff where conformal invariance is explicitly broken at the UV scale. 
Moving away from the UV brane into the bulk corresponds in the 4D dual to 
running down from the UV scale to lower energy scales. Since the bulk is AdS
the 4D dual gauge theory quickly becomes conformal at energies below the UV scale.
This implies that the symmetry breaking terms in the CFT at the UV scale must be irrelevant 
operators~\cite{arp,rz,pv}. 

An important consequence of introducing a UV cutoff on the CFT is that the source field 
$\phi_0$ now becomes dynamical. Not only is a kinetic term for the source field 
always induced by the CFT but one can directly add an explicit kinetic term for the 
source field at the UV scale or equivalently on the UV brane. Thus in the presence of a 
UV brane the generating functional becomes
\begin{eqnarray}
\label{adsbdyrel}
  \int{\cal D}\phi_0~e^{-S_{UV}[\phi_0]} \int_{\Lambda_{UV} }{\cal D}
    \phi_{CFT} ~e^{-S_{CFT}[\phi_{CFT}] - \int d^4 x ~\phi_0  
     {\cal O}}\nonumber \\
  =\int{\cal D}\phi_0 ~e^{-S_{UV}[\phi_0]}\int_{\phi_0}
    {\cal D}\phi~ e^{-S_{bulk}[\phi]}~,
\end{eqnarray}
where $S_{UV}$ is the UV Lagrangian for the source field $\phi_0$ and the 
source field is the UV boundary value of the bulk field i.e. $\phi_0 =\Phi \big |_{y=0}$. 

Next we add an IR brane at $y=\pi R$ which compactifies the fifth dimension and produces
Kaluza-Klein states. In the CFT the conformal symmetry is now broken at the IR scale, 
$\Lambda_{IR}=\Lambda_{UV} e^{-\pi kR}$, causing particle bound states of the CFT to appear. 
However, unlike the breaking associated with the UV brane, the breaking at the IR scale is 
${\it spontaneous}$~\cite{arp,rz}. 
This can simply be understood by noting that the scalar fluctuations of the
metric (\ref{adsmetric}) contain a massless (radion) field with a wavefunction profile
that localizes the mode towards the IR brane. The radion is clearly associated with 
the presence of the IR brane since it remains in the spectrum even if the UV brane is removed.
Therefore at the IR scale the CFT must contain a massless particle which is interpreted 
as the Nambu-Goldstone boson of a spontaneously broken conformal symmetry. 
This so called dilaton is therefore the dual interpretation of the radion. 

Of course this assumes that the interbrane separation is not stabilized. When a brane stabilization 
mechanism is included (such as the Golberger-Wise mechanism), the conformal symmetry is explicitly 
broken leading to a massive radion. The mass is still typically lighter than the IR scale so that the 
radion becomes a pseudo-Nambu-Goldstone boson. This situation is analogous to that which occurs 
in QCD where massless pions are the Nambu-Goldstone bosons of the spontaneously broken 
chiral symmetry at the IR scale $\Lambda_{QCD}$. The chiral symmetry is explicitly broken 
by the quark masses leading to massive pions. Indeed the AdS/CFT correspondence suggests that 
QCD may be the holographic description of a bulk (string) theory.

Thus, the 4D dual interpretation of a slice of AdS not only contains a dual 
CFT with a UV cutoff, but also a dynamical source field $\phi_0$ with UV 
Lagrangian $S_{UV}[\phi_0]$. In particular note that the source field is 
an elementary (point-like) state all the way up to the UV scale, while 
particles in the CFT sector are only effectively point-like below the 
IR scale but are composite above the IR scale. The interaction between the 
elementary (source) sector and the CFT sector then occurs via the source term
$\phi_0 {\cal O}$. We will see that all the features of the 5D warped bulk can be 
understood in terms of the interaction between these two sectors.

\subsection{Holography of Scalar Fields}

As a simple application of the AdS/CFT correspondence in a slice of AdS$_5$
we shall investigate in more detail the dual theory corresponding 
to a bulk scalar field $\Phi$ with boundary mass terms. The qualitative 
features will be very similar for other spin fields. In order to obtain 
the correlation functions of the dual theory we first need to compute the 
on-shell bulk action $S_{eff}$. According to (\ref{genwf}) the bulk scalar 
solution is given by
\begin{equation}
\label{bulkscsoln}
   \Phi(p,z)= \Phi(p) A^{-2}(z)\left[J_\alpha(i q) -\frac{J_{\alpha\pm 1}
    (iq_1)}{Y_{\alpha\pm 1}(i q_1)}Y_\alpha(i q)\right]~,
\end{equation}
where $z=(e^{ky}-1)/k$, $A(z)=(1+kz)^{-1}$, $q=p/(kA(z))$ and $\Phi(p,z)$ is 
the 4D Fourier transform of $\Phi(x,z)$. The $\pm$ refers to the two branches 
associated with $b=b_\pm=2\pm\alpha$. Substituting this solution into the 
bulk scalar action and imposing the IR boundary condition, $(\partial_5 - b k)\Phi \big |_{\pi R}=0$
leads to the on-shell action
\begin{eqnarray}
\label{effS}
    S_{eff} &=& \frac{1}{2}\int \frac{d^4 p}{(2\pi)^4} \left[
     A^3(z)\Phi(p,z)\left(\Phi'(-p,z)- b k \,A(z)\Phi(-p,z)\right)\right]
     \bigg|_{z=z_0}\nonumber \\
    &=&\frac{k}{2}\int \frac{d^4 p}{(2\pi)^4} F(q_0,q_1) \Phi(p) \Phi(-p)~,
\end{eqnarray}
where 
\begin{eqnarray}
    F(q_0,q_1) &=& \mp~iq_0 \left[ J_{\nu \mp 1}(iq_0) - 
Y_{\nu \mp1} (iq_0)\; \frac{J_{\nu}(iq_1)}{Y_{\nu}(iq_1)}
\right]\nonumber\\
&& \qquad\qquad\qquad\times\left[ J_{\nu}(iq_0) - Y_{\nu} (iq_0)\; 
\frac{J_{\nu}(iq_1)}{Y_{\nu}(iq_1)} \right]~,
\end{eqnarray}
and $\nu\equiv\nu_\pm=\alpha\pm 1$. The dual theory two-point function of the operator 
$\cal O$ sourced by the bulk field $\Phi$ is contained in the self-energy $\Sigma(p)$ obtained by
\begin{eqnarray}
\label{sigp}
      \Sigma(p) &=& \int d^4x\, e^{-ip\cdot x} \frac{\delta^2 S_{eff}}
     {\delta(A^2(z_0) \Phi(x,z_0))\delta(A^2(z_0)\Phi(0,z_0))}~,\nonumber\\
     &=& \frac{k}{g_\phi^2}
     \frac{q_0 (I_\nu(q_0)K_\nu(q_1)-I_\nu(q_1)K_\nu(q_0))}
     {I_{\nu\mp 1}(q_0)K_\nu(q_1)+I_\nu(q_1)K_{\nu\mp 1}(q_0)}~,
\end{eqnarray}
where a coefficient $1/g_\phi^2$ has been factored out in front of the
scalar kinetic term in (\ref{scalaraction}), so that $g_\phi$ is a 5D expansion 
parameter with dim$[1/g_\phi^2]=1$.

The behaviour of $\Sigma(p)$ can now be studied for various momentum limits
in order to obtain information about the dual 4D theory.
When $A_1\equiv A(z_1)\rightarrow 0$ the effects of the conformal symmetry
breaking (from the IR brane) are completely negligible. The leading 
nonanalytic piece in $\Sigma(p)$ is then interpreted as the pure 
CFT correlator $\langle \cal O \cal O \rangle$ that would be obtained 
in the string AdS/CFT correspondence with 
$A_0\equiv A(z_0)\rightarrow \infty$. However in a slice of AdS the 
poles of $\langle {\cal O}{\cal O}\rangle$ determines the pure CFT
mass spectrum with a nondynamical source field $\phi_0$.
These poles are identical to the poles of $\Sigma(p)$ 
since $\Sigma(p)$ and $\langle \cal O \cal O \rangle$ only differ by 
analytic terms. Hence the poles of the correlator $\Sigma(p)$ correspond to 
the Kaluza-Klein spectrum of the bulk scalar fields 
with Dirichlet boundary conditions on the UV brane.

There are also analytic terms in $\Sigma(p)$. In the string version of the 
AdS/CFT correspondence these terms are subtracted away by adding appropriate 
counterterms. However with a finite UV cutoff (corresponding to the scale
of the UV brane) these terms are now interpreted as kinetic (and higher 
derivative terms) of the source field $\phi_0$, so that the source becomes 
dynamical in the holographic dual theory. The source field can now mix with 
the CFT bound states and therefore the self-energy $\Sigma(p)$ must be resummed
and the modified mass spectrum is obtained by inverting the whole
quadratic term $S_{UV} + S_{eff}$. In the case with no UV boundary action 
$S_{UV}$, this means that the zeroes of (\ref{sigp}) are identical with 
the Kaluza-Klein mass spectrum (\ref{scKK}) corresponding to (modified) 
Neumann conditions for the source field. In both cases (either Dirichlet or
Neumann) these results are consistent with the fact that the 
Kaluza-Klein states are identified with the CFT bound states.

At first sight it is not apparent that there are an infinite number of 
bound states in the 4D dual theory required to match the infinite number of 
Kaluza-Klein modes in the 5D theory. How is this possible in the 4D gauge 
theory? It has been known since the early 1970's that the two-point 
function in large-$N$ QCD can be written as~\cite{thooft, wittenN}
\begin{equation}
    \langle {\cal O}(p){\cal O}(-p)\rangle = \sum_{n=1}^\infty 
      \frac{F_n^2}{p^2+m_n^2}~,
\end{equation}
where the matrix element for ${\cal O}$ to create the $n$th meson with mass
$m_n$ from the vacuum is $F_n=\langle 0| {\cal O}|n \rangle \propto 
\sqrt{N}/(4\pi)$. In the large $N$ limit the intermediate states are one-meson 
states and the sum must be infinite because we know that the two-point 
function behaves logarithmically for large $p^2$. Since the 4D 
dual theory is a strongly-coupled SU(N) gauge theory that is conformal 
at large scales, 
it will have this same behaviour. This clearly has the same qualitative 
features as a Kaluza-Klein tower and therefore a dual 5D interpretation 
could have been posited in the 1970's!

To obtain the holographic interpretation of the bulk scalar field, recall
that the scalar zero mode can be localized anywhere in the bulk with 
$-\infty <b< \infty$ where $b\equiv b_\pm=2\pm \alpha$ and 
$-\infty < b_- < 2$ and $2 < b_+ <\infty$. Since $b_\pm =1\pm \nu_\pm$ 
we have $-1<\nu_-<\infty$
and $1<\nu_+<\infty$. The $\nu_-$ branch corresponds to $b_- <2$, while
the $\nu_+$ branch corresponds to $b_+> 2$. Hence the $\nu_- (\nu_+)$ branch 
contains zero modes which are localized on the UV (IR) brane.

\subsubsection{$\nu_-$ branch holography}
We begin first with the $\nu_-$ branch. In the limit
$A_0 \rightarrow\infty$ and $A_1\rightarrow 0$ one obtains
\begin{equation}
\label{bmsig}
    \Sigma(p)\simeq -\frac{2k}{g_\phi^2}\left[\frac{1}{\nu}
     \left(\frac{q_0}{2}\right)^2 + \left(\frac{q_0}{2}\right)^{2\nu+2} 
       \frac{\Gamma(-\nu)}{\Gamma(\nu+1)}+\dots\right]~,
\end{equation}
where the expansion is valid for noninteger $\nu$. The expansion for integer
$\nu$ will contain logarithmns. Only the leading analytic term has been 
written in (\ref{bmsig}). The nonanalytic term
is the pure CFT contribution to the correlator $\langle \cal O\cal O\rangle$.
Formally it is obtained by rescaling the fields by an amount $A_0^{\nu+1}$
and taking the limit
\begin{equation}
   \langle {\cal O}{\cal O}\rangle = \lim_{A_0\rightarrow\infty} (\Sigma(p) +
    {\rm counterterms})=\frac{1}{g_\phi^2}\frac{\Gamma(-\nu)}{\Gamma(\nu+1)} 
     \frac{p^{2(\nu+1)}}{(2k)^{2\nu+1}}~.
\end{equation}
Since
\begin{equation}
   \langle {\cal O}(x) {\cal O}(0)\rangle = \int \frac{d^4p}{(2\pi)^4} 
     e^{ipx}\langle {\cal O} {\cal O}\rangle~,
\end{equation}
the scaling dimension of the operator $\cal O$ is
\begin{equation}
    {\rm dim}\,{\cal O} = 3+\nu_- = 4-b_-=2+\sqrt{4+a}~.
    \label{dimOm}
\end{equation}

If $A_0$ is finite then the analytic term in (\ref{bmsig}) becomes the kinetic term 
for the source field $\phi_0$. Placing the UV brane at $z_0=0$ with $A_0=1$ 
leads to the dual Lagrangian below the cutoff scale $\Lambda \sim k$
\begin{equation}
\label{numL}
    {\cal L}_{4D} = -Z_0 (\partial\phi_0)^2 + 
      \frac{\omega}{\Lambda^{\nu_-}}~\phi_0 {\cal O}+{\cal L}_{CFT}~,
\end{equation}
where $Z_0, \omega$ are dimensionless constants. This Lagrangian describes 
a massless dynamical source field $\phi_0$ interacting with the CFT via the 
mixing term $\phi_0{\cal O}$. This means that the mass eigenstate in the dual 
theory will be a mixture of the source field and CFT particle states. The coupling 
of the mixing term is irrelevant for $\nu_- >0~(b_-<1)$, marginal if $\nu_-=0~(b_-=1)$ 
and relevant for $\nu_-<0~(b_->1)$. This suggests the following dual interpretation 
of the massless bulk zero mode. When the coupling is irrelevant ($\nu_- >0$), 
corresponding to a UV brane localized bulk zero mode, the mixing can be neglected 
at low energies, and hence to a very good approximation the bulk zero mode is 
dual to the massless 4D source field $\phi_0$. However for relevant $(-1 <\nu_- <0)$ 
or marginal couplings $(\nu_-=0)$ the mixing can no longer be neglected. In this 
case the bulk zero mode is no longer UV-brane localized, and the dual interpretation 
of the bulk zero mode is a part elementary, part composite mixture of the source field 
with massive CFT particle states. 

The first analytic term in (\ref{bmsig}) can be matched to the wavefunction
constant giving $Z_0 = 1/(2\nu g_\phi^2 k)$. However at low energies 
the couplings in ${\cal L}$ will change. The low energy limit $q_1\ll 1$ 
for $\Sigma(p)$ (and noninteger $\nu$) leads to
\begin{equation}
    \Sigma(p)_{IR}\simeq -\frac{2k}{g_\phi^2}\left[(1-A_1^{2\nu_-})
    \left(\frac{q_0}{2}\right)^2\frac{1}{\nu}+\ldots\right]~,
\label{irsig}
\end{equation}
where $A_1=e^{-\pi kR}$. Notice that there is no nonanalytic term because 
the massive CFT modes have decoupled. The analytic term has now also 
received a contribution from integrating out the massive CFT states. 
Note that when $\nu_->0$ the $A_1$ contribution to $Z_0$ is negligible 
and the kinetic term has the correct sign. On the other hand for relevant 
couplings the $A_1$ term now dominates the $Z_0$ term. The kinetic term
still has the correct sign because $\nu_- <0$.

The features of the couplings in (\ref{numL}) at low energies can be neatly 
encoded into a renormalisation group equation. If we define a dimensionless 
running coupling $\xi(\mu) = \omega/\sqrt{Z(\mu)}(\mu/\Lambda)^{\gamma}$, 
which represents the mixing between the CFT and source sector with a 
canonically normalized kinetic term, then it will 
satisfy the renormalisation group equation~\cite{cp}
\begin{equation}
\label{rge}
    \mu\frac{d\xi}{d\mu} = \gamma\,\xi + \eta
      \frac{N}{16\pi^2}\xi^3+\dots~,
\end{equation}
where $\eta$ is a constant and we have replaced $1/(g_\phi^2 k) = 
N/(16\pi^2)$. The first term arises from the scaling of the coupling
of the mixing term $\phi_0{\cal O}$ (i.e. $\gamma =\nu_-$), 
and the second term arises from the CFT contribution to the wavefunction 
constant $Z_0$ (i.e. the second term in (\ref{bmsig})). The solution of the 
renormalisation group equation for an initial condition $\xi(M)$ at the scale 
$M\sim\Lambda$ is
\begin{equation}
\label{rgesoln}
     \xi(\mu)= \left(\frac{\mu}{M}\right)^\gamma
     \left\{\frac{1}{\xi^2(M)}+\eta\frac{N}{16\pi^2\gamma}
     \left[1-\left(\frac{\mu}{M}\right)^{2\gamma}\right]\right\}^{-1/2}~.
\end{equation}
When $\gamma<0$, the constant $\eta>0$ and the renormalisation group equation 
(\ref{rge}) has a fixed point at $\xi_\ast \sim 4\pi\sqrt{-\gamma/(\eta N)}$,
which does not depend on the initial value $\xi(M)$. This occurs when 
$-1<\nu_-< 0$ and therefore since  $\xi_\ast$ is nonnegligible the mixing 
between the source and the CFT cannot be neglected. 

In the opposite limit, $\gamma>0$, the solution (\ref{rgesoln}) for 
$M\sim\Lambda$ becomes $\xi(\mu)\sim 4\pi\sqrt{\gamma/N}(\mu/M)^\gamma$, 
where the solution (\ref{rgesoln}) has been matched to the low energy value 
$Z(ke^{-\pi kR})=1/(2\gamma g_\phi^2 k)(1-e^{-2\gamma\pi kR})$ arising from 
(\ref{irsig}) (with $\gamma=\nu_-$). Thus when $\nu_- >0$ the mixing between 
the source and CFT sector quickly becomes irrelevant at low energies. 

\subsubsection{$\nu_+$ branch holography}
Consider the case $\nu=\nu_+ >1$. In the limit 
$A_0 \rightarrow\infty$ and $A_1\rightarrow 0$ we obtain for noninteger $\nu$
\begin{equation}
\label{apUVcor}
 \Sigma(p)\simeq -\frac{2k}{g_\phi^2}\left[ (\nu-1) +  \left(\frac{q_0}{2}
  \right)^2\frac{1}{(\nu-2)}+ \left(\frac{q_0}{2}\right)^{2\nu-2}
  \frac{\Gamma(2-\nu)}{\Gamma(\nu-1)}\right]~,
\end{equation}
where only the leading analytic terms have been written. The nonanalytic
term is again the pure CFT contribution to the correlator 
$\langle {\cal O}{\cal O}\rangle$ and gives rise to the scaling
dimension
\begin{equation}
   {\rm dim}\,{\cal O} = 1+\nu_+ = b_+= 2+\sqrt{4+a}~.
   \label{dimOp}
\end{equation}
This agrees with the result for the $\nu_-$ branch. At low energies 
$q_1\ll 1$ one obtains
\begin{equation}
\label{IRcorr}
  \Sigma(p)_{IR} \simeq -\frac{2k}{g_\phi^2}\left[ (\nu-1) + 
   \left(\frac{q_0}{2}\right)^2 \frac{1}{(\nu-2)}  - \nu(\nu-1)^2 \; 
   \frac{A_1^{2\nu}}{A_0^{2\nu}}\,\left(\frac{2}{q_0}\right)^2\right]~,
\end{equation}
where the large-$A_0$ limit was taken first. 
We now see that at low energies the nonanalytic term has a pole at $p^2=0$
with the correlator
\begin{equation}
   \langle {\cal O}{\cal O}\rangle = \frac{8k^3}{g_\phi^2}\nu_+(\nu_+-1)^2
    e^{-2\nu_+\pi kR}\frac{1}{p^2}~,
\end{equation}
where $A_0=1$ and $A_1=e^{-\pi kR}$. This pole indicates that the CFT has 
a massless scalar mode at low energies! What about the massless 
source field? As can be seen from (\ref{apUVcor}) and (\ref{IRcorr}) the 
leading analytic piece is a constant term which corresponds to a mass term 
for the source field~\cite{pv}. This leads to the dual Lagrangian below
the cutoff scale $\Lambda\sim k$
\begin{equation}
      {\cal L}_{4D} = -{\widetilde Z}_0(\partial\phi_0)^2 + m_0^2 \phi_0^2+ 
    \frac{\chi}{\Lambda^{\nu_+-2}} \phi_0 {\cal O} +{\cal L}_{CFT}~,
\end{equation}
where ${\widetilde Z}_0,\chi$ are dimensionless parameters and $m_0$ is a mass 
parameter of order the curvature scale $k$. The bare parameters 
${\widetilde Z}_0$ and $m_0$ can be determined from (\ref{apUVcor}).
Thus, the holographic interpretation is perfectly consistent. There is a 
massless bound state in the CFT and the source field $\phi_0$ 
receives a mass of order the curvature scale and decouples. In the bulk the 
zero mode is always localized towards the IR brane. Indeed for $\nu_+>2$ the 
coupling between the source field and the CFT is irrelevant and therefore the 
mixing from the source sector is negligible. Hence to a good approximation 
the mass eigenstate is predominantly the massless CFT bound state. When 
$1\leq \nu_+ \leq 2$ the mixing can no longer be neglected and the mass 
eigenstate is again part elementary and part composite.

The scalar field example of holography can also be used to describe the holography
of other bosonic fields, such as the graviton ($b=0$) and the gauge boson ($b=1$). 
For the graviton the dual operator is the energy-momentum tensor $T_{\mu\nu}$
with scaling dimension 4 and the holographic theory is similar to that of the 
$\nu_-$ branch of the scalar field. The gauge boson has a dual operator $J_\mu$
with scaling dimension 3 and again the holographic theory resembles the 
$\nu_-$ branch of the scalar field. The holography of fermion fields is qualitatively 
similar to that of the scalar field, although the spinor nature of the fermions 
causes subtle differences. The detailed holographic picture of bulk fermions 
can be found in Ref.~\cite{cp}. For a zero-mode fermion field $\psi_\pm^{(0)}$ the 
dual fermionic operator has scaling dimension $3/2+|c\pm 1/2|$, so just like
the scalar field example, a range of behaviour is encountered by varying the bulk
mass parameter $c$.

\subsection{The Holographic Basis}
We have seen that the 4D dual interpretation of bulk fields in a slice of AdS
consists of an elementary (source) sector interacting with a composite
sector of CFT bound states. The interaction occurs via the source term 
$\phi_0 {\cal O}$ which mixes the two sectors to produce the corresponding
mass eigenstates. In the 5D bulk theory the mass eigenstates are Kaluza-Klein states 
obtained from solving the equations of motion with appropriate boundary conditions. 
Therefore this 4D dual picture can be nontrivially checked by explicitly diagonalising the two 
sectors to reproduce the Kaluza-Klein mass eigenstates.

To represent the mixing taking place between the elementary and composite sectors, 
the 5D action can be decomposed into these two sectors by expanding the 5D bulk field 
$\Phi(x^\mu,y)$ directly in terms of a source field $\varphi^s(x^\mu)$ and a tower of CFT bound 
states $\varphi^{(n)}(x^\mu)$, with the associated wavefunctions $g^s(y)$ and $g_\varphi^{(n)}(y)$:
\begin{equation}
      \Phi(x^\mu,y)=\varphi^s(x^\mu) g^s(y) +\sum_{n=1}^\infty\varphi^{(n)}(x^\mu)g_\varphi^{(n)}(y)~.
\label{holobasis}
\end{equation}
This expansion is referred to as the {\it holographic basis}~\cite{bg1}. As we noted earlier
from the correlator $\Sigma(p)$, the pure CFT spectrum arises from imposing a Dirichlet 
condition at the UV boundary, $\Phi(x^\mu,y)|_{y=0}=0$, and the usual (modified) Neumann 
condition at the IR boundary. The corresponding CFT profiles $g_\varphi^{(n)}(y)$ are then 
given by (\ref{scalarKKprofile}) which satisfy the boundary conditions 
$g_\varphi^{(n)}(y)\big\vert_0 =0$ and $(\partial_5-b k)g_\varphi^{(n)}(y)\big\vert_{\pi R}=0$. 
The source profile is instead taken to be~\cite{bg1}
\begin{equation}
g^s(y)=N_s e^{(4-\Delta) k y}
=\begin{cases}
\sqrt{\frac{2(b-1)k}{e^{2(b-1)\pi k R}-1}~}e^{bky}   \quad\quad~{\rm for} \quad b<2~,\\\\
\sqrt{\frac{2(3-b)k}{e^{2(3-b)\pi k R}-1}}~e^{(4-b)k y} \quad  {\rm for} \quad  b>2~,
\end{cases}
\label{fs}
\end{equation}
where the normalization $N_s$ is chosen to obtain a canonical kinetic term
and $\Delta= 2+|b-2|$ is the scaling dimension of $\cal O$.
These profiles have behaviour that is consistent with the mixing inferred from 
the operator dimension $\Delta$ (\ref{dimOm}) and (\ref{dimOp}). Also note that 
regardless of the basis used, the bulk field $\Phi(x^\mu,y)$ can always be shown
to satisfy the usual Neumann boundary conditions. 

With the holographic basis (\ref{holobasis}) defined, we can now decompose the bulk action 
(\ref{scalaraction}) and examine the elementary/composite mixing in the holographic theory.
By construction this will produce mixing between the source $\varphi^s$ and the CFT fields 
$\varphi^{(n)}$. We will explcitly check that upon diagonalizing this system we will indeed 
reproduce the mass eigenstates derived from the usual Kaluza-Klein procedure.
Inserting the expansion (\ref{holobasis}) into the action (\ref{scalaraction}) (assuming for simplicity 
a real scalar field), gives
\begin{equation}
S=S(\varphi^s)+S(\varphi^{(n)})+S_{mix}~,
\label{Lmix}
\end{equation}
where
\begin{eqnarray}
S(\varphi^s)&=&\int d^4x \left[-\frac{1}{2}(\partial_\mu \varphi^s)^2-\frac{1}{2}M_s^2(\varphi^s)^2\right], \\
S(\varphi^{(n)})&=&\int d^4x \sum_{n=1}^\infty\left[-\frac{1}{2}(\partial_\mu \varphi^{(n)})^2-\frac{1}{2}M_n^2(\varphi^{(n)})^2\right],\\
S_{mix}&=&\int d^4x \sum_{n=1}^{\infty}\left[-z_n\partial_\mu \varphi^s \partial^\mu \varphi^{(n)}-\mu_n^2\varphi^s \varphi^{(n)} \right].
\end{eqnarray}
where $M_n$ are the CFT masses determined from imposing Dirichlet (Neumann) conditions 
on the UV (IR) boundary and $M_s^2$ is defined to be
\begin{equation}
M_s^2=\frac{e^{2(2-b)\pi k R}-1}{e^{2(3-b)\pi k R}-1}4 (b-2)(b-3)k^2~.
\label{sourcemass}
\end{equation}
We see that the two sectors mix in a nontrivial way via kinetic mixing $z_n$ and mass mixing $\mu_n^2$, 
both of which can be computed from wavefunction overlap integrals:
\begin{eqnarray}
z_n&=&\int_0^{\pi R}  dy~e^{-2 k y}g^s g_\varphi^{(n)}, \label{kineticmix}\\
\mu_n^2 &=&\int_0^{\pi R} dy~e^{-4 k y}\left[ \partial_5 g^s \partial_5 g_\varphi^{(n)} + g^sg_\varphi^{(n)}\left(
ak^2 +2 b k   \left(\delta(y)-\delta(y-\pi R)\right)\right)\right]~.\nonumber\\
\label{massmix}
\end{eqnarray}
The kinetic mixing $z_n \neq 0$, which means that the functions $g^s(y)$ and $g_\varphi^{(n)}(y)$ form a nonorthogonal basis.

The system can also be represented more compactly in matrix notation:
\begin{equation}
{\cal L}=\frac{1}{2}\vec{ \varphi}^{\rm T} {\bf Z} \Box \vec{\varphi}-\frac{1}{2}\vec{\varphi}^{\rm T}
{\bf M}^2\vec{\varphi}~,
\end{equation}
where $\vec{\varphi}^{\rm T}=(\varphi^s, \varphi^1, \varphi^2, \cdots)$ and the mixing matrices are defined as
\begin{equation}
{\bf Z}= \left( \begin{array}{ccccc} 1 & z_1 & z_2 & z_3 &\cdots \\ z_1 & 1 & 0 & 0 & \cdots\\
z_2 & 0 & 1 & 0 & \cdots \\ z_3 & 0 & 0 & 1 & \cdots\\ \vdots & \vdots & \vdots & \vdots & \ddots \end{array} \right),\label{kmix}
\end{equation}
\begin{equation}
{\bf M}^2=  \left( \begin{array}{ccccc} M_s^2 & \mu_1^2 & \mu_2^2 & \mu_3^2 &\cdots \\ \mu_1^2 & M_1^2 & 0 & 0 & \cdots\\
\mu_2^2 & 0 & M_2^2 & 0 & \cdots \\ \mu_3^2 & 0 & 0 & M_3^2 & \cdots \\ \vdots & \vdots & \vdots & \vdots & \ddots \end{array} \right).\label{mmix}
\end{equation}

This system can be diagonalized by proceeding in three steps. First we perform an orthogonal rotation in field 
space, $\vec{\varphi}\rightarrow {\rm \bf U} \vec{\varphi}$ which diagonalizes the kinetic portion of the 
Lagrangian. Second, although the resulting kinetic action is diagonal, we must additionally canonically normalize 
the action. This is done via a nonorthogonal diagonal matrix 
${\rm \bf T} = {\rm diag}(1/\sqrt{{\rm eigenvalue}({\bf Z})})$. Altogether, we have
\begin{eqnarray}
{\bf Z} &\rightarrow &{\bf T}~{\bf U}~{\bf Z}~{\bf U}^{\rm T}~{\bf T}={\bf 1}~, \\
{\bf M}^2 &\rightarrow &{\bf T}~{\bf U}~{\bf M}^2~{\bf U}^{\rm T}~{\bf T}={\bf M'}^2~. \label{mp}
\end{eqnarray}
Third, the transformations that diagonalize the kinetic terms will create a more complicated mass matrix 
${\bf M'}^2$ than initially appears in (\ref{mmix}). Another orthogonal field rotation, 
$\vec{\varphi}\rightarrow {\rm \bf V} {\rm \bf T}^{-1} {\rm \bf U} \vec{\varphi}$, must be performed 
which diagonalizes the mass Lagrangian,
\begin{equation}
{\bf M}^2 \rightarrow  {\bf V}~{\bf T}~{\bf U}~{\bf M}^2~{\bf U}^{\rm T}~{\bf T}~{\bf V}^{\rm T}={\bf m}^2~.
\end{equation}
When this is done the diagonalized system can be shown~\cite{bg1} to exactly match the Kaluza-Klein mass eigenbasis:
\begin{equation}
{\bf m}^2= \left( \begin{array}{ccccc} 0 & 0 & 0 & 0 &\cdots \\0  & m_1^2 & 0 & 0 & \cdots\\
0 & 0 & m_2^2 & 0 & \cdots \\ 0 & 0 & 0 & m_3^2 & \cdots\\ \vdots & \vdots & \vdots & \vdots & \ddots \end{array} \right).
\label{massbasis}
\end{equation}
This provides a nontrivial confirmation that the holographic basis indeed describes the mixing
between the elementary (source) and composite CFT sectors.

Finally, we can write the mass eigenstates in terms of the source and CFT fields to see precisely how much 
each mass eigenstate is elementary and composite. Defining $\vec{\phi}^{\rm T}=(\phi^{(0)}, \phi^1, \phi^2, \cdots)$, 
we have
\begin{equation}
\vec{\phi}={\bf V}~{\bf T}^{-1}~{\bf U}~\vec{\varphi}~.
\label{tran}
\end{equation}
Notice that the transformation ${\bf T}$ is not orthogonal, but rather simply a scaling of the fields. Thus, the 
mass eigenstates cannot be written as an orthogonal combination of source and CFT fields. It is still possible 
to characterize the source/CFT content for any given mass eigenstate by examining the corresponding eigenvector.

A nontrivial check of the holographic basis is the existence of a massless zero mode, 
which is true if $\det {\bf M'}^2=0$. It is straightforward to compute this determinant:
\begin{eqnarray}
\det {\bf M'}^2 \propto M_s^2-\sum_{n=1}^\infty \frac{\mu_n^4}{M_n^2}~.
\label{C0}
\end{eqnarray}
On the $(-)$ branch $\det {\bf M'}^2=0$  is trivially satisfied since the source is massless and there is in fact no 
mass mixing. On the $(+)$ branch, there is nontrivial mass mixing as well as kinetic mixing between the source 
and CFT sectors. However, it can be shown that 
$\det {\bf M'}^2=0$ implying that there is indeed a massless eigenstate on the ($+$) branch.

\subsubsection{Partial Compositeness}
The holographic basis correctly describes the elementary/composite mixing of the 4D dual theory
so that mass eigenstates are seen to be composed of part elementary, part composite fields. 
The eigenvectors can be directly obtained by equating the Kaluza-Klein (\ref{scalardecomp}) and holographic 
(\ref{holobasis}) expansions of the bulk field $\Phi(x^\mu,y)$:
\begin{equation}
\sum_{n=0}^\infty\phi^{(n)}(x^\mu)f_\phi^{(n)}(y)=\varphi^s(x^\mu)g^s(y)+\sum_{n=1}^\infty\varphi^{(n)}(x^\mu)g_\varphi^{(n)}(y)~.
\end{equation}
Using the orthonormal condtion (\ref{normcondition}), we can write the mass eigenstate in terms of the source and CFT fields:
\begin{equation}
\phi^{(n)}(x^\mu)=v^{ns}\varphi^s(x^\mu)+\sum_{n=1}^\infty v^{nm}\varphi^{(m)}(x^\mu)~,
\end{equation}
where
\begin{eqnarray}
v^{ns} & = &\int_0^{\pi R} dy\, e^{-2ky} f_\phi^{(n)}(y) g^s(y)~, \label{v1}\\
v^{nm} & = &\int_0^{\pi R} dy\, e^{-2ky} f_\phi^{(n)}(y) g_\varphi^{(m)}(y)~. \label{v2}
\end{eqnarray}

In particular, for the massless mode $\phi^{(0)}(x^\mu)$, the integrals can be performed analytically. 
Consider first the ($-$) branch, $b<2$. Since $g^s(y)=f_\phi^{(0)}(y)$, the eigenvector takes a very simple 
form with $v^{0s}=1$, $v^{0n}=z_n$, where
\begin{equation}
       z_n=-\frac{2k N_s N_n^{CFT}}{\pi M_n^2 Y_\alpha(\frac{M_n}{k})}~,
\end{equation}
and $N_s, N_n^{CFT}$ are normalization constants.
On the ($+$) branch, the source wavefunction (\ref{fs}) is different from $f_\phi^{(0)}(y)$, but it is still straightforward 
to compute the zero mode eigenvector. Consider $v^{0s}$:
\begin{eqnarray}
v^{0s}&=&\sqrt{\frac{(3-b)}{e^{2(3-b)\pi k R}-1}}\sqrt{\frac{(b-1)}{e^{2(b-1)\pi k R}-1}}(e^{2 \pi k R}-1)~,\\
&\simeq&\begin{cases}
\sqrt{(3-b)(b-1)} \quad\quad\quad\quad\quad~{\rm for} \quad 2<b<3~,\\\\
\sqrt{(b-3)(b-1)}e^{-(b-3)\pi k R}   \quad  {\rm for} \quad  b>3~.
\end{cases}
\label{v0+}
\end{eqnarray}
This matches our expectation from the dependence of the dimension of the CFT operator ${\cal O}$ on $b$. 
For $2<b<3$ there is a relevant coupling between the source and CFT sectors, reflected by the fact that the 
source yields an order one contribution to the massless mode in (\ref{v0+}). On the other hand, the source 
contribution to the zero mode content is exponentially suppressed for $b>3$, consistent with our knowledge 
that the source/CFT interaction is irrelevant for large values of $b$.

Similarly the coefficient $v^{0n}$ for $b>2$ is found to be
\begin{equation}
v^{0n}=\frac{-2k N_s N^{CFT}_n}{\pi M_n^2 Y_\alpha(\frac{M_n}{k})}=z_n-\frac{\mu^2_n}{M_n^2}~.
\end{equation}
For the first composite state, which has an exponentially light mass, one obtains $v^{01}\sim 1$
for $b>3$. On the other hand, for the higher composite modes $n>1$, $v^{0n}$ is exponentially suppressed. 
Along with (\ref{v0+}), this tells us that on the (+) branch for $b>3$, the zero mode is effectively the first CFT 
bound state $\phi^{(0)}(x^\mu) \sim \varphi^1(x^\mu)$. Finally, consider massive eigenmodes. On the $(-)$ branch, 
these modes are purely composite and contain no source field. Explicitly, since $g^s(y)=f_\phi^{(0)}(y)$, $v^{ns}=0$ 
by (\ref{normcondition}). However, the massive eigenmodes do become partly elementary on the $(+)$ branch, 
since $v^{ns}\neq 0$.

Finally we end with examples of bulk fields and detail their partial compositeness. The scalar field
theory actually mimics other bosonic field theories such as the graviton $(b=0)$ and the
gauge boson $(b=1)$. A complete analysis of all cases in given in Ref.~\cite{bg1}. Similarly a 
holographic basis for fermions can be constructed~\cite{bg2}, which leads to equivalent results. 
\\
\\
\noindent
$\bullet$ {\bf Graviton}\\
We have seen that the zero mode graviton $h^{(0)}_{\mu\nu}(x^\rho)$ is localized towards the 
UV brane with the profile (\ref{gravprofile}). This suggests that the graviton mode is essentially
a source field and part of the elementary sector. The transformation which diagonalizes the system 
is extremely close to the unit matrix: ${\bf V}{\bf T}^{-1}{\bf U}\simeq {\bf 1}$, so that the massless eigenstate can be written as
\begin{equation}
h_{\mu\nu}^{(0)}(x^\rho)\simeq h_{\mu\nu}^s(x^\rho)+\sin\theta_g \,h_{\mu\nu}^{(1)(CFT)}(x^\rho)+\cdots,
\end{equation}
where
$\sin\theta_g\simeq\theta_g \simeq 2.48~e^{-\pi k R}\simeq 10^{-15}$. Thus we see that
the graviton is effectively equivalent to the source field and is purely elementary. Instead the 
Kaluza-Klein states are purely composite. In particular the first Kaluza-Klein mode decomposes as
\begin{equation}
h_{\mu\nu}^{(1)}(x^\rho)\simeq \cos\theta_g h_{\mu\nu}^{(1)(CFT)}(x^\rho)+\cdots,
\end{equation}
where $\cos\theta_g \sim 1-\theta_g^2$. The higher Kaluza-Klein modes can similarly be written in terms of the CFT states.
\\
\\
\noindent
$\bullet$ {\bf Gauge boson}\\
The zero mode $A^{(0)}_\mu(x^\nu)$ has a flat profile (\ref{gaugeprofile}) and is not localized 
in the bulk. Thus we expect it to have a nontrivial composite mixture in the dual theory. 
The transformation matrix which diagonalizes the gauge field action is
\begin{equation}
\left( \begin{array}{c} A_\mu^{(0)}  \\ A_\mu^{(1)} \\ A_\mu^{(2)} \\ \vdots \end{array} \right)
= \left( \begin{array}{rrrr} 1 & -0.19 & 0.13 &\cdots \\
 0 & -0.98 & -0.03 & \cdots \\
  0 & 0.01 & -0.99 &\cdots \\
\vdots & \vdots & \vdots & \ddots
\end{array} \right)
 \left( \begin{array}{c} A_\mu^s \\ A_\mu^{(1)(CFT)} \\ A_\mu^{(2)(CFT)} \\ \vdots \end{array} \right).
\label{tranamu}
\end{equation}
The zero-mode gauge field is primarily an elementary field. The massive eigenstates, on the other hand, are 
comprised of purely composite fields, with no elementary mixture. An approximate analytic expression can again be 
written for the lowest two states, leading to
\begin{eqnarray}
   A_\mu^{(0)}(x^\nu)  &\simeq& A_\mu^s(x^\nu) +\sin\theta_A \,A_\mu^{(1)(CFT)}(x^\nu)+\dots~,\nonumber\\
      A_\mu^{(1)}(x^\nu)  &\simeq& \cos\theta_A A_\mu^{(1)(CFT)}(x^\nu)+\dots~,
 \end{eqnarray}
where $\sin\theta_A \simeq -1.13/\sqrt{\pi k R}$. In fact, the mixing between the elementary gauge field 
and the corresponding CFT current $J_\mu^{CFT}$ is marginal since $\Delta_J=3$, explaining why the zero 
mode is primarily elementary. The situation is analogous to the (elementary) photon of QED mixing with the 
(composite) spin-1 mesons of QCD associated with cutoff scale $\Lambda_{QCD}$.
The physical photon is a partly composite admixture of QCD bound states.
\\
\\
\newpage
\noindent
$\bullet$ {\bf Light fermions}\\
For light fermions, such as the electron, the compositeness is completely negligible, as was the case for the 
graviton. When $c>1/2$ one can show that~\cite{bg2}
\begin{equation}
    \Psi_\pm^{(0)}(x^\mu)\simeq \psi_\pm^s(x^\mu) + \omega_1 e^{-(c-\frac{1}{2})\pi k R}\, 
    \lambda_\pm^{(1)}(x^\mu)+\dots~,
\end{equation}
where $\omega_1$ is an $O(1)$ coefficient independent of $k$ and $R$ and $\lambda_\pm^{(1)}$
is the first pure CFT fermion state. In the holographic basis we see that the light fermions are 
essentially the source field and therefore are purely elementary.
\\
\\
\noindent
$\bullet$ {\bf Left-handed top and bottom quarks}\\
We now consider a nontrivial case, that of the left-handed top and bottom $Q_{L3}=\left(t_L , b_L\right)$ where there is appreciable mixing between the elementary and composite sectors. The zero mode is mildly localized on the UV brane, and we will take for concreteness $c=0.4$. Transforming from the holographic basis to the mass eigenbasis, we determine the content of each mode:~\cite{bg2}
\begin{equation}
\left( \begin{array}{c} Q_{L3}^{(0)}  \\ Q_{L3}^{(1)} \\ Q_{L3}^{(2)} \\ \vdots \end{array} \right)
= \left( \begin{array}{rrrr} 
 1       &  -0.484   &  0.290    &  \cdots \\
 0       &  -0.874   & -0.200    &  \cdots \\
 0       &  -0.035   &  0.934    &  \cdots \\
\vdots   & \vdots    & \vdots    &  \ddots   
\end{array} \right)
\left( \begin{array}{c} Q_{L3}^{s(1)} \\ Q_{L3}^{(1)(CFT)} \\ Q_{L3}^{(2)(CFT)}  \\ \vdots \end{array}\right).
\label{Q3} 
\end{equation}
We see that the zero mode  contains a significant mixture of CFT bound states. Notice also that the massive modes are purely composite. This case has many of the same features as the gauge boson, which is flat in the bulk and couples marginally to the CFT. 
\\
\\
\noindent
$\bullet$ {\bf Right-handed top quark}\\
Consider now the case of the right-handed top quark $t_R$ which is exponentially peaked on the IR brane. Different values are taken in the literature for the mass $c_R$, but in nearly all cases $c_R<1/2$ in order to obtain an $O(1)$ Yukawa coupling. If $-1/2<c_R<1/2$, the mixing is qualitatively similar to that of $Q_{L3}$ just considered. In particular, the zero mode will be mostly elementary, while the Kaluza-Klein modes are purely composites. When $c_R\sim -1/2$, the massless mode is approximately half elementary and half CFT bound states. This is consistent with the scaling dimension of the dual operator, which takes its lowest value at this point. 

Now consider $c_R<-1/2$. On this branch, it can be shown that the source field marries with a new elementary field, picking up a mass of order $k$, and that there is an ultra-light mode in the CFT 
spectrum~\cite{bg2}. We therefore expect the right-handed top quark to be primarily a composite state. 
Taking $c=-0.7$ for concreteness, the transformation matrix becomes~\cite{bg2}
\begin{equation}
\left( \begin{array}{c} t_{R}^{(0)}  \\ t_{R}^{(1)} \\ t_{R}^{(2)} \\ t_{R}^{(3)} \\ \vdots \end{array} \right)
= \left( \begin{array}{rrrrr} 
  0.9796 & \sim -1    &  \sim 0   & \sim 0 &   \cdots \\
  -0.1816 &  \sim 0   & \sim -1     & \sim 0 &   \cdots \\
  0.0514 &  \sim 0   &  \sim 0   & \sim - 1 &   \cdots \\
  0.0471 &  \sim 0   & \sim 0    & \sim 0 &   \cdots \\
\vdots   & \vdots    & \vdots    & \vdots &   \ddots   
\end{array} \right)
\left( \begin{array}{c} t_{R}^{s} \\ t_{R}^{CFT(1)} \\ t_{R}^{CFT(2)}  \\ t_{R}^{CFT(3)} \\ \vdots \end{array}\right).
\label{tR} 
\end{equation}  
This shows that the right-handed top quark is approximately a 50/50 mixture of the source field and the first CFT composite state. Furthermore the Kaluza-Klein modes now contain some elementary component, 
which differs from the case when $c>-1/2$. 

\subsubsection{Gauge symmetries}
Finally note that local symmetries in the bulk, such as gauge symmetries 
or general coordinate invariance, also have a 4D dual interpretation~\cite{arp,ad}. 
After compactification, a 5D gauge boson (or graviton) leads to a massless 4D zero mode 
plus an infinite tower of massive Kaluza-Klein states. We have seen that the IR localized 
Kaluza-Klein states are interpreted as CFT bound states, while the massless gauge boson 
(or graviton) is a field of the elementary (source) sector. Since the CFT contains no massless 
field, the bulk gauge symmetry 
appears as a global symmetry of the bound state spectrum. Adding the massless 
elementary (source) gauge field with a perturbative gauge coupling, ``weakly gauges" the 
global symmetry of the CFT. 
Therefore, the holographic dual of a bulk gauge theory with local symmetry group $G$ broken 
to $H$ on the UV brane is a CFT in which a subgroup $H$ of the global symmetry group $G$ 
of the CFT is weakly gauged by the source gauge fields of the elementary sector.

\newpage
\subsection{SUMMARY}
\label{AdSCFTsummary}
The AdS/CFT dictionary for a slice of AdS$_5$:
\begin{center}
\begin{tabular}{c}
slice of AdS$_5$ \\
(5D gravity)
\end{tabular}
\begin{tabular}{c}
{\footnotesize\sc DUAL}\\[-2mm]
$\quad\Longleftrightarrow\quad$
\end{tabular}
\begin{tabular}{c}
4D elementary (source) sector \\
+\\
strongly-coupled 4D CFT\\
(spontaneously broken in IR)
\end{tabular}
\end{center}

\vspace{0.5cm}
\noindent
$\bullet$ {\bf Zero modes $(m_0=0)$}
\begin{center}
UV brane localized field
\begin{tabular}{c}
{\footnotesize\sc DUAL}\\[-2mm]
$\quad\Longleftrightarrow\quad$
\end{tabular}
$|\phi^{(0)}\rangle \simeq |\varphi^s\rangle + \epsilon |\varphi_{CFT}\rangle
\qquad(\epsilon \ll 1)$
\end{center}
\begin{center}
IR brane localized field
\begin{tabular}{c}
{\footnotesize\sc DUAL}\\[-2mm]
$\quad\Longleftrightarrow\quad$
\end{tabular}
$|\phi^{(0)}\rangle \simeq \epsilon\,|\varphi^s\rangle + |\varphi_{CFT}\rangle
\qquad(\epsilon \ll 1)$
\end{center}

\vspace{0.5cm}
\noindent
$\bullet$ {\bf Kaluza-Klein modes $(m_n\neq 0)$}
\begin{center}
$\phi^{(n)}(x^\mu)$
\begin{tabular}{c}
{\footnotesize\sc DUAL}\\[-2mm]
$\quad\Longleftrightarrow\quad$
\end{tabular}
\begin{tabular}{c}
CFT bound states!\\
($|\phi^{(n)}\rangle \simeq \epsilon\, |\varphi^s\rangle +  |\varphi_{CFT}\rangle
\qquad(\epsilon \ll 1)$
\end{tabular}
\end{center}

\noindent
$\bullet$ {\bf Bulk mass, $m_\Phi$}
\begin{center}
\begin{tabular}{c}
\begin{tabular}{|c|c|}
\hline
  & {\rm mass}\\
\hline
$\phi^{(0)}$ &  $a$\\
$\psi^{(0)}_\pm$ & $c$ \\
$A_\mu^{(0)}$ & 0 \\
$ h_{\mu\nu}^{(0)}$ & 0\\
\hline
\end{tabular}
\end{tabular}
\begin{tabular}{c}
{\footnotesize\sc DUAL}\\[-2mm]
$\quad\Longleftrightarrow\quad$
\end{tabular}
\begin{tabular}{c}
\begin{tabular}{|c|c|}
\hline
 &  dim $\cal O$ \\
\hline
$\phi^{(0)}$ &  $2+\sqrt{4+a}$\\
$\psi^{(0)}_\pm$ & $\frac{3}{2}+|c\pm \frac{1}{2}|$ \\
$A_\mu^{(0)}$ & 3\\
$ h_{\mu\nu}^{(0)}$ & 4\\
\hline
\end{tabular}
\end{tabular}
\end{center}

\vspace{0.5cm}
\noindent
$\bullet$ {\bf Symmetries}
\begin{center}
\begin{tabular}{c}
{\it Bulk gauge symmetry} $G$,\\ 
{\it broken to} $H$ {\it on UV brane}
\end{tabular}
\begin{tabular}{c}
{\footnotesize\sc DUAL}\\[-2mm]
$\quad\Longleftrightarrow\quad$
\end{tabular}
\begin{tabular}{c}
{\it CFT global symmetry} $G$, \\
{\it with weakly gauged} \\ {\it subgroup} $H$
\end{tabular}
\end{center}

\newpage
\section{Dual 4D Description of the Standard Model in the Bulk}

Using the AdS/CFT dictionary we can now give a 4D dual description of the 
Standard Model in the bulk. Since the Higgs field is confined to the IR brane, it
is interpreted as a pure CFT bound state in the dual 4D theory. This means 
that the RS1 solution to the hierarchy problem is holographically identical to 
the way 4D composite Higgs models~\cite{gk} solve the problem via a low-scale cutoff. 
The Higgs mass is quadratically divergent but only sensitive to the strong-coupling scale 
$\Lambda_{IR}=\Lambda_{UV} e^{-\pi kR}$, which is hierarchically smaller than 
$\Lambda_{UV}$. The identification is:
\begin{center}
{\bf 5D}:\quad
 $m_H^2= (M_5 e^{-\pi k R})^2$ 
$\quad\Longleftrightarrow\quad$
{\bf 4D}:\quad $m_H^2 = \Lambda_{IR}^2$
\end{center}

The remaining Standard Model fields propagate in the bulk. In the holographic basis 
we have seen that the bulk gauge bosons are mostly elementary states but with a 
sizeable admixture of CFT bound states. Therefore the standard model gauge group 
$SU(3)\times SU(2)_L \times U(1)_Y$ is a global symmetry of the CFT, which is weakly 
gauged by the gauge bosons of the elementary (source) sector. Similarly the bulk standard 
model fermions are also admixtures of elementary and CFT  fields. To obtain a large top Yukawa 
coupling the top quark was localized near the IR brane, so in the dual 4D theory the 
top quark is (predominantly) a composite state of the CFT. The remaining fermions are 
localized to varying degrees towards the UV brane, with the lightest fermions being the most
elementary particle states in the dual theory. Thus, the Standard Model in the warped 5D bulk 
is dual to a 4D strongly-coupled CFT interacting with a 4D elementary sector where the mass 
eigenstates are an admixture of elementary and composite states. This picture is not
too dissimilar from the elementary states of QED, such as the photon, mixing with
the bound states of QCD, such as the $\rho$ meson, to form the mass eigenstates.

\subsubsection{Yukawa couplings}
\begin{figure}
\begin{center}
\includegraphics[width=0.35\textwidth,height=0.19\textheight]{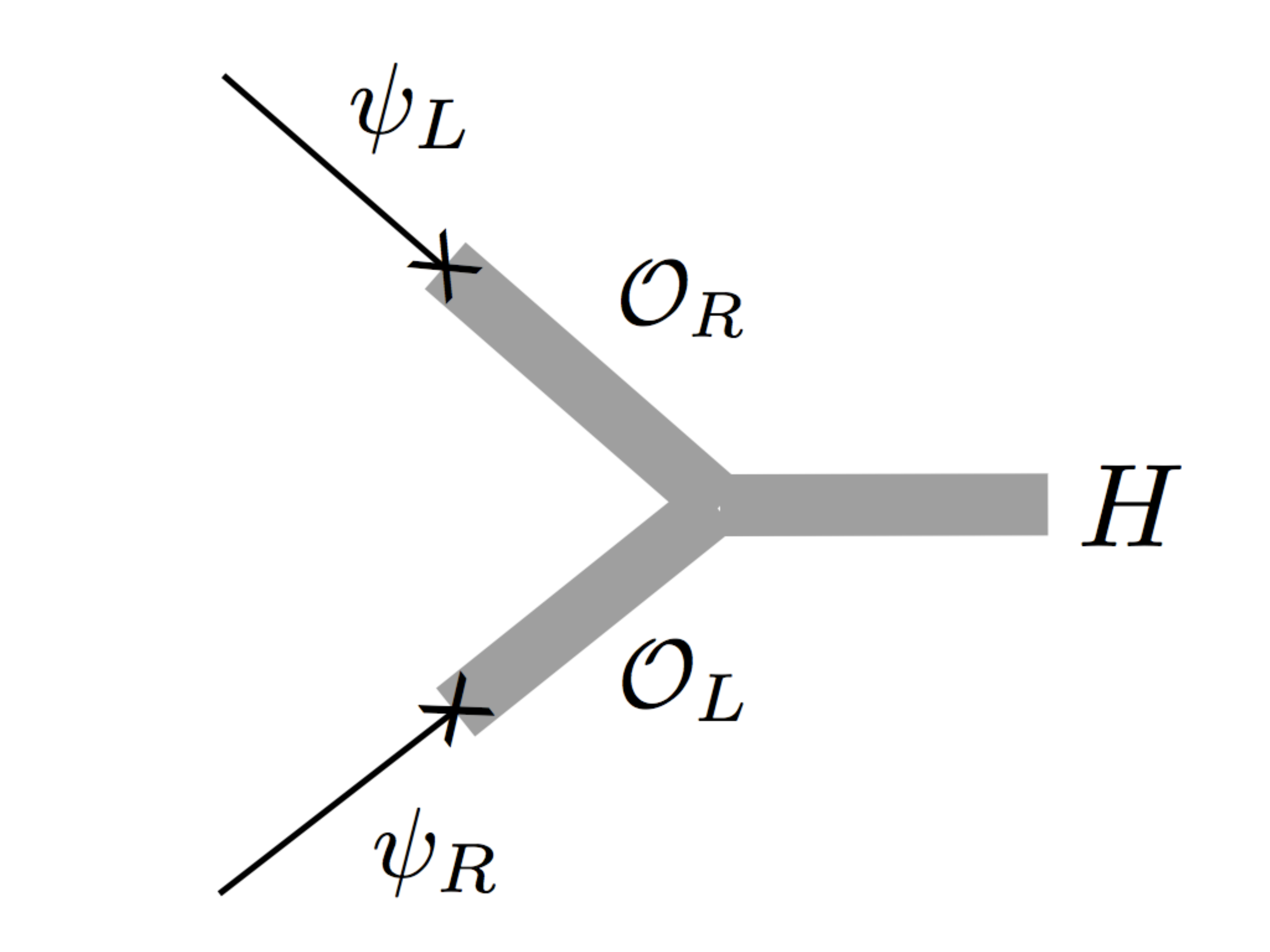}
\end{center}
\caption{\it The three-point Yukawa coupling vertex in the 4D dual theory
when the fermions are predominantly elementary (source) fields.}
\label{Yukcoup}
\end{figure}

The Yukawa coupling hierarchies can also be understood from the dual 4D theory.
Consider first an electron (or light fermion) with $c>1/2$. In the dual 4D 
theory the electron is predominantly an elementary field. The dual 4D 
Lagrangian is obtained from analysing $\Sigma(p)$ for fermions, where the CFT
induces a kinetic term for the source field $\psi_L^{(0)}$. 
It is given by~\cite{cp}
\begin{equation}
\label{dualfermL}
     {\cal L}_{4D} = {\cal L}_{CFT} + Z_0 \bar\psi_L^{(0)} i\gamma^\mu
     \partial_\mu\psi_L^{(0)} + \frac{\omega}{\Lambda^{|c+\frac{1}{2}|-1}} 
     (\bar\psi_L^{(0)} {\cal O}_R +h.c.)~,
\end{equation}
where $Z_0, \omega$ are dimensionless couplings and 
dim ${\cal O}_R = 3/2 +|c+1/2|$. The source field $\psi_L^{(0)}$ pertains to 
the left-handed electron $e_L$ and a similar Lagrangian is written for the 
right-handed electron $e_R$. At energy scales $\mu<k$ we have a renormalisation 
group equation like (\ref{rge}) for the mixing parameter $\xi$ but with $\gamma=|c+1/2|-1$.
Since $c>1/2$ the first term in (\ref{rge}) dominates and the coupling 
$\xi$ decreases in the IR. In particular at the TeV scale ($ke^{-\pi kR}$) 
the solution (\ref{rgesoln}) gives
\begin{equation}
     \xi({\rm TeV})\sim \sqrt{c-\frac{1}{2}} \frac{4\pi}{\sqrt{N}} 
    \left(\frac{k e^{-\pi kR}}{k}\right)^{c-\frac{1}{2}}
    = \sqrt{c-\frac{1}{2}} \frac{4\pi}{\sqrt{N}}e^{-(c-\frac{1}{2})\pi kR}~.
    \label{xiIRsoln}
\end{equation}
The actual physical Yukawa coupling $\lambda$ follows from the three-point 
vertex between the physical states. Since both $e_L$ and $e_R$ are
predominantly elementary they can only couple to the composite Higgs
via the mixing term in (\ref{dualfermL}). This is depicted in 
Fig.~\ref{Yukcoup}. In a large-$N$ gauge theory the matrix
element $\langle 0| {\cal O}_{L,R}|\Psi_{L,R}\rangle \sim \sqrt{N}/(4\pi)$,
and the vertex between three composite states 
$\Gamma_3\sim 4\pi/\sqrt{N}$~\cite{wittenN}.
Thus if each of the elementary fields $e_L$ and $e_R$ mixes in the same way 
with the CFT so that $c_{eL}=-c_{eR}\equiv c$ then
\begin{equation}
     \lambda \propto  \langle 0|{\cal O}_{L,R}| 
       \Psi_{L,R}\rangle^2~\Gamma_3~\xi^2({\rm TeV})
    = \frac{4\pi}{\sqrt{N}} (c-1/2) e^{-2(c-\frac{1}{2})\pi kR}~.
\end{equation}
This agrees precisely with the bulk calculation (\ref{yukcoup})
where $\lambda^{(5)}_{ij} k \sim 4\pi/\sqrt{N}$.

Similarly we can also obtain the Yukawa coupling for the top quark
with $c\lesssim -1/2$ in the dual theory. For this value of $c$, the top quark is 
mostly a CFT bound state in the dual theory and we can neglect the mixing 
with the CFT. As in the scalar field example this follows from the fact 
that the two point function $\langle {\cal O}_R \bar{\cal O}_R\rangle$ now has 
a massless pole. The CFT will again generate a mass term for the massless 
source field, so that the only massless state in the dual theory is the CFT bound 
state. The dual Lagrangian is given by~\cite{cp}
\begin{eqnarray}
    {\cal L}_{4D} &=& {\cal L}_{CFT} 
    + Z_0~\bar\psi_L^{(0)} i\gamma^\mu\partial_\mu\psi_L^{(0)} 
    + {\tilde Z}_0~\bar\chi_R i\gamma^\mu\partial_\mu\chi_R\nonumber\\
    &&~+ d~k~({\bar\chi}_R \psi_L^{(0)} + h.c. )
    +~\frac{\omega}{\Lambda^{|c+\frac{1}{2}|-1}}~({\bar\psi}_L^{(0)} 
    {\cal O}_R +h.c.)~,
\end{eqnarray}
where $Z_0, {\tilde Z}_0, d, \omega$ are dimensionless constants.
The fermion $\psi_L^{(0)}$ pertains to $t_L$ and a similar Lagrangian is 
written for $t_R$. Just as in the scalar case this dual Lagrangian is 
inferred from the behaviour of $\Sigma(p)$ for fermions.
The CFT again induces a kinetic term for the source field $\psi_L^{(0)}$ but 
also generates a Dirac mass term of order the curvature scale $k$ with a 
new elementary degree of freedom $\chi_R$. Hence the elementary source field 
decouples from the low energy spectrum and the mixing term is no longer
relevant for the Yukawa coupling. Instead the physical Yukawa coupling
will arise from a vertex amongst three composite states so that
$\lambda_t \sim \Gamma_3 \sim 4\pi/\sqrt{N} \sim \lambda^{(5)} k$, and 
consequently there is no exponential suppression in the Yukawa coupling. 
This is again consistent with the bulk calculation.

\subsubsection{The GIM mechanism}

\begin{figure}
\begin{center}
\includegraphics[width=0.3\textwidth,height=0.18\textheight]{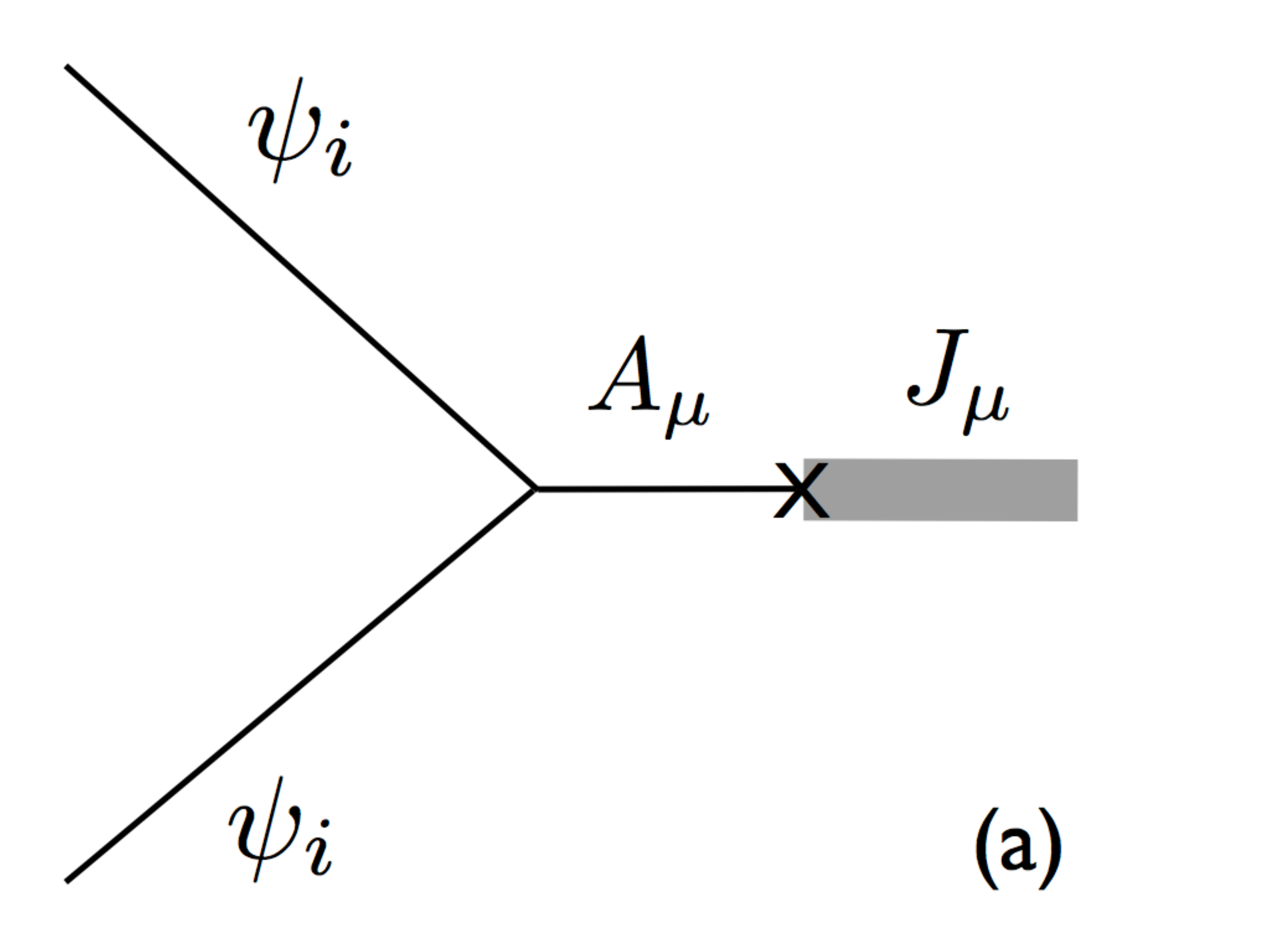}
\includegraphics[width=0.3\textwidth,height=0.18\textheight]{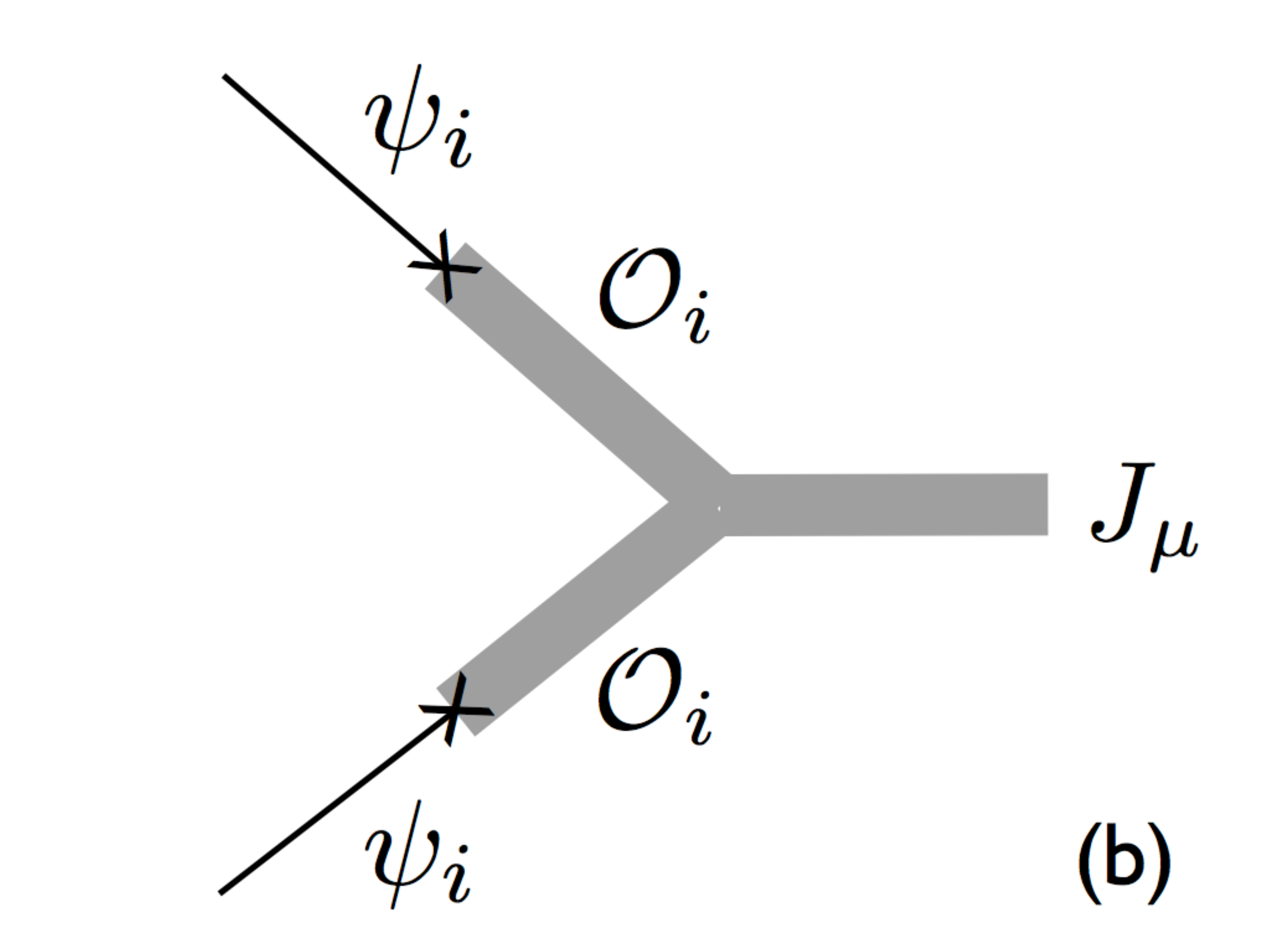}
\end{center}
\caption{\it The gauge interaction between the light fermions and the vector current 
in the 4D dual theory: (a) An elementary vertex between the fermion and gauge source fields. 
(b) A three-point vertex between the fermion and vector currents. }
\label{GIMfig}
\end{figure}

Next we would like to see how the GIM-like mechanism arises in the dual 4D theory.
Consider the gauge interaction between the light fermion zero modes and the gauge 
boson Kaluza-Klein modes. Using the holographic basis we know that the light fermion zero modes
are almost pure elementary (source) states, while gauge boson Kaluza-Klein modes are pure
CFT bound states. Therefore, in the dual 4D theory the gauge interaction must be between
source fermion fields and the vector current of the CFT. This can occur in two ways as 
depicted in Fig~\ref{GIMfig}.

The first contribution arises from the fermion and gauge boson source states in the elementary 
sector which interact in a flavor universal way due to gauge invariance. The gauge boson source then couples 
marginally to the CFT current. The dual 4D Lagrangian has the form
\begin{equation}
   {\cal L}_{4D} = {\cal L}_{CFT}  + Z_A F_{\mu\nu}^2 + \omega_A  A_\mu J^\mu,
\end{equation}
where $Z_A,\omega_A$ and dim $J_\mu = 3$ for the CFT vector current $J_\mu$.
At energy scales $\mu < k$ the renormalization group equation for the dimensionless mixing 
parameter $\xi_A(\mu) = \omega_A/\sqrt{Z_A(\mu)}$ is similar to (\ref{rge}) except that $\gamma=0$. 
The mixing therefore runs logarithmically and at the IR scale $(k e^{-\pi k R})$ is given by
\begin{equation}
    \xi_A({\rm IR})\sim \frac{4\pi}{\sqrt N} \frac{1}{\sqrt{\pi kR}}.
\end{equation}
The actual gauge coupling follows from the three-point source vertex with 4D coupling $g$. 
Using the fact that in a large $N$ gauge theory 
$\langle 0|J_\mu| A_\mu^{(n)}\rangle \sim \sqrt{N}/(4\pi)$ we obtain
\begin{equation}
      g^{(n)} \propto g \, \xi_A({\rm IR})\, \langle 0|J_\mu | A_\mu^{(n)}\rangle= \frac{g}{\sqrt{\pi k R}}.
\end{equation}
Thus we see that this contribution is (flavor) universal. Note that the coupling is also suppressed
compared to the massless gauge bosons.

The second contribution to the gauge coupling arises from the fermion source-operator mixing 
(\ref{xiIRsoln}) and the three-point vertex between the fermion and vector currents 
$\Gamma_{\Psi\Psi J}$. Since in a large $N$ gauge theory we have 
$\Gamma_{\Psi\Psi J} \sim 4\pi/\sqrt{N}$ the gauge coupling becomes
\begin{equation}
      \Delta g_i^{(n)} \sim \xi^2({\rm IR})   \langle 0|{\cal O}_{L,R}| \Psi_{i+}\rangle^2 
      \langle 0|J_\mu| A_\mu^{(n)}\rangle \Gamma_{\Psi\Psi J} \sim (c_i-1/2) e^{(1-2c_i)\pi k R}~.
\end{equation}
We see that the nonuniversal piece is exponentially suppressed for $c_i>1/2$ and proportional 
to the Yukawa coupling, just like the minimal flavor violation hypothesis~\cite{mfv}. 
This is because the mixing between the CFT and source fermions is irrelevant (due to the large 
anomalous dimensions of CFT operators), so that the coupling via the CFT fermion and vector 
currents is suppressed at low energies. This compares with the gauge field source which couples 
marginally to the CFT current. Therefore for light fermions, $c_i > 1/2$, the universal contribution 
dominates\cite{flavor2}, and the SM fermions primarily interact through the three-point ``source" vertex. 
A more detailed analysis in the holographic basis is given in Ref.~\cite{bg2,Contino:2006nn}.

For heavy quarks the situation is different. The dominant interaction is no longer between elementary 
(source) states because, as we have seen, the third generation quarks can contain a significant 
composite admixture. Therefore for $-3/2<c_i<1/2$ the three-point composite vertex can no longer be 
neglected, leading to sizable couplings (both universal and nonuniversal). This is why heavy quarks, 
like $t_R$, are likely to be important signals of new physics at the LHC, for instance from Kaluza-Klein 
gluons~\cite{lhc}, or flavor violation~\cite{flavor}.

\section{Electroweak Symmetry Breaking}

So far we have said very little about electroweak symmetry breaking and
the Higgs mass. If the Higgs is confined to the IR (or TeV) brane then the 
tree-level Higgs mass parameter is naturally of order 
$\Lambda_{IR}=\Lambda_{UV} e^{-\pi kR}$. 
Since there are fermions and gauge bosons in the bulk the effects of their
corresponding Kaluza-Klein modes must be sufficiently suppressed. This 
requires $k e^{-\pi kR} \sim {\cal O}(\rm TeV)$ and since, in RS1,
$\Lambda_{UV} \sim 10 k$ we have $\Lambda_{IR}\sim {\cal O}(10~{\rm TeV})$. 
Consequently a modest amount of fine-tuning would be required to obtain a 
physical Higgs mass of ${\cal O}(100)$ GeV, as suggested by electroweak 
precision data~\cite{higgs}. 

There are two ways to address this problem. The first approach is to invoke
a symmetry to keep the Higgs mass naturally lighter than the IR cutoff scale.
This can be either the spontaneous breaking of a global symmetry, or as we will
see later, supersymmetry. In the second approach there is no Higgs boson and 
electroweak symmetry is broken by strong dynamics. This leads to so-called
Higgsless models.

However before we discuss these two approaches it is important to note that 
after electroweak symmetry breaking the Higgs sector must give rise to weak-boson
masses satisfying the relation, $m_W=m_Z \cos\theta_w$, where $\theta_w$ is the 
weak-mixing angle, to better than $1\%$ accuracy. This is equivalent to preventing 
excessive contributions to the Peskin-Takeuchi $T$ parameter~\cite{peskintakeuchi}. 
This is enforced in the usual 4D standard model by assuming that the Higgs sector 
is invariant under an unbroken global $SU(2)$ custodial symmetry. In particular, 
ignoring gauge couplings, the $SO(4)$ global symmetry of the Higgs Lagrangian 
is broken by the Higgs vacuum expectation value down to an $SO(3)$ 
(or $SU(2)$) global symmetry. 

In the 4D dual picture the (unspecified) strong dynamics underlying the Higgs 
sector must therefore contain a global $SU(2)$ custodial symmetry after electroweak 
symmetry breaking. This would seem to be achieved with a minimal Higgs potential 
confined to the IR brane. However, the underlying strong dynamics of the 4D dual theory 
is also responsible for the Kaluza-Klein states of bulk fields. After electroweak symmetry 
breaking, with just the standard model gauge group in the bulk, these states will not be
invariant under an $SU(2)$ global symmetry. Indeed exchange of gauge boson Kaluza-Klein 
modes leads to excessive contributions to the $T$ parameter~\cite{TKK}. Instead the bulk gauge 
group must be enlarged to $SU(2)_L\times SU(2)_R\times U(1)_{B-L}$ which is then broken 
to $SU(2)_{L+R}\times U(1)_{B-L}$ on the IR brane, with the $SU(2)_{L+R}$ playing the 
role of the custodial symmetry~\cite{adms}. Of course to obtain the usual standard 
model gauge bosons, the bulk gauge symmetry is broken to $SU(2)_L\times U(1)_Y$ 
on the UV brane. Thus by the AdS/CFT dictionary we have weakly gauged the 
$SU(2)_L\times U(1)_Y$ part of the CFT global symmetry $SU(2)_L\times SU(2)_R\times U(1)_{B-L}$,
which is then spontaneously broken to $SU(2)_{L+R}\times U(1)_{B-L}$ at the IR scale.
Almost all models of electroweak symmetry breaking in a slice of AdS$_5$ 
assume this form of custodial protection.

\subsection{The Higgs as a pseudo Nambu-Goldstone Boson}
Motivated by the fact that the dimensional reduction of a five-dimensional gauge 
boson $A_M=(A_\mu,A_5)$ contains a scalar field $A_5$, one can suppose that 
the Higgs boson is part of a higher-dimensional gauge field~\cite{higgsxd}. 
In a slice of AdS$_5$ the $A_5$ terms in the gauge boson kinetic term of the bulk 
Lagrangian (\ref{gaugeaction}) are
\begin{equation}
     -\frac{1}{2}\int d^4 x\, dy~e^{-2ky} \left[(\partial A_5)^2 -2
       \eta^{\mu\nu}\partial_\mu A_5 \partial_5 A_\nu \right] + \dots~. 
\end{equation}
In particular notice that the higher-dimensional gauge symmetry prevents a 
tree-level mass for $A_5$. However if the zero mode of the $A_5$ scalar field plays
any role in addressing the hierarchy problem it must be localized near the IR brane. 
The solution for $A_5$ can be obtained by adding a gauge fixing term that cancels the 
mixed $A_5 A_\nu$ term~\cite{rsch,cnp}. This gives the zero mode solution 
$A_5^{(0)}$ with $y$ dependence proportional to $e^{+2 k y}$, which when substituted 
back into the action leads to
\begin{equation}
     -\frac{1}{2}\int d^4 x\, dy~e^{+2ky} (\partial_\mu A_5^{(0)}(x))^2 
     + \dots~.
\end{equation}
Hence with respect to the flat 5D metric the massless scalar mode 
$A_5^{(0)}$ is indeed localized towards the IR brane and therefore
can play the role of the Higgs boson.

\begin{figure}
\begin{center}
\includegraphics[width=0.5\textwidth,height=0.25\textheight]{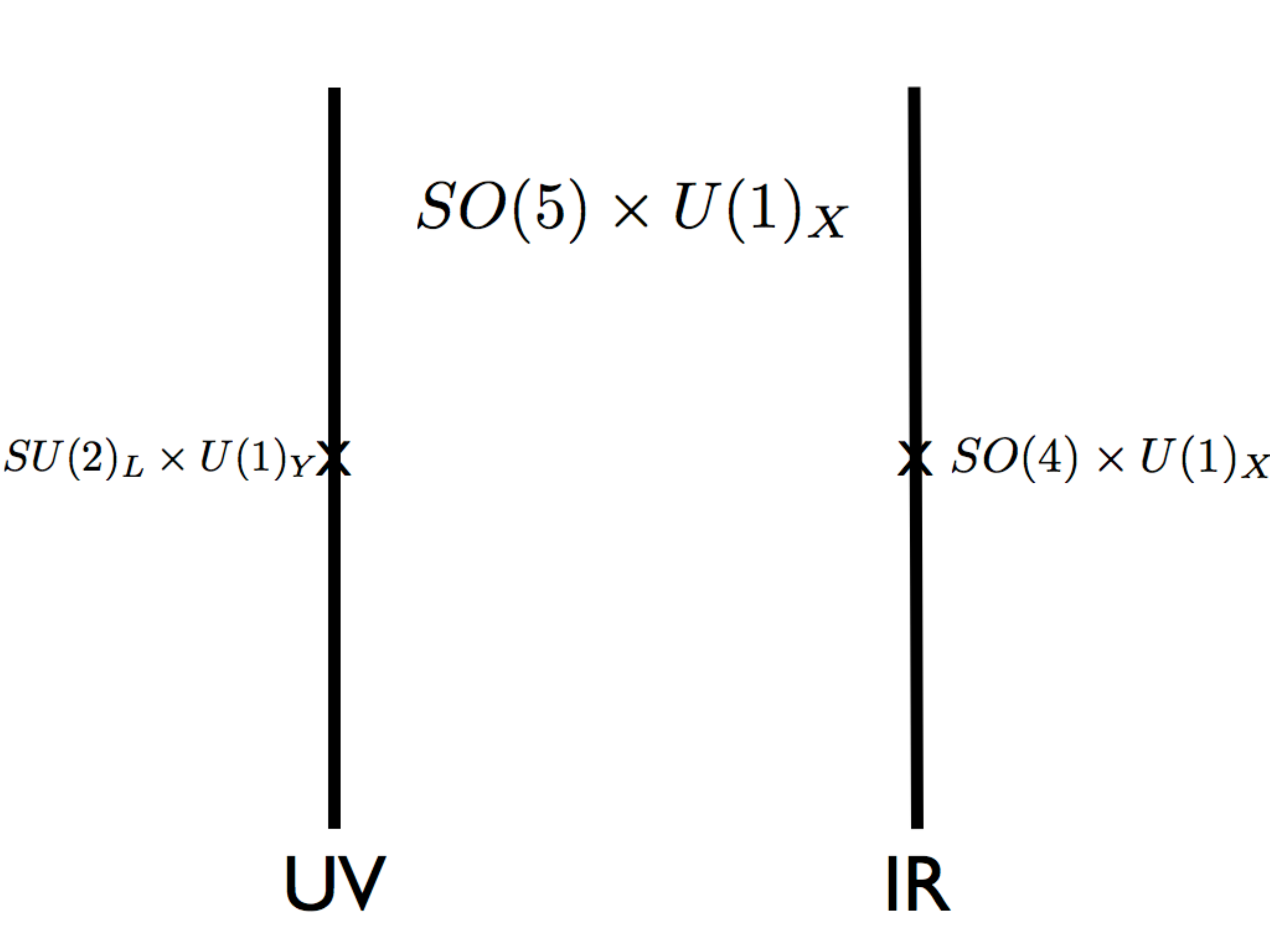}
\end{center}
\caption{\it A schematic diagram of the gauge symmetry breaking pattern
in the composite Higgs model.}
\label{compositeHiggsfig}
\end{figure}

To obtain a realistic model one assumes an SO(5)$\times$U(1)$_X$ bulk gauge 
symmetry for the electroweak sector~\cite{cnp}. On the IR brane this symmetry is 
spontaneously broken by boundary conditions to SO(4)$\times$U(1)$_X$. This leads to four 
Nambu-Goldstone bosons transforming as the ${\bf 4}$ of SO(4), (or a real bidoublet of 
SU(2)$_L\times$SU(2)$_R$), that are identified with the scalar fields in the Standard Model 
Higgs doublet. On the UV boundary the bulk gauge symmetry is reduced to the 
standard model electroweak gauge group $SU(2)_L\times U(1)_Y$, where hypercharge 
$Y$ is defined as $Y=X+T_3^R$ (with $T_3^R$ the third component of the $SU(2)_R$ 
isospin). The setup is depicted in Figure~\ref{compositeHiggsfig}. The bulk fermions must 
also fill out representations of SO(5). It turns out that 
to avoid large corrections to the $Z{\bar b}_L b_L$ coupling, the boundary symmetry is 
enlarged to $O(4)$ (to include discrete transformations) and the top quarks are 
embedded into either the fundamental $({\bf 5})$ or antisymmetric $({\bf 10})$ 
representations of SO(5)~\cite{acdp}.  For example, under $SU(2)_L\times U(1)_Y$  
the fundamental representation decomposes as 
${\bf 5} = {\bf 2}_{7/6}+ {\bf 2}_{1/6}+{\bf 1}_{2/3}$, with ${\bf 2}_{7/6}$ 
containing a fermion with electromagnetic charge $Q=5/3$. The lowest Kaluza-Klein excitation 
of this fermion can be detected at the LHC and represents a smoking-gun signal of the 
model~\cite{cdp}.

A Higgs mass is then generated because the SO(5) gauge 
symmetry is explicitly broken in the fermion sector, in particular by the top 
quark. At one loop this generates an effective potential and electroweak 
symmetry is broken dynamically via top-quark loop corrections~\cite{acp}. 
This effect is finite and arises from the Hosotani mechanism with
nonlocal operators in the bulk~\cite{hosotani}. An unbroken 
$O(3)=SU(2)\times P_{LR}$ custodial symmetry (with $P_{LR}$ representing a 
$L\leftrightarrow R$ discrete symmetry) guarantees that the Peskin-Takeuchi 
parameter $T=0$. The important point however is that radiative corrections to 
the Higgs mass depend on $k e^{-\pi k R}$ and not on $\Lambda_{UV} e^{-\pi kR}$. 
Together with the accompanying one-loop factor $\frac{1}{16\pi^2}$ this 
guarantees a light Higgs mass of order $m_{Higgs} \lesssim 140$ GeV.
Furthermore this model can be shown to pass stringent electroweak precision 
tests without a significant amount of fine-tuning~\cite{ac,cdp}.

Using the AdS/CFT dictionary the Higgs as a pseudo Nambu-Goldstone 
boson scenario has a simple 4D dual interpretation. The bulk SO(5) gauge 
symmetry is interpreted as an SO(5) {\it global} symmetry of the CFT that is 
spontaneously broken down to SO(4) at the IR scale by the (unknown) strong 
dynamics of the CFT. The electroweak gauge bosons weakly gauge the global 
symmetry. The nature of the Nambu-Goldstone bosons can be obtained by
examining the two point function $\langle J_\mu J_\nu\rangle$ of the global 
symmetry current $J_\mu$. In the limit of $p\ll k e^{-\pi k R}$ one finds that~\cite{rz, bg1}
\begin{eqnarray}
\langle J_\mu J_\nu\rangle (p)& \simeq & (p^2\eta_{\mu\nu}-p_\mu p_\nu) \frac{1}{g_5^2 k} 
\left[\log{(i p/2k)}+\gamma-\frac{\pi}{2}\frac{Y_1\left(i p e^{\pi k R}/k\right)}{J_1\left( i p e^{\pi k R}/k\right)}\right] 
\nonumber\\
&\simeq &  (p^2\eta_{\mu\nu}-p_\mu p_\nu)\frac{2(k e^{- \pi k R})^2}{g_5^2 k}\frac{1}{p^2}+\dots,
\end{eqnarray}
where a Dirichlet condition has been imposed at the IR brane corresponding to the
breaking of the global symmetry by the CFT dynamics. We see that there 
is a massless pole corresponding to the exchange of the Nambu-Goldstone mode 
$A_5$. Since it is associated with the global symmetry current of the CFT the Higgs
is interpreted as a composite state in the dual theory~\cite{bg1, acp}. Consequently 
this model is also referred to as the {\it composite Higgs} model. 

To break electroweak symmetry, an effective Higgs potential is generated 
at one loop by explicitly breaking the SO(5) symmetry in the elementary 
(fermion) sector and transmitting it to the CFT. The top quark plays the 
major role in breaking this symmetry. It must be localized near the IR brane 
to obtain a large overlap with the Higgs field and therefore a large Yukawa 
coupling. This means that the top quark will have a sizeable degree of 
compositeness compared to the light fermions. Electroweak symmetry breaking 
therefore crucially depends on the heaviness of the top quark.

In summary, identifying the Higgs scalar field as a pseudo Nambu-Goldstone
boson is equivalent to a 4D composite Higgs model. Although the idea of a 
composite Higgs boson is not new~\cite{earlyCH}, the gravity dual description
provides a new and calculable framework to address the gauge hierarchy, 
fermion mass hierarchies and flavor problems. The partially composite 4D 
model is consistent with electroweak precision tests, and leads to a predictive 
scenario for the electroweak symmetry breaking sector that can be tested at the 
LHC~\cite{ac,cdp}.

\subsection{The warped Higgsless model}
\begin{figure}
\begin{center}
\includegraphics[width=0.5\textwidth,height=0.25\textheight]{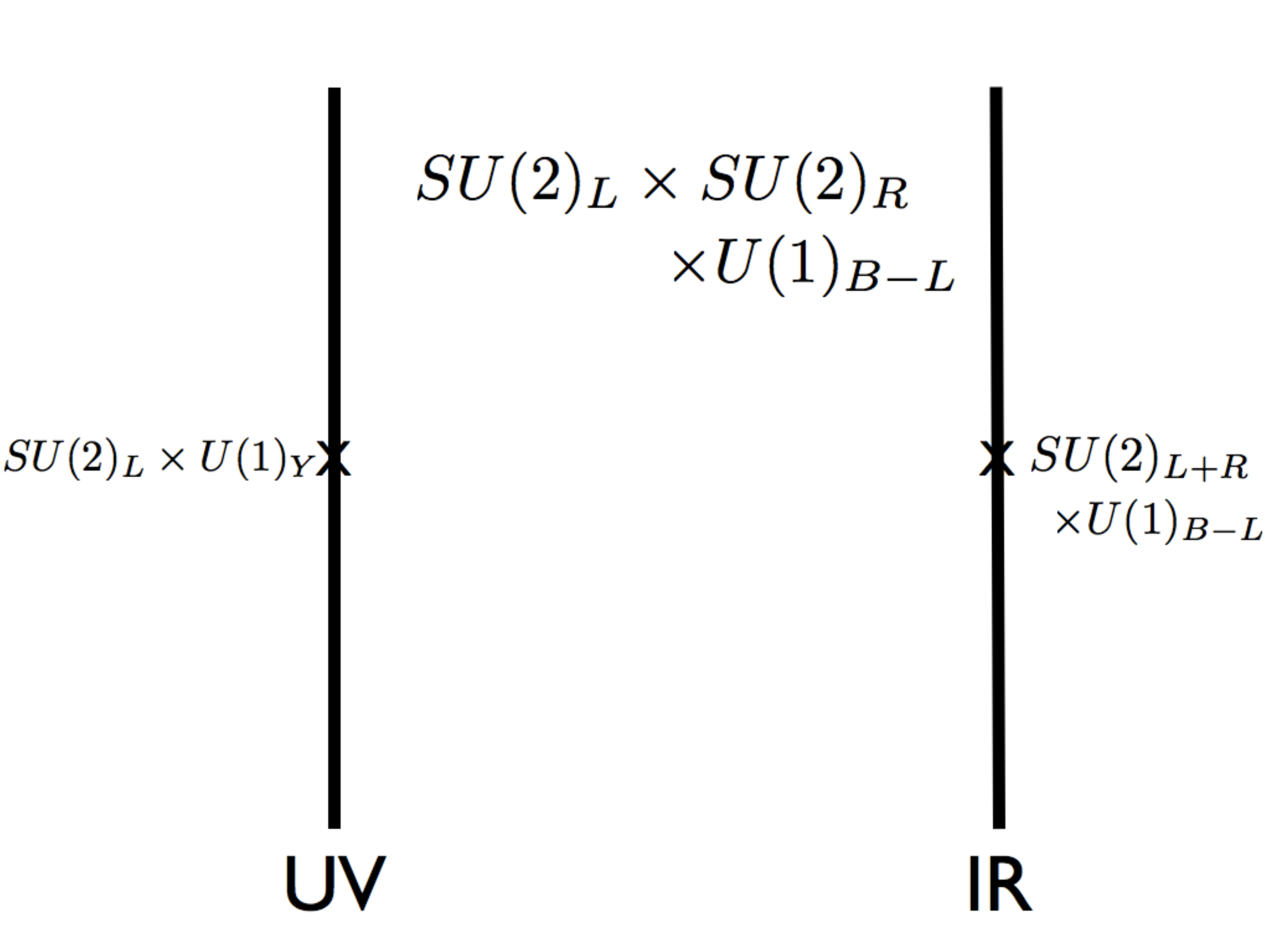}
\end{center}
\caption{\it A schematic diagram of the gauge symmetry breaking pattern
in the warped Higgsless model.}
\label{higgslessfig}
\end{figure}

In the composite Higgs model we have seen that the IR breaking of a bulk gauge 
symmetry leads to four Nambu-Goldstone bosons, three of which eventually become the
longitudinal components of the $W,Z$ gauge bosons, while the fourth remains as a Higgs boson.
A radiatively-generated Higgs potential is then used to break the electroweak symmetry.
However a more economical possibility is to directly break electroweak symmetry by IR 
boundary conditions, thereby eliminating the need for a Higgs boson. 
To ensure custodial protection of the $T$ parameter the bulk gauge group only needs to 
be $SU(2)_L\times SU(2)_R\times U(1)_{B-L}$. On the UV brane this symmetry is broken to 
the electroweak gauge group $SU(2)_L\times U(1)_Y$, while boundary conditions on the 
IR brane are imposed to break the symmetry to $SU(2)_{L+R}\times U(1)_{B-L}$ 
(see Figure~\ref{higgslessfig}). This leads to masses for the $W,Z$ bosons directly without 
any Higgs field. These models are therefore referred to as the warped $\it Higgsless$ 
models~\cite{higgsless}.

The 4D dual interpretation follows from the AdS/CFT dictionary. Since the UV brane preserves
an $SU(2)_L\times U(1)_Y$ symmetry, the massless $W, Z$ gauge bosons are elementary fields
and weakly gauge the CFT global symmetry $SU(2)_L\times SU(2)_R\times U(1)_{B-L}$. 
The (unknown) strong dynamics of the CFT is responsible for breaking the global symmetry 
$SU(2)_L\times SU(2)_R \rightarrow SU(2)_{L+R}$. This produces the requisite three 
Nambu-Goldstone modes (``pions") which are eaten by the elementary $W,Z$ gauge bosons to 
become massive. This 4D dual description is similar to the idea of technicolor~\cite{tcref}, where in 
analogy with QCD, a nonzero technifermion condensate $\langle {\bar Q}  Q\rangle = \langle {\bar Q}_L Q_R
+ {\bar Q}_R Q_L \rangle$ breaks electroweak symmetry with the composite technipions becoming the
longitudinal components of the elementary $W,Z$ bosons. 

The minimal warped Higgsless model is however ruled out by electroweak precision tests.
The custodial symmetry does protect the $T$ parameter but the problem lies with the $S$ parameter
where it is found that at tree level~\cite{bpr,ccgt}
\begin{equation}
     S\simeq 1.15 ; \quad T=0.
\end{equation}
The large value of the $S$ parameter can be tuned away by carefully choosing the profiles of bulk fermions.
If the light fermion profiles are almost flat $(c\simeq 1/2)$ then their coupling to the gauge boson 
Kaluza-Klein modes is vanishingly small thereby suppressing the contributions to the $S$ 
parameter~\cite{adms,fermionccgt}. However a flat profile is difficult to reconcile with a heavy top quark which
requires a profile localized towards the IR brane. This can also lead to a large deviation in the 
observed $Z {\bar b}_L b_L$ coupling~\cite{fermionccgt}. These issues can be addressed by further 
complicating the model and introducing a separate brane specifically for the third generation~\cite{ccgmt}.
Nevertheless the idea that there is no Higgs boson remains a logical possibility that will be tested at the
LHC.

\subsection{Emergent electroweak symmetry breaking}
In the Higgsless model the electroweak gauge bosons obtain their mass from the strong dynamics
associated with the IR scale. As usual, the massive $W,Z$ bosons originate 
from massless gauge bosons, or a fundamental $SU(2)_L\times U(1)_Y$ gauge symmetry. However, 
an alternative viewpoint is to assume that there is no fundamental gauge symmetry and that the 
$W,Z$ bosons originate from massive states associated with conformal dynamics at the IR scale.
This is a radical departure from the usual paradigm where now $SU(3)_C\times U(1)_Q$,
associated with QCD and electromagnetism, are the only fundamental gauge symmetries in the 
Standard Model.

\begin{figure}
\begin{center}
\includegraphics[width=0.5\textwidth,height=0.25\textheight]{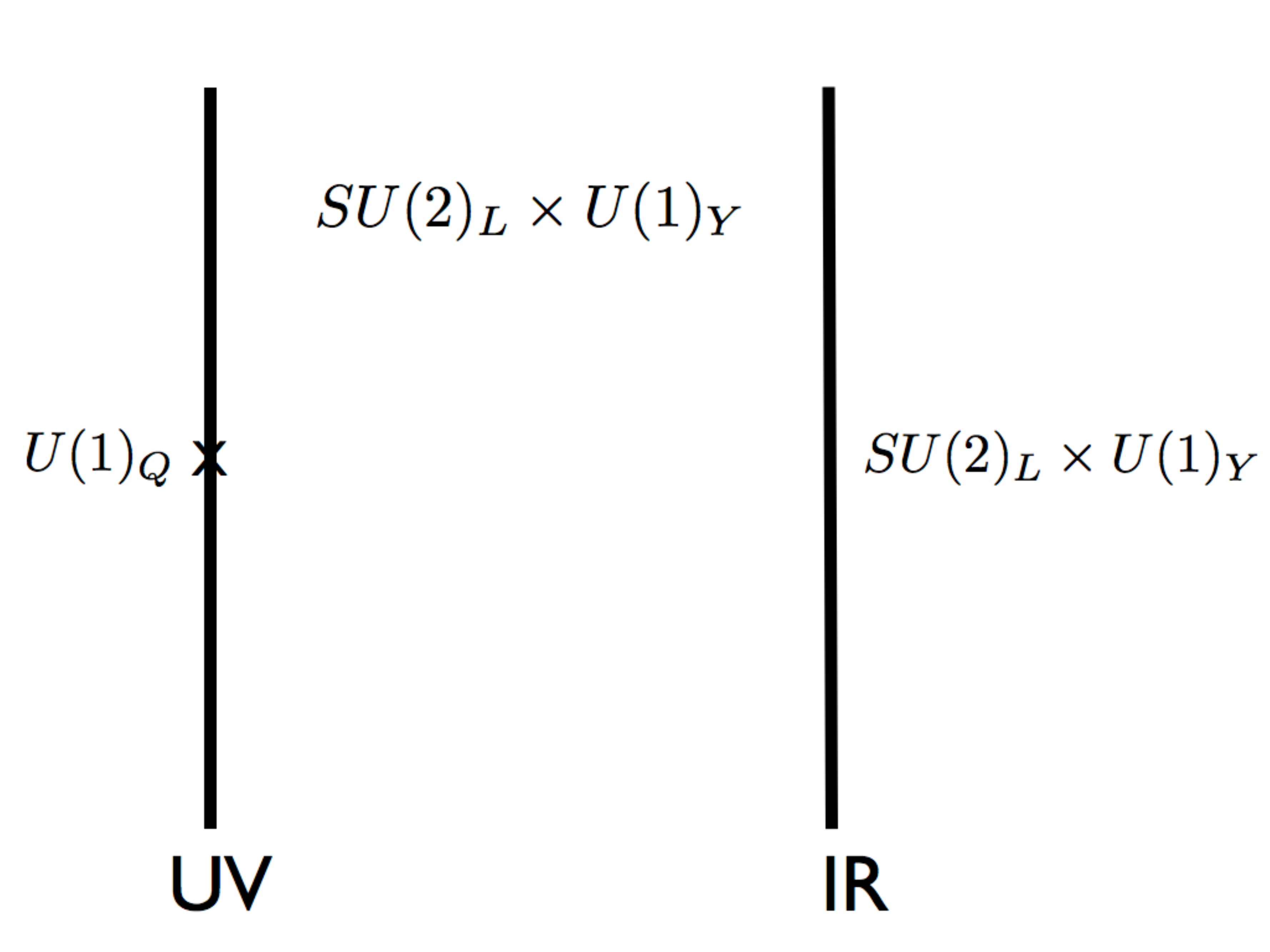}
\end{center}
\caption{\it A schematic diagram of the gauge symmetry breaking pattern
in the emergent electroweak symmetry breaking model.}
\label{emergentfig}
\end{figure}

It is straightforward to implement this scenario in the warped bulk~\cite{cgw}. Ignoring QCD, the 
minimal bulk gauge group required is $SU(2)_L\times U(1)_Y$. Since the IR brane is not responsible for 
electroweak symmetry breaking this symmetry is also preserved on the IR brane. However on the 
UV brane, boundary conditions are chosen to break this symmetry down to the electromagnetic 
gauge group $U(1)_Q$, as shown in Figure~\ref{emergentfig}. Note that the original massless gauge 
bosons decouple from the low-energy spectrum, and the usual massive $W,Z$ gauge bosons are 
identified with the first Kaluza-Klein states. The lowest-lying Kaluza-Klein modes can be sufficiently 
separated from the rest of the Kaluza-Klein tower, by adding appropriate boundary kinetic terms. 
The $W,Z$ boson profiles are localized towards the IR brane while the (massless) photon has its usual flat profile.

In the 4D dual description we see that the electromagnetic gauge group $U(1)_Q$ weakly
gauges the $SU(2)_L\times U(1)_Y$ global symmetry of the CFT. The $SU(2)_L$ plays the role
of a custodial symmetry in the sense that the $W^{1,2,3}$-bosons are degenerate in mass 
before electroweak mixing takes place. Therefore it is not surprising to find that the 
rho parameter, $\rho=1$. In addition since the $W,Z$ bosons are localized towards the IR brane,
they are identified as composite states of the CFT, while the photon is an elementary state. In this 
way we see that when conformal symmetry is broken, massive composite $W,Z$ bosons 
emerge at the IR scale. In other words electroweak symmetry breaking is an emergent phenomena.
Similar ideas invoking composite $W,Z$ bosons were previously considered in Refs.~\cite{compWZrefs}.

The 5D gravity description allows the $S$ parameter to be computed and it is found to be 
consistent with electroweak precision tests since the Kaluza-Klein mass scale is sufficiently heavy. 
In addition fermion masses can also be generated by allowing a common bulk profile with flavor 
dependent mass terms on the UV boundary~\cite{cgw,cgs}. 
Since the bulk gauge symmetry is simply the electroweak gauge group, this avoids having to introduce 
exotic fermion representations. Interestingly the nonuniversality of the gauge couplings 
is proportional to the square of the ratio of the fermion mass to IR scale, so that the light fermion gauge 
couplings are universal at the per mille level. Large deviations do occur for the top quark coupling and 
this causes some tension in the electroweak precision tests~\cite{cgs}.
Finally, note that the emergent model contains no Higgs boson and 
therefore is another example of a ``Higgsless" model, albeit radically different from technicolor 
or the warped Higgsless model since there is also no Higgs mechanism. This exemplifies the utility 
of the warped extra dimension in providing a general framework to study diverse models of electroweak 
symmetry breaking.

\section{Supersymmetric Models in Warped Space}

Supersymmetry elegantly solves the hierarchy problem because quadratic
divergences to the Higgs mass are automatically cancelled thereby
stabilising the electroweak sector. However this success must be tempered
with the fact that supersymmetry has to be broken in nature. 
In order to avoid reintroducing a fine-tuning in the Higgs mass, 
the soft mass scale cannot be much larger than the TeV scale. Hence
one needs an explanation for why the supersymmetry breaking scale is low.
Since in warped space hierarchies are easily generated, the warp factor 
can be used to explain the scale of supersymmetry breaking, instead of 
the scale of electroweak breaking. This is one motivation for studying 
supersymmetric models in warped space. Thus, new possibilities 
open up for supersymmetric model building, and in particular for the 
supersymmetry-breaking sector. Moreover by the AdS/CFT correspondence these 
new scenarios have an interesting blend of supersymmetry and compositeness 
that lead to phenomenological consequences at the LHC.

A second motivation arises from the fact that electroweak precision data
favours a light (compared to the TeV scale ) Higgs boson mass~\cite{higgs}. 
As noted earlier the Higgs boson mass in a generic warped model without any 
symmetry is near the IR cutoff (or from the 4D dual perspective the Higgs 
mass is near the compositeness scale). Introducing supersymmetry 
provides a simple reason for why the Higgs boson mass is light and below the 
IR cutoff of the theory.

\subsection{Supersymmetry in a Slice of AdS}
It is straightforward to incorporate supersymmetry in a slice of
AdS~\cite{gp,bagger}. The amount of supersymmetry allowed in five dimensions 
is determined by the dimension of the spinor representations.
In five dimensions only Dirac fermions are allowed by the Lorentz algebra,
so that there are eight supercharges which corresponds from the 4D
point of view to an ${\cal N}=2$ supersymmetry. This means that all bulk 
fields are in ${\cal N}=2$ representations. At the massless level only half 
of the supercharges remain and the orbifold breaks the bulk supersymmetry to 
an ${\cal N}=1$ supersymmetry. 

Consider an ${\cal N}=1$ (massless) chiral multiplet 
$(\phi^{(0)},\psi^{(0)})$ in the bulk. We have seen that the zero mode 
bulk profiles of $\phi^{(0)}$ 
and $\psi^{(0)}$ are parametrized by their bulk mass parameters $a$ and $c$, 
respectively. Since supersymmetry treats the scalar and fermion components 
equally, the bulk profiles of the component fields must be the same. 
It is clear that in general this is not the case except when $1\pm\sqrt{4+a} = 1/2 -c$ 
(assuming $c_L=-c_R\equiv c$), as follows from the exponent of the zero mode profiles 
in Section~\ref{profilesummary}. This leads to the condition that 
\begin{equation}
\label{acond}
   a=c^2+c-15/4~, 
\end{equation}
and the one remaining mass parameter $c$
determines the profile of the chiral multiplet to be
\begin{equation}
      \left(\begin{tabular}{c} $\phi^{(0)}$\\ $\psi^{(0)}$
            \end{tabular}\right) \propto e^{(\frac{1}{2}-c)k y}~.
            \label{cmprofile}
\end{equation}
Thus for $c > 1/2$ ($c< 1/2$) the chiral supermultiplet is localized 
towards the UV (IR) brane. It can be shown that the scalar boundary mass,
that was tuned to be $b=2-\alpha$, follows from the invariance under a 
supersymmetry transformation~\cite{gp} when (\ref{acond}) is satisfied.

Similarly a gauge boson with bulk profile $A_\mu^{(0)}(y) \propto 1$ 
and a gaugino with bulk profile 
$\lambda^{(0)}(y)\propto e^{(\frac{1}{2}-c_\lambda)ky}$ can be combined into
an ${\cal N}=1$ vector multiplet only for $c_\lambda = 1/2$. Of course
this means that the gaugino zero-mode profile is flat like the gauge boson.
At the massive level the on-shell field content of an ${\cal N}=2$ vector
multiplet is $(A_M, \lambda_i,\Sigma)$ where $\lambda_i$ is a 
symplectic-Majorana spinor (with $i=1,2$) and $\Sigma$ is a real scalar in 
the adjoint representation of the gauge group. Invariance under
supersymmetry transformations requires that $\Sigma$ have bulk and boundary
mass terms with $a=-4$ and $b=2$, respectively. So if $\Sigma$ is even
under the orbifold symmetry, then these values will ensure a scalar 
zero mode.

Finally, a graviton with bulk profile $h_{\mu\nu}^{(0)}(y)\propto e^{-k y}$ and 
a gravitino with bulk profile $\psi_\mu^{(0)}(y)\propto 
e^{(\frac{1}{2}-c_\psi)ky}$ can be combined into
an ${\cal N}=1$ gravity multiplet only for $c_\psi = 3/2$. In this case 
the gravitino zero-mode profile is localized on the UV brane.

\subsection{The Warped MSSM}

\begin{figure}
\begin{center}
\includegraphics[width=0.5\textwidth,height=0.25\textheight]{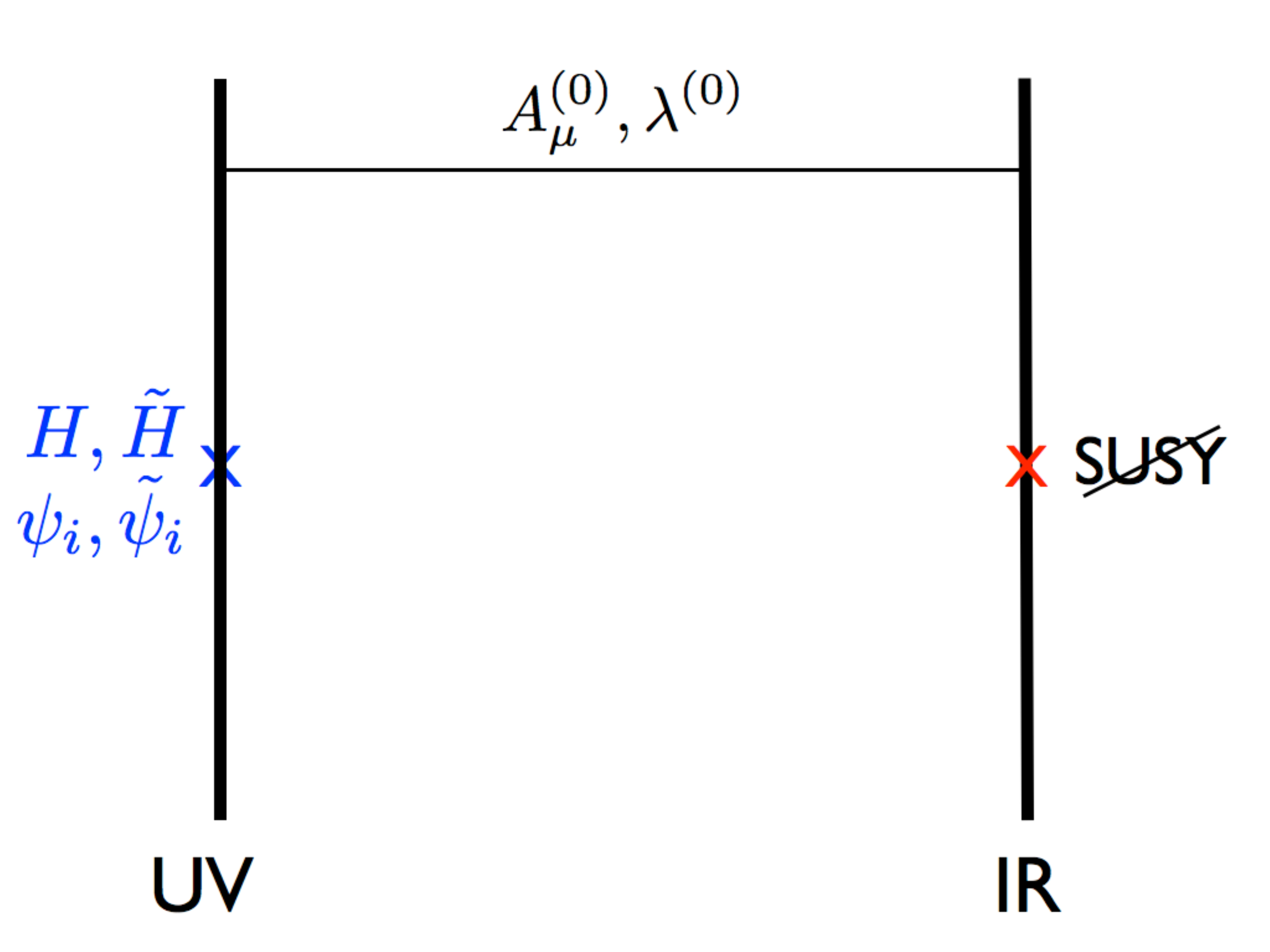}
\end{center}
\caption{\it A schematic diagram of the warped MSSM model. The matter
and Higgs superfields are confined to the UV brane.}
\label{wMSSMfig}
\end{figure}

In the warped MSSM the warp factor is used to naturally generate TeV scale 
soft masses~\cite{gp1}. The setup is depicted in Figure~\ref{wMSSMfig}. 
The UV (IR) scale is identified with the 
Planck (TeV) scale. The IR brane is associated with the scale of 
supersymmetry breaking, while the bulk and UV brane are supersymmetric. 
At the massless level the particle content is identical to the 
MSSM. The matter and Higgs superfields are assumed to be confined on the UV 
brane. This naturally ensures that all higher-dimension operators associated 
with proton decay and FCNC processes are sufficiently suppressed. In the bulk
there is only gravity and the Standard Model gauge fields. These are 
contained in an ${\cal N}=1$ gravity multiplet and vector multiplet,
respectively.

Supersymmetry is broken by choosing different IR brane boundary conditions 
between the bosonic and fermionic components of the bulk superfields. 
Instead of Neumann boundary conditions on the IR brane, the superpartners 
are chosen to have Dirichlet conditions. For example, if gauginos
have Dirichlet conditions, while the gauge bosons have Neumann conditions,
then supersymmetry will be broken.
The gaugino zero mode is no longer massless and receives a mass
\begin{equation}
\label{gaugem}
    m_\lambda \simeq \sqrt{\frac{2}{\pi kR}}~k\,e^{-\pi kR}~.
\end{equation}
Since the theory has a U(1)$_R$ symmetry this is actually a Dirac mass 
where the gaugino zero mode pairs up with a Kaluza-Klein mode~\cite{gp1}.
The Kaluza-Klein mass spectrum of the gauginos also shifts relative to 
that of the gauge bosons by an amount $-\frac{1}{4}\pi k e^{-\pi kR}$.
Similarly for the gravity multiplet the gravitino is assumed to
have Dirichlet boundary conditions on the IR brane, while the graviton has Neumann
boundary conditions. The gravitino zero mode then receives a mass
\begin{equation}
\label{gravm}
    m_{3/2} \simeq \sqrt{8}~k\,e^{-2\pi kR}~,
\end{equation}
while the Kaluza-Klein modes are again shifted by an amount similar to
that of the vector multiplet. 

Assuming $k e^{-\pi kR}=$ TeV then the gaugino mass (\ref{gaugem}) is
$m_\lambda \simeq 0.24$ TeV while the gravitino mass (\ref{gravm}) is
$m_{3/2}\simeq 3\times 10^{-3}$ eV. Even though both the gaugino and 
gravitino are bulk fields the difference in their supersymmetry breaking 
masses follows from their coupling to the IR brane, which is where
supersymmetry is broken. The gaugino is not localized in the bulk and 
couples to the IR brane with an ${\cal O}(1)$ coupling. Hence it receives 
a TeV scale mass. On the other hand the gravitino is localized on the UV 
brane and its coupling to the TeV brane is exponentially suppressed. 
This explains why the gravitino mass is much smaller than the gaugino mass.

\begin{figure}
\begin{center}
\includegraphics[width=0.35\textwidth,height=0.25\textheight]{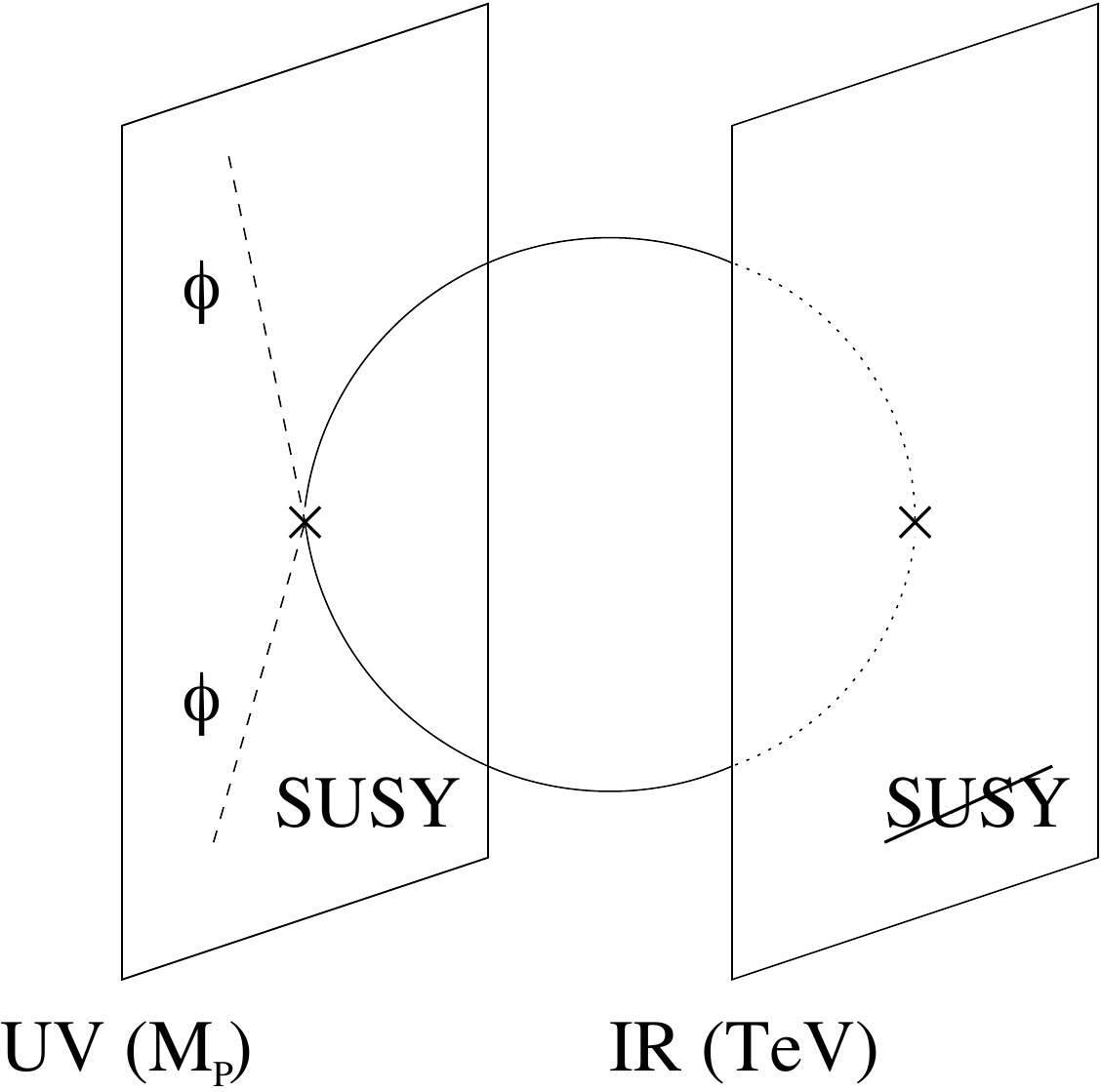}
\end{center}
\caption{\it The transmission of supersymmetry breaking in the warped MSSM 
to UV-brane localized matter fields via bulk gauge interactions which couple 
directly to the IR brane.}
\label{loopwMSSM}
\end{figure}

The scalars on the UV brane will obtain a supersymmetry breaking
mass at one loop via gauge interactions with the bulk vector multiplets. 
The gravity interactions with the gravity multiplet are negligible. 
A one-loop calculation leads to the soft mass spectrum
\begin{equation}
\label{softm}
     {\widetilde m}_j^2 \propto \frac{\alpha_i}{4\pi} ({\rm TeV})^2~,
\end{equation}
where $\alpha_i= g_i^2/(4\pi)$ are individual gauge contributions 
corresponding to the particular gauge quantum numbers of the particle state.
The exact expressions are given in Ref.~\cite{gp1}.
Unlike loop corrections to the usual 4D supersymmetric soft masses, 
the masses in (\ref{softm}) are finite. Normally UV divergences in a 
two-point function arise when the two spacetime points coincide. But 
the spacetime points in the 5D loop diagram can never coincide, because 
the two branes are assumed to be a fixed distance apart, and therefore the 
5D one-loop calculation leads to a finite result (see Figure~\ref{loopwMSSM}).
This is similar to the cancellation of divergences in the Casimir effect~\cite{casimir}. 
Since the contribution to the scalar masses is due to gauge interactions 
the scalar masses are naturally flavor diagonal. This means that the 
right-handed slepton is the lightest scalar particle since it has the smallest
gauge coupling dependence. The lightest supersymmetric particle
will be the superlight gravitino.

\subsubsection{The dual 4D interpretation}
We can use the AdS/CFT dictionary to obtain the dual 4D interpretation 
of the warped MSSM. Clearly the matter and Higgs fields confined
to the UV brane are elementary fields external to the CFT. This is also
true for the zero modes of the gravity multiplet since it is localized 
towards the UV brane. However, the bulk gauge field zero modes 
are partly composite since they are not localized. The Kaluza-Klein states,
which are bound states of the CFT and localized near the IR brane, 
do not respect supersymmetry.
Therefore at the TeV scale not only is conformal symmetry broken by the CFT
but also supersymmetry. This requires some (unknown) nontrivial IR dynamics
of the CFT, but the point is that supersymmetry is dynamically broken.
Since the CFT is charged under the Standard Model gauge group, the gauginos 
(and gravitinos) will receive a tree-level supersymmetry breaking mass,
while the squarks and sleptons will receive their soft mass at one loop.
In some sense this model is very similar to 4D gauge-mediated models
except that there is no messenger sector since the CFT, responsible for
supersymmetry breaking, is charged under the Standard Model gauge group.

In particular the bulk gaugino mass formula (\ref{gaugem}) can be understood 
in the dual theory. Since the gaugino mass is of the Dirac type the gaugino 
(source) field must marry a fermion bound state to become massive. This 
occurs from the mixing term ${\cal L} = \omega \lambda {\cal O}_\psi$.
Since $c_\lambda=1/2$, we have from Section~\ref{AdSCFTsummary} 
that dim$\,{\cal O}_\psi=5/2$ and therefore $\omega$ is dimensionless. 
This means that the mixing term coupling runs logarithmically so that at low 
energies the solution of (\ref{rge}) is
\begin{equation}
     \xi^2(\mu)\sim \frac{16\pi^2}{N \log\frac{k}{\mu}}~.
\end{equation}
Thus at $\mu=k e^{-\pi kR}$ we obtain the correct factor in (\ref{gaugem})
since the Dirac mass $m_\lambda \propto \xi\,
\langle 0| {\cal O}_\psi|\Psi\rangle$, where in the large-$N$ limit the 
matrix element for ${\cal O}_\psi$ to create a bound state fermion 
is $\langle 0| {\cal O}_\psi|\Psi\rangle \sim \sqrt{N}/(4\pi)$~\cite{wittenN}.

Thus, in summary we have the dual picture
\begin{equation}
\begin{tabular}{c}
    5D warped\\ MSSM
\end{tabular}
\begin{tabular}{c}
{\footnotesize\sc DUAL}\\[-2mm]
$\quad\Longleftrightarrow\quad$
\end{tabular}
\begin{tabular}{l}
4D MSSM $\oplus$ gravity \\
$\oplus$ strongly coupled 4D CFT
\end{tabular}
\end{equation}

\noindent
The warped MSSM is a very economical model of dynamical
supersymmetry breaking in which the soft mass spectrum is calculable and 
finite, and unlike the usual 4D gauge-mediated models does not require 
a messenger sector. The soft mass TeV scale is naturally explained and the
scalar masses are flavor diagonal. In addition, gauge coupling unification 
occurs with logarithmic running~\cite{pomarol} arising primarily from the 
elementary (supersymmetric) sector as in the usual 4D MSSM\cite{gns}.

\subsection{The ``Single-Sector" Supersymmetric Model}
In the warped MSSM the standard model fermions were all assumed to be confined 
to the UV (or Planck) brane. The warp factor was therefore only used to explain
the scale of supersymmetry breaking. However by placing the standard model fermions 
in the bulk the warp factor can also simultaneously address the fermion mass hierarchy.
This leads to so called ``single-sector" models of supersymmetry breaking~\cite{ggg}.

\begin{figure}
\begin{center}
\includegraphics[width=0.5\textwidth,height=0.25\textheight]{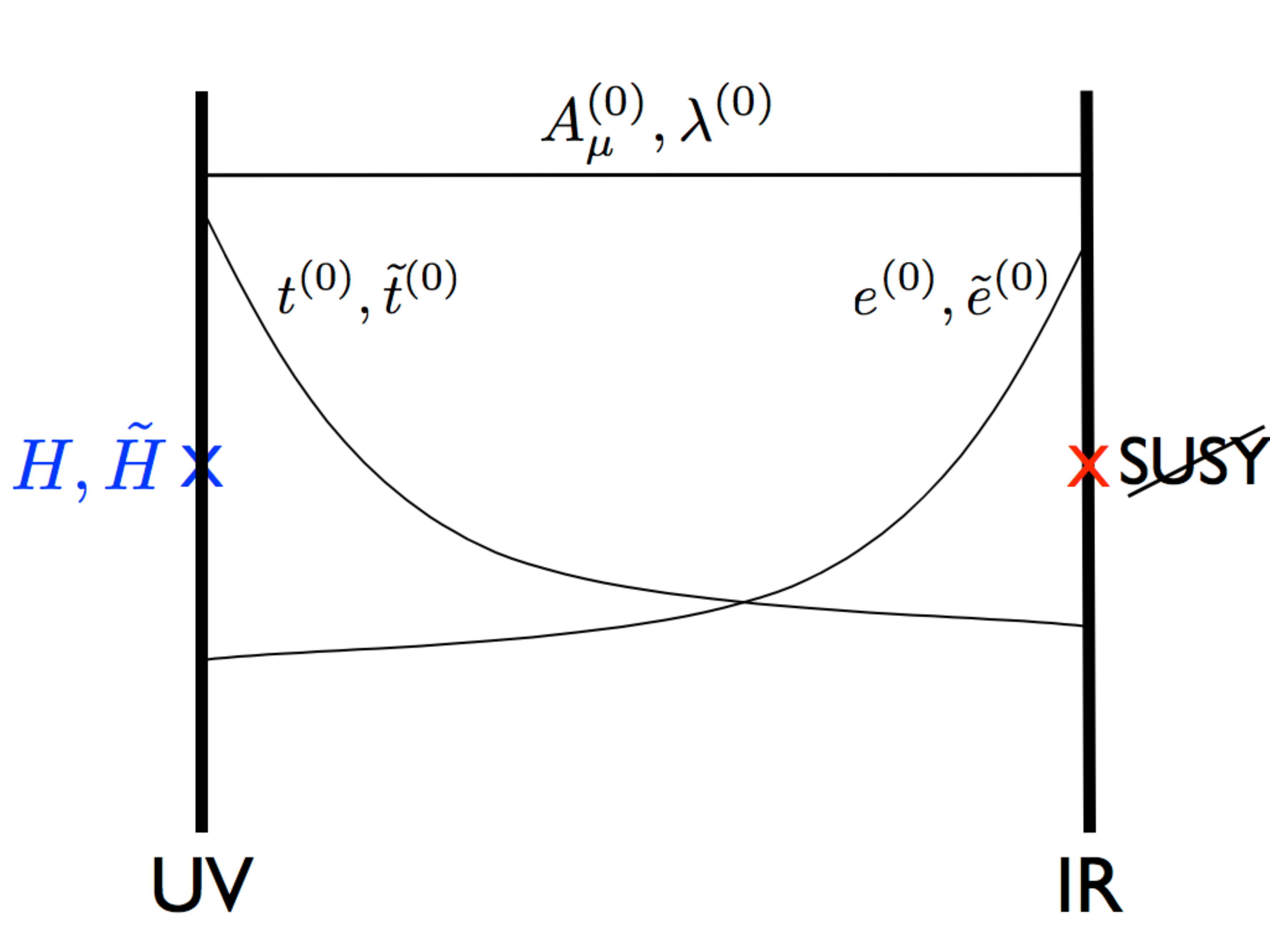}
\end{center}
\caption{\it A schematic diagram of the 5D gravity dual of single-sector supersymmetry 
breaking models. Note that the Higgs and top quark supermultiplets are now UV localized, 
while the light fermions, such as the electron, are IR localized.}
\label{singlesectorfig}
\end{figure}

Since the model is supersymmetric, the Higgs boson can now be confined to the UV 
brane and supersymmetry is used to solve the gauge hierarchy problem. The overlap 
of the bulk fermion wavefunctions with the UV localized Higgs boson can therefore be used
to generate the fermion mass hierarchy. In particular, the top quark must be localized near 
the UV brane, whereas the light fermions are localized near the IR brane. This setup
is depicted in Figure~\ref{singlesectorfig}. This fermion ``geography" is completely 
opposite to that encountered earlier when the Higgs boson was assumed to be confined 
to the IR brane. 

Furthermore, the fermions are part of chiral supermultiplets, and therefore their particular 
localization in the bulk also determines the corresponding localization of their superpartner, 
as shown in (\ref{cmprofile}). The IR brane is used to break supersymmetry and therefore 
superpartners which are IR localized will receive the largest soft masses. This leads to a 
distinctive soft-mass spectrum that is related to the fermion mass spectrum: light fermions 
have heavy superpartners, while heavy fermions have light superpartners.

In Ref.\cite{ggg}, inspired by flux-background solutions of type IIB supergravity, the IR brane 
was replaced by a metric background that deviates from AdS and softly breaks supersymmetry 
at the IR scale. For a fermion localized with a bulk mass parameter $c_i$, the corresponding 
scalar superpartner receives a mass~\cite{ggg}
\begin{equation}
         {\widetilde m_i} \propto 
\begin{cases}
 m_{IR} &    ${\rm for}$~c_i\leq \frac{1}{2}~, \\
e^{(\frac{1}{2}-c_i)\pi k R} \,m_{IR} &  ${\rm for}$~c_i>\frac{1}{2}~,
 \end{cases}
\end{equation}
where $m_{IR}\equiv k e^{-\pi k R}$. 
Thus we see that for light fermions localized near the IR brane $(c_i< \frac{1}{2})$,
the corresponding superpartners are much heavier than the stop, whose superpartner
the top quark must have $c_i>\frac{1}{2}$.

However it turns out that to explain the top quark mass, the corresponding stop masses 
are too light. A secondary source of supersymmetry breaking is needed to give tree-level 
gaugino masses which then generate a gauge-mediated contribution to the third generation squark 
and slepton masses. In Ref.~\cite{ggg} the IR scale was chosen to be $m_{IR}= 35$ TeV, which
leads to sparticle masses of order 10 TeV for the first two generations, while the gauginos
and third generation sparticles obtain masses of order the TeV scale.
The LSP is the gravitino, since as expected from its UV localized profile, receives a tiny
supersymmetry-breaking mass from the IR scale. This spectrum is similar to that considered in 
Refs.~\cite{Dimopoulos:1995mi, Pomarol:1995xc} and is also reminiscent of the ``more minimal" 
supersymmetric standard model~\cite{mmssm}, where heavy first two generation scalar fields 
were considered to ameliorate flavor problems. There are also flavor constraints 
which restrict the soft-mass spectrum and lead to restrictions on the range of $c_i$ 
values~\cite{ggg,sword}. The fermion mass hierarchy is therefore achieved from a combination 
of wavefunction overlap and mild hierarchies in the 5D Yukawa couplings.
Interestingly single-sector supersymmetry breaking models in warped throat backgrounds 
have been considered in Ref.~\cite{Benini:2009ff,Franco:2009wf,Craig:2009hf}.

\subsubsection{The dual 4D interpretation}
The AdS/CFT dictionary can be used to obtain the holographic dual description of this 
model. Since the Higgs boson and top quark are UV localized they are elementary states,
while the light fermions which are IR localized are composite states of the dual strong 
dynamics. Again the (unknown) dual strong dynamics is responsible for breaking
supersymmetry and conformal symmetry.

Interestingly, the 4D dual description is remarkably similar to models constructed directly 
in four dimensions.  In Refs.~\cite{Arkani-Hamed:1997fq,Luty:1998vr} models are 
explicitly constructed where, for example, the first two generations of the MSSM arise
as composite states $(P \bar U)$ of a strongly coupled gauge theory, with $P,\bar U$ 
charged under the confining gauge group (see Ref.~\cite{Luty:1998vr}).
The fields $\bar U$ acquire large $F$-terms, so that the composites $(P \bar U)$
directly feel the supersymmetry breaking.  The first and second generation scalars obtain large 
masses, whereas the fermion composites remain massless due to chiral symmetries.  
The $\bar U$ fields also carry Standard Model charges, and therefore they communicate 
supersymmetry breaking to the rest of the MSSM through gauge mediation. Since 
supersymmetry breaking is directly transmitted to the MSSM without invoking a messenger 
sector, these models are referred to as ``single-sector" models of supersymmetry breaking.

Thus, the dual picture can be summarized as follows
\begin{equation}
\begin{tabular}{c}
    5D warped\\ ``single-sector" model
    \end{tabular}
\begin{tabular}{c}
{\footnotesize\sc DUAL}\\[-2mm]
$\quad\Longleftrightarrow\quad$
\end{tabular}
\begin{tabular}{l}
4D ``more-minimal" SSM $\oplus$ gravity\\
$\oplus$ composite 1st, 2nd generations\\
 $\oplus$ strongly-coupled 4D CFT
\end{tabular}
\end{equation}

\subsection{The Partly Supersymmetric Standard Model}

Besides solving the hierarchy problem the supersymmetric standard 
model has two added bonuses. First, it successfully 
predicts gauge coupling unification and second, it 
provides a suitable dark matter candidate. Generically, however, there 
are FCNC and CP violation problems arising from the soft mass Lagrangian, 
as well as the gravitino and moduli problems in cosmology~\cite{cosmology}. 
These problems stem from the fact that the soft masses are of order 
the TeV scale, as required for a natural solution to the hierarchy problem.
Of course clever mechanisms exist that avoid these problems but 
perhaps the simplest solution would be to have all scalar masses at the 
Planck scale while still naturally solving the hierarchy problem. In the 
partly supersymmetric standard model~\cite{gp2} this is precisely what 
happens while still preserving the successes of the MSSM.

\begin{figure}
\begin{center}
\includegraphics[width=0.5\textwidth,height=0.25\textheight]{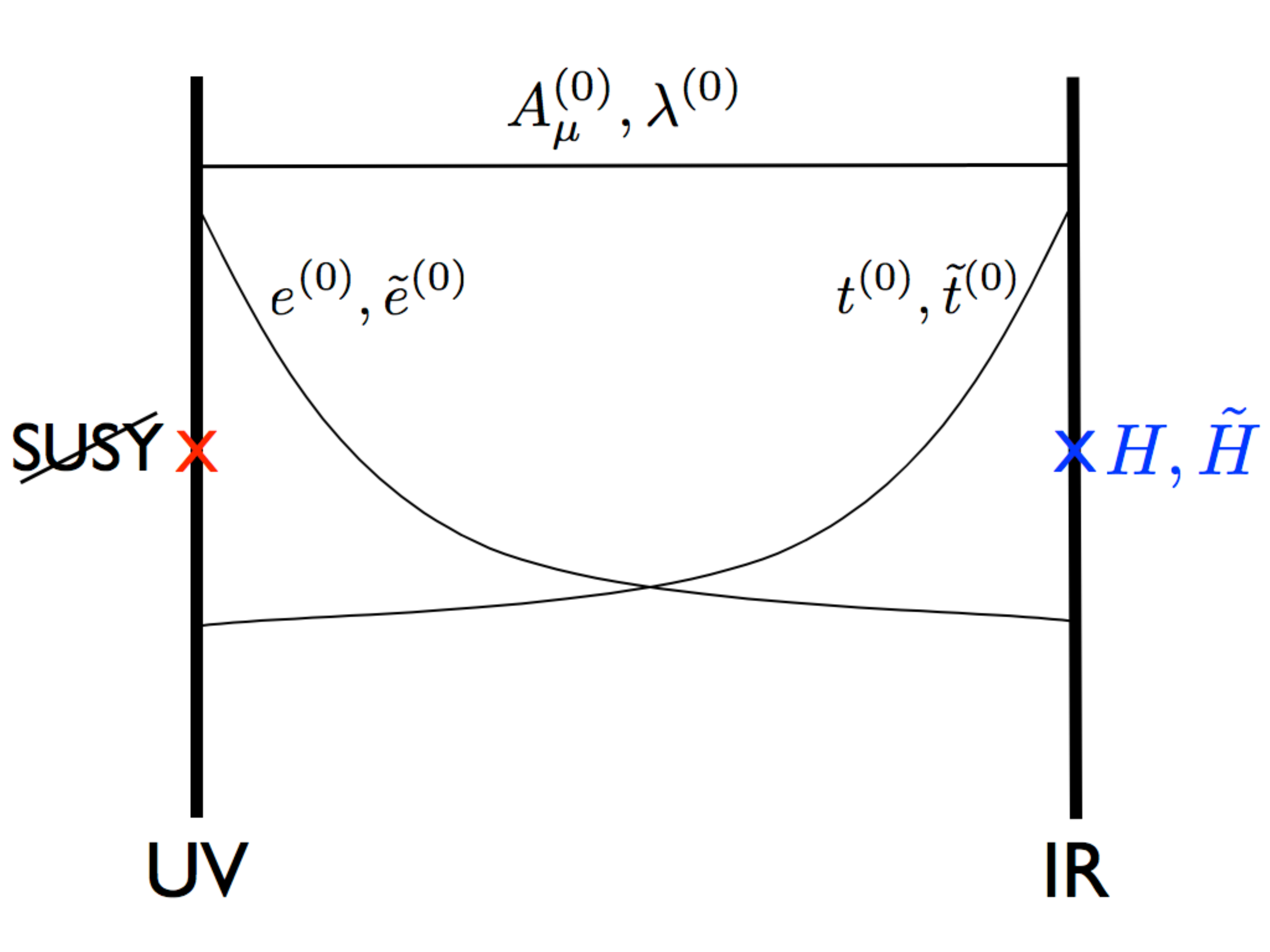}
\end{center}
\caption{\it A schematic diagram of the partly supersymmetric standard model.}
\label{partialsusyfig}
\end{figure}

In 5D warped space the setup of the model is as follows. 
Supersymmetry is assumed to be broken on the UV brane while it is 
preserved in the bulk and the IR brane. The vector, matter, and
gravity superfields are in the bulk while the Higgs superfield is 
confined to the IR brane (see Figure~\ref{partialsusyfig}).
On the UV brane the supersymmetry breaking
can be parametrized by a spurion field $\eta=\theta^2 F$, where
$F\sim M_P^2$. In the gauge sector
we can add the following UV brane term
\begin{equation}
      \int d^2\theta~\frac{\eta}{M_P^2}\frac{1}{g_5^2}W^\alpha W_\alpha
     \delta(y) + h.c.
\end{equation} 
This term leads to a gaugino mass for the zero mode $m_\lambda\sim M_P$, 
so that the gaugino decouples from the low energy spectrum. The gravitino
also receives a Planck scale mass via a UV brane coupling 
and decouples from the low energy spectrum~\cite{luty}. Similarly
a supersymmetry breaking mass term for the squarks and sleptons can be added
to the UV brane
\begin{equation}
     \int d^4\theta~\frac{\eta^\dagger\eta}{M_P^4}~k~S^\dagger S~\delta(y)~,
\end{equation}
where $S$ denotes a squark or slepton superfield. This leads to a soft 
scalar mass ${\widetilde m}\sim M_P$, so that the squark and slepton zero 
modes also decouple from the low energy spectrum. 

\begin{figure}
\begin{center}
\includegraphics[width=0.4\textwidth,height=0.28\textheight]{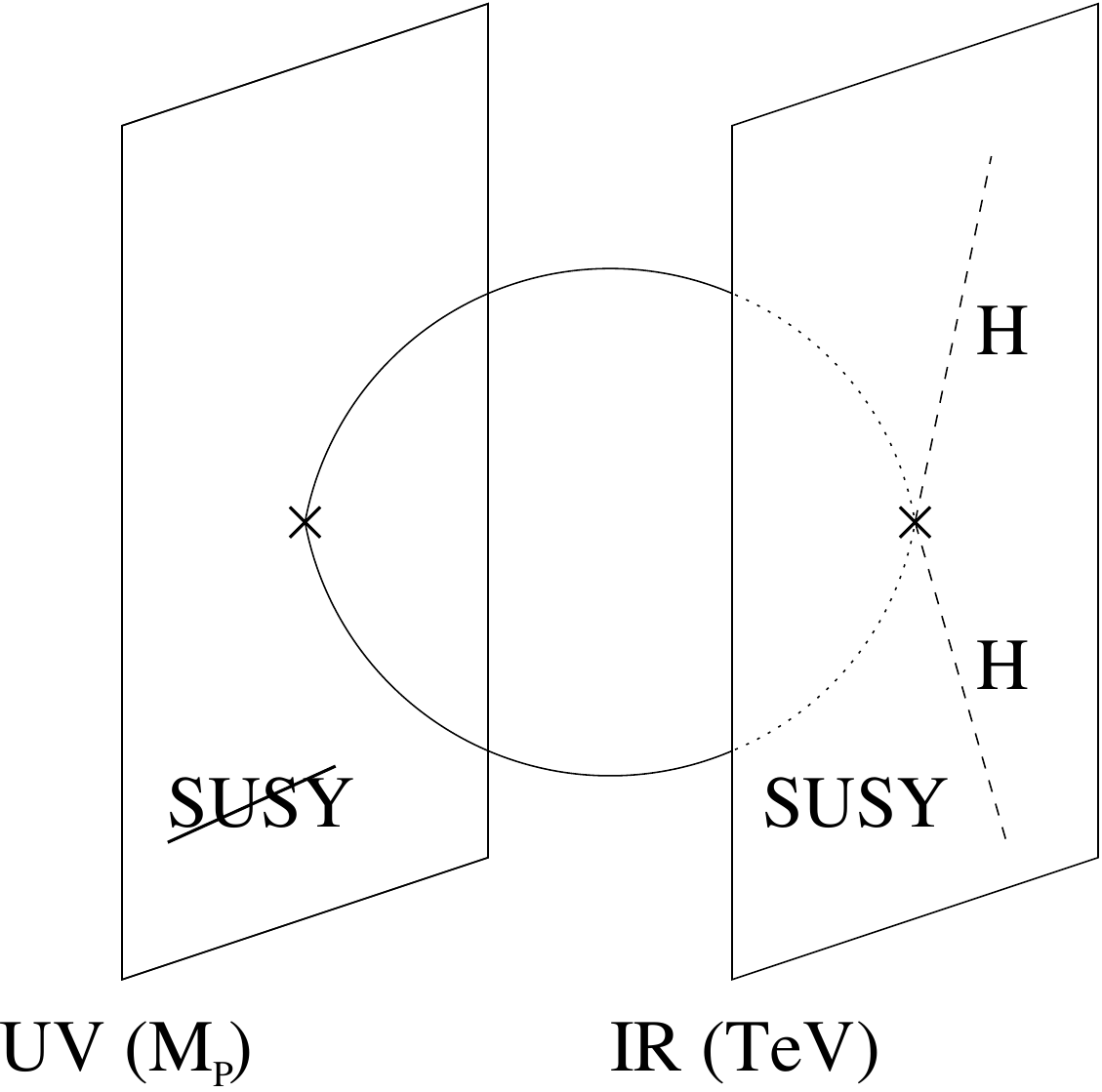}
\end{center}
\caption{\it The transmission of supersymmetry breaking in the partly
supersymmetric standard model to the supersymmetric Higgs sector via bulk 
gauge interactions which couple directly to the UV brane.}
\label{looppartialsusyfig}
\end{figure}

The Higgs sector is different because the Higgs lives on the IR
brane and there is no direct coupling to the UV brane. Hence, at tree-level
the Higgs mass is zero, but a (finite) soft Higgs mass will be induced at 
one loop via the gauge interactions in the bulk of order
\begin{equation}
     m_H^2 \sim \frac{\alpha}{4\pi} (k e^{-\pi k R})^2 \ll M_P^2~.
\end{equation}
As noted earlier the finiteness is due to the fact that the two 5D spacetime 
points on the UV and IR branes can never coincide (see Figure~\ref{looppartialsusyfig}).
Thus, we see that because of the warp factor the induced Higgs soft mass is 
much smaller than the scale of supersymmetry breaking at the Planck scale. 
So while at the massless level the gauginos, squarks and sleptons have received 
Planck scale masses, the Higgs sector remains (approximately) supersymmetric. 
In summary, at the massless level the particle spectrum consists of the Standard 
Model gauge fields and matter (quarks and leptons) plus a Higgs scalar and Higgsino. 
This is why the model is referred to as {\it partly} supersymmetric. 

At the massive level the Kaluza-Klein modes are also approximately
supersymmetric. This is because they are localized towards the IR brane 
and have a small coupling to the UV brane. So the Planck scale 
supersymmetry breaking translates into an order TeV scale splitting
between the fermionic and bosonic components of the Kaluza-Klein 
superfields. 

Given that there are no gauginos, squarks or sleptons in the low energy 
spectrum it may seem puzzling how the quadratic divergences cancel in this
model. Normally in the supersymmetric standard model the quadratic 
divergences in the Higgs mass are cancelled by a superpartner contribution 
of the opposite sign. However in the partly supersymmetric standard model
there are no superpartners at the massless level. Instead what happens is
that the difference between the Kaluza-Klein fermions and bosons sums
up to cancel the zero mode quadratic divergence. 
Thus, the Kaluza-Klein tower is responsible for keeping the Higgs mass 
natural even though supersymmetry is broken at the Planck scale.

\subsubsection{Higgs sector possibilities}
The motivation for making the Higgs sector supersymmetric is that
the Higgs mass is induced at loop level and therefore the Higgs mass is
naturally suppressed below the IR cutoff. In addition the supersymmetric
partner of the Higgs, the Higgsino, provides a suitable dark matter 
candidate~\cite{gp2,mm}. However since the Higgsino is a fermion,
gauge anomalies could be generated and these must be cancelled.
This leads us to consider the following three possibilities:

{\bf (i) Two Higgs doublets:}
As in the MSSM we can introduce two Higgs doublet superfields $H_1$ and 
$H_2$, so that the gauge anomaly from the two Higgsinos cancel amongst 
themselves. In this scenario we can add the following superpotential on 
the IR brane
\begin{equation}
\label{twoHW}
     \int d^2\theta \left( y_d H_1 Q d + y_u H_2 Q u 
         + y_e H_1 L e +\mu H_1 H_2 \right)~.
\end{equation}
Thus the quarks and leptons receive their masses in the usual way.
In addition the $\mu$ term in (\ref{twoHW}) is naturally of order the 
TeV scale so that there is no $\mu$ problem. The IR brane is approximately
supersymmetric and the supersymmetric mass $\mu$ has a  natural
TeV value. This is unlike the MSSM where the natural scale of $\mu$
is $M_P$ and leads to phenomenological problems.

{\bf (ii) One Higgs doublet:}
At first this possibility seems to be ruled out since one massless Higgsino
gives rise to a gauge anomaly. However starting with a bulk Higgs
${\cal N}=2$ hypermultiplet $H=(H_1,H_2)$ with bulk mass parameter $c_H=1/2$ 
that consists of two ${\cal N}=1$ chiral
multiplets $H_{1,2}$ we can generate a Higgsino Dirac mass and only
one Higgs scalar doublet in the low energy spectrum. The trick is to use
mixed boundary conditions where $H_1$ has Neumann (Dirichlet) boundary
conditions on the UV (IR) brane and vice versa for $H_2$. This leads
to a $\mu$ term
\begin{equation}
    \mu\simeq \sqrt{\frac{2}{\pi kR}}~k~e^{-\pi kR}~,
\end{equation}
which is similar to the gaugino mass term (\ref{gaugem}) obtained in the 
warped MSSM.
In this case the $\mu$-term is naturally suppressed below the TeV scale by
the factor $1/\sqrt{\pi kR}$. Only one Higgs scalar remains in the low 
energy spectrum because the twisted boundary conditions localize
one Higgs scalar doublet towards the UV brane where it obtains a Planck 
scale mass, and the other Higgs scalar is localized towards the IR brane
where it obtains a mass squared $\mu^2$. 

{\bf (iii) No Higgs doublet--Higgs as a Slepton:}
No anomalies will occur if the Higgs is considered to be the 
superpartner of the tau (or other lepton). This idea is not
new and dates back to the early days of supersymmetry~\cite{fayet}.
The major obstacle in implementing this possibility in the MSSM is that
the gauginos induce an effective operator $\frac{g^2}{m_\lambda}\nu\nu h h$
that leads to neutrino masses of order 10 GeV which are experimentally
ruled out. However in the partly supersymmetric model $m_\lambda \sim M_P$ 
and neutrinos masses are typically of order $10^{-5}$ eV. This
is phenomenologically acceptable and at least makes this a viable possibility.
However the stumbling block is to generate a realistic 
spectrum of fermion masses without introducing abnormally large
coefficients~\cite{gp2}.

\subsubsection{Electroweak symmetry breaking}
In this model electroweak symmetry breaking can be studied and calculated
using the 5D bulk propagators. Consider, for simplicity, a one Higgs doublet
version of the model. The scalar potential is 
\begin{equation}
\label{higgspot}
    V(h) =\mu^2 |h|^2 + \frac{1}{8} (g^2+g'^2) |h|^4 + V_{gauge}(h) 
    + V_{top}(h)~,
\end{equation}
where $V_{gauge}(h)$ and $V_{top}(h)$ are one-loop contributions to the
effective potential arising from gauge boson and top quark loops, 
respectively. The first two terms in (\ref{higgspot}), which 
arise at tree-level, are monotonically increasing giving rise to a
minimum at $\langle |h| \rangle = 0$ and therefore do not break 
electroweak symmetry. This is why we need to calculate the one-loop 
contributions. The one-loop gauge contribution is given by
\begin{equation}
   V_{gauge}(h) = 6\int_0^\infty \frac{dp}{8\pi^2}~p^3~\log
    \left[\frac{1+ g^2 |h|^2 G_B(p)}{1+ g^2 |h|^2 G_F(p)}\right]~,
\end{equation}
where $G_{B,F}(p)$ are the boson (fermion) gauge propagators in the bulk 
whose expressions can be found in Ref.~\cite{gp2}. The contribution to
the effective potential from $V_{gauge}$ is again monotonically increasing.
However there is also a sizeable contribution from top quark loops (due to
the large top Yukawa coupling) given by
\begin{equation}
   V_{top}(h) = 6\int_0^\infty \frac{dp}{8\pi^2}~p^3~\log
    \left[\frac{1+ p^2 y_t^2 |h|^2 G_B^2(p)}{1+ p^2 y_t^2 |h|^2 G_F(p)}
    \right]~.
\end{equation}
This contribution generates a potential that monotonically decreases with
$|h|$, destabilising the vacuum and thus triggering electroweak symmetry 
breaking. In order for this to occur the top quark needs to be
localized near the IR brane with a bulk mass parameter $c_t \simeq -0.5$.
Since the top quark ${\cal N}=1$ chiral multiplet is localized near the 
IR brane, the top squark will only receive a TeV scale soft mass and
consequently will remain in the low energy spectrum. In fact this radiative
breaking of electroweak symmetry due to a large top Yukawa coupling 
is similar to that occuring in the usual MSSM. As in the MSSM the value of
the Higgs mass is very model dependent but if no large tuning of parameters 
is imposed one obtains a light Higgs boson with mass 
$m_{Higgs}\lesssim 120$ GeV.

Note, however that the partly supersymmetric model can be improved to
more completely address the little hierarchy between the IR scale and the electroweak scale.
This involves extensions~\cite{ArkaniHamed:2004fb,Sundrum:2009gv} of the partly 
supersymmetric model that allow for the possibility of a light gaugino which helps to suppress
all the dominant Higgs radiative corrections in the little hierarchy. 
In addition relevant deformations of the UV theory caused by D-terms on the UV boundary
can also be avoided by embedding the SM gauge group into a non-Abelian gauge group
or the use of exact discrete symmetries such as charge-conjugation invariance~\cite{Sundrum:2009gv}.
Furthermore stable non-supersymmetric warped throats have been 
constructed in Ref.~\cite{Kachru:2009kg}, suggesting that the idea of partial supersymmetry can be realized
in string theory.

\subsubsection{Dual 4D interpretation}
The dual 4D interpretation of the partly supersymmetric model follows
from applying the rules of the AdS/CFT dictionary. 
Supersymmetry is broken at the Planck scale in the dual 4D theory
and is approximately supersymmetric at the IR scale. 
Thus from a 4D point of view supersymmetry is really 
just an accidental (or emergent) symmetry at low energies. At the massless level the 
Higgs is confined on the IR brane and the top quark is localized towards 
the TeV brane so both of these states are CFT composites and supersymmetric
at tree level. The compositeness of the Higgs and stop explains why these 
states are not sensitive to the UV breaking of supersymmetry. These states 
are ``fat'' with a size of order TeV$^{-1}$, and are transparent to
high momenta or short wavelength probes that transmit the breaking of 
supersymmetry. At one loop TeV-scale supersymmetry 
breaking effects arise from the small mixing with the elementary source 
fields, which directly feel the Planck scale supersymmetry breaking. The bulk 
gauge fields are partly composite and the light fermions which are localized 
to varying degrees near the UV brane are predominantly elementary fields. 
Since the light fermion superpartners are predominantly source fields they 
obtain Planck scale soft masses. 

Thus the dual picture can be summarized as follows
\begin{equation}
\begin{tabular}{c}
    5D partly\\ supersymmetric SM
\end{tabular}
\begin{tabular}{c}
{\footnotesize\sc DUAL}\\[-2mm]
$\quad\Longleftrightarrow\quad$
\end{tabular}
\begin{tabular}{l}
4D SM $\oplus$ gravity \\
$\oplus$ composite Higgsino, stop\\
$\oplus$ strongly coupled 4D CFT
\end{tabular}
\end{equation}

\noindent
The partly supersymmetric standard model is a natural model
of high-scale supersymmetry breaking. Supersymmetry is realized in the
most economical way. Only the Higgs sector and top quark are supersymmetric
and composite, while all other squarks and sleptons have Planck scale masses. 
The Higgsino is the dark matter candidate and even gauge 
coupling unification is achieved~\cite{psgu}.

\section{Conclusion}\label{conclusion}
Warped models in a slice of AdS$_5$ provide a new framework 
to simultaneously address the gauge hierarchy problem and fermion mass
hierarchies in the Standard Model. The warp factor can be used to either 
stabilize the electroweak scale in a nonsupersymmetric way or instead 
explain why the scale of supersymmetry breaking is near the TeV scale. 

Remarkably, by the AdS/CFT correspondence, 5D warped models are 
dual to strongly-coupled 4D gauge theories. The Higgs localized on the IR brane 
is dual to a composite Higgs. The corresponding Higgs boson mass can be light,
compared to the IR cutoff, by using either a global symmetry and treating 
the Higgs as a pseudo Nambu-Goldstone boson or using supersymmetry to make
only the Higgs sector supersymmetric. Alternatively, Higgsless models
can be constructed representing dual 5D models of technicolor, or even 
emergent models where electroweak gauge symmetry is not even fundamental. 

The good news is that these models are testable at the LHC (and an eventual 
linear collider), so it will be an exciting time to discover whether Nature makes 
use of the fifth dimension in this novel way. If not, there is no bad news, because 
the warped fifth dimension literally provides a new theoretical framework for 
studying the dynamics of strongly-coupled 4D gauge theories and this will be an 
invaluable tool for many years to come.

\section*{Acknowledgements}
I would like to thank Csaba Cs\'aki and Scott Dodelson, the TASI 2009 organizers,
for the kind invitation to present these lectures, the University of Colorado staff for 
their support and hospitality as well as the students for their enthusiastic participation 
and contribution to the stimulating atmosphere of the school. I am especially grateful 
to Alex Pomarol for collaboration on the topics which formed the basis of these lectures. 
I also thank Kaustubh Agashe, Brian Batell, Roberto Contino, Yanou Cui, Joel Giedt, 
Yasunori Nomura, Erich Poppitz and Raman Sundrum for discussions on topics covered 
in these lectures. I also thank the Masters students in the School of Physics at the University 
of Melbourne who read through a preliminary version of this manuscript. This work was 
supported by the Australian Research Council.


\end{document}